\documentclass[11pt]{amsart}

\usepackage{amsmath,amssymb,amsthm,amscd,mathrsfs}
\usepackage{latexsym,amsmath,amssymb,graphicx}
\usepackage[all]{xy} 
\usepackage{multirow}
\usepackage{color}
\usepackage{epsf}
\usepackage{youngtab}
\xyoption{v2} \xyoption{2cell} \xyoption{dvips}
\CompileMatrices \numberwithin{equation}{section}

\numberwithin{equation}{section}

\newcommand{\be}{\begin{equation}}
\newcommand{\ee}{\end{equation}}

\newcommand{\IP}{\mathbb{P}}

\newcommand{\wC}{{\widetilde C}}

\newcommand\IZ{\mathbb {Z}}

\newcommand\IQ{\mathbb {Q}}
\newcommand{\IC}{\mathbb{C}}

\newcommand{\IR}{\mathbb{R}}

\newcommand{\ba}{\begin{array}}
\newcommand{\ea}{\end{array}}

\newcommand{\wH}{{\widetilde H}}

\newcommand{\CX}{{\mathcal X}}

\newcommand{\CB}{{\mathcal B}}
\newcommand{\CK}{{\mathcal K}}

\newcommand{\cf}{{\mathcal F}}

\newcommand{\bal}{\begin{aligned}}
\newcommand{\eal}{\end{aligned}}

\newcommand{\wcV}{{\widetilde {\mathcal V}}}
\newcommand{\wcH}{{\widetilde {\mathcal H}}}

\newcommand{\wX}{{\widetilde X}}

\newcommand{\rk}{{\rm{rk}}}

\newcommand{\longto}{\longrightarrow}

\newcommand{\ch}{{\mathrm{ch}}}

\newcommand{\CO}{{\mathcal O}}

\newcommand{\CH}{{\mathcal H}}

\newcommand{\CM}{{\mathcal M}}

\newcommand{\wCB}{{\widetilde {\mathcal B}}}

\newcommand{\wS}{{\widetilde S}}

\newcommand{\wT}{{\widetilde T}}

\newcommand{\wCD}{{\widetilde{\mathcal D}}}

\newcommand{\PT}{\operatorname{PT}}
\newcommand{\pP}{\operatorname{PT}}

\CompileMatrices \LaTeXdiagrams \UseAllTwocells
\newdimen\tableauside\tableauside=1.0ex
\newdimen\tableaurule\tableaurule=0.4pt
\newdimen\tableaustep
\def\phantomhrule#1{\hbox{\vbox to0pt{\hrule height\tableaurule width#1\vss}}}
\def\phantomvrule#1{\vbox{\hbox to0pt{\vrule width\tableaurule height#1\hss}}}
\def\sqr{\vbox{%
  \phantomhrule\tableaustep
  \hbox{\phantomvrule\tableaustep\kern\tableaustep\phantomvrule\tableaustep}%
  \hbox{\vbox{\phantomhrule\tableauside}\kern-\tableaurule}}}
\def\squares#1{\hbox{\count0=#1\noindent\loop\sqr
  \advance\count0 by-1 \ifnum\count0>0\repeat}}
\def\tableau#1{\vcenter{\offinterlineskip
  \tableaustep=\tableauside\advance\tableaustep by-\tableaurule
  \kern\normallineskip\hbox
    {\kern\normallineskip\vbox
      {\gettableau#1 0 }%
     \kern\normallineskip\kern\tableaurule}%
  \kern\normallineskip\kern\tableaurule}}
\def\gettableau#1 {\ifnum#1=0\let\next=\null\else
  \squares{#1}\let\next=\gettableau\fi\next}
\begin{document}

\title{Vertical D4-D2-D0 bound states on K3 fibrations and modularity}
\author[V. Bouchard, T. Creutzig, D.-E. Diaconescu, C. Doran, C. Quigley, A. Sheshmani]{Vincent Bouchard${}^1$, 
Thomas Creutzig${}^1$, Duiliu-Emanuel Diaconescu${}^2$, Charles Doran$^{1,3}$, Callum Quigley${}^{1,4}$,  Artan Sheshmani${}^{5,6}$}
\address{${}^1$ Department of Mathematical and Statistical Sciences, University of Alberta,
Edmonton, AB, Canada}
\address{${}^2$ NHETC, Rutgers University,
Piscataway, NJ 08854-0849 USA}
\address{${}^3$ Physics Department, University of Maryland, College Park, MD, USA}
\address{${}^4$ Department of Physics, University of Toronto, Toronto, ON, Canada}
\address{${}^5$ Department of Mathematics, Ohio State University, Columbus, OH, USA}
\address{${}^6$ Kavli Institute for the Physics and
Mathematics of the Universe (WPI),The University of Tokyo Institute for Advanced
Study, The University of Tokyo, Kashiwa, Chiba 277-8583, Japan}
\date{}
\maketitle

\begin{abstract} An explicit formula is derived for the generating function of vertical D4-D2-D0 bound states on smooth K3 fibered 
Calabi-Yau threefolds, generalizing previous results of 
Gholampour and Sheshmani.  It is also shown that this formula satisfies 
strong modularity properties, as predicted by string theory. 
This leads to a new construction of vector valued modular forms which exhibits some of the features of a generalized Hecke transform. 
\end{abstract}

\tableofcontents

\section{Introduction}\label{intro}

The main goal of this paper is to derive an
explicit formula for 
all BPS degeneracies of vertical D4-D2-D0 bound states on 
K3 fibered Calabi-Yau threefolds $X$, subject to certain genericity 
conditions. Counting 
D4-D2-D0 BPS states is a natural problem in the context of IIA and M-theory compactifications, as well as  IIA/heterotic duality. This problem 
is closely related to black hole entropy
\cite{DH,Vafa:1995zh, Vafa:1997gr,Maldacena:1997de, Ooguri:2004zv, prec_counting,Denef:2007vg}, as well as BPS algebras 
\cite{BPS_strings,Alg_BPS}. Further ramifications include connections 
with quantum two dimensional Yang-Mills 
theory and topological strings  \cite{Vafa:2004qa,Aganagic:2004js,Aganagic:2005wn,Jafferis:2007ti}, 
as well as wallcrossing 
phenomena \cite{Nishinaka:2011nn,Nishinaka:2011is,Nishinaka:2013mba}. 
As pointed out in \cite{Vafa:1997gr, E_strings, Quantum_elliptic} 
and studied in more depth in \cite{vertical}, a different connection 
with topological strings is provided by the Fourier-Mukai transform 
on elliptic fibrations. 
This problem is also firmly  rooted in gauge theory, 
since it can be regarded as a string theoretic generalization of
Vafa-Witten theory \cite{VW} on a K3 surface. 
Exact results for this case have been obtained in 
\cite{E_strings, VW_SUN}. 
At the same time, this problem has received a fair amount of attention in the mathematical literature \cite{DT_twodim, Gen_DT_twodim, Toda_flops, G_S_Toda, Toda_An, Toda_Ptwo}, in the framework of 
Dolandson-Thomas invariants of stable sheaves on Calabi-Yau threefolds

As shown in \cite{VW,M5_genus,Farey_tail,Denef:2007vg}, on physics grounds the generating function for D4-D2-D0 
BPS indices is expected to have strong modularity properties. 
This has been confirmed by many explicit computations for stable torsion-free sheaves supported on rational 
surfaces \cite{Betti_Hilbert, KY_betti_I, KY_betti_II, KY_S_duality, KY_euler, GL_theta, GNY, JM_BPS, JM_rank_three, JM_rational,GJK,Toda_An,Toda_Ptwo}. These results can be placed in the context of Donaldson-Thomas theory 
if such sheaves are viewed as two dimensional torsion modules 
supported on rigid divisors in Calabi-Yau threefolds. 
For two dimensional sheaves supported on the fibers of a K3 pencil, 
modularity has been shown in  certain cases in \cite{DT_twodim}, as discussed in more detail below.

\subsection{Partition functions} 
The geometric framework for this paper is that of lattice polarized 
K3 fibered projective Calabi-Yau threefolds satisfying certain natural genericity assumptions. 
In particular all fibers are assumed to be irreducible 
algebraic K3 surfaces with at most simple nodal singularities. 
In addition, one also assumes the existence of a lattice 
polarization defined as in \cite{GW_NL,NL_YZ} in terms of an auxiliary smooth K3 fibration obtained by extremal transitions. 
Very briefly, a lattice polarization is defined 
by specifying a sublattice $\Lambda$ of the Picard group of 
$X$ satisfying certain conditions listed in Section 
\ref{setup}. This is a sublattice of the middle cohomology lattice 
$H^2(S,\IZ)$ of a smooth generic K3 surface such that 
the restriction of the intersection form to $\Lambda$ 
is a non-degenerate bilinear form of signature $(1, {\rk \Lambda} -1)$.  
Moreover, for the class of threefolds under consideration,  $\Lambda$ will be naturally identified with 
the Picard  lattice of any sufficiently generic smooth K3 fiber of $X$. As explained in Section \ref{verticalKthree}, Poincar\'e duality then leads to a natural identification of the lattice of vertical 
curve classes on $X$ with the dual lattice $\Lambda^\vee$. 

For further reference note that
$\Lambda$ is naturally identified with 
a finite index sublattice of $\Lambda^\vee$ using the given bilinear 
pairing. Since $\Lambda^\vee$ is contained in 
$\Lambda_\IQ= \Lambda \otimes_\IZ \IQ$, the same bilinear pairing induces a $\IQ$-valued pairing on $\Lambda^\vee$ which will be denoted by 
$(d, d')\mapsto d\cdot d'\in \IQ$. Moreover, as the symmetric bilinear form is diagonalizable over $\IR$, 
any element $d\in \Lambda$ admits a canonical 
decomposition $d = d_++d_-$, with $d_+, d_-\in \Lambda_\IR= \Lambda \otimes_\IZ \IR$, such that 
$\left(d_+\right)^2 \geq 0$, $\left(d_-\right)^2\leq 0$
and $d_+\cdot d_-=0$.

In this context, the main object of study of this work is the thermal  
partition function  
constructed by Denef and Moore and \cite{Denef:2007vg} 
for D4-D2-D0 bound states, which is reviewed in Section \ref{verticalKthree}. Alternative constructions in M-theory fivebranes have been carried out in 
\cite{M5_genus,Farey_tail}. The approach of \cite{Denef:2007vg} will be adopted in this paper since it is based on D-brane effective actions, hence 
can be readily generalized to higher rank theories. 
The framework employed in \cite{Denef:2007vg} is 
an Euclidean 
IIA compactification on $X$ where the time direction is periodic 
with period $T$. Given the decomposition 
$H^2(X,\IZ)\simeq \Lambda \oplus \IZ\langle D\rangle$, the  background $B$-field will be an element $-B\in  \Lambda_\IR$. 
Similarly, the background 
RR 3-form field will be given by 
$C_3 =  -Cdt/T$, $C \in \Lambda_\IR$,
  where $0\leq t < T$  
is the Euclidean time coordinate. The vertical components 
have been set to zero since they have trivial restriction to the K3 fibers, and the signs are included for later convenience. 
Let also $C_1=C_0dt/T$ be the expectation value of the RR 
1-form, with $C_0\in \IR$, and  $\tau = C_0 + iT / g_s$ where $g_s$ is the IIA string coupling.

The discrete topological invariants of vertical D4-D2-D0 configurations 
on $X$  are triples 
$\gamma=(r, d, n)$ where $r,n\in \IZ$ are D4 and anti-D0 multiplicities respectively, while $d\in \Lambda^\vee$ is the D2 charge. 
The BPS indices counting such states  
will be denoted by $\Omega(\gamma)$. By analogy with 
\cite{Denef:2007vg}, it is shown in Section \ref{BPSindices} that the thermal partition function 
can be written as a finite sum 
\be\label{eq:strgenfctC} 
Z_{BPS} (X,r; \tau, {\bar \tau}, B,C) = \sum_{\delta \in \Lambda^\vee/r\Lambda} Z_{BPS}(X,r,\delta; \tau) 
 \Theta^*_{ r,\delta} (\tau, {\bar \tau}; C,B)
\ee 
over equivalence classes $\delta \in \Lambda^\vee/ r\Lambda$. 
For each such equivalence class, 
\[
Z_{BPS}(X,r, \delta;\tau)= \sum_{n} \Omega(\gamma) e^{2\pi i \tau (n + d^2/2r -1)}. 
\]
and 
\[
\Theta^*_{r, \delta}(\tau, {\bar \tau}; C,B) =\sum_{\alpha \in \Lambda} e^{-2 \pi i \tau (d+r\alpha+rB/2)_-^2/2r -2\pi i {\bar \tau} (d+r\alpha+rB)_+^2/2r+2\pi i (d + r\alpha+rB/2)\cdot C}.
\]
where $d\in \Lambda^\vee$ is any representative of $\delta$. Note that the above sum is actually the complex conjugate of a Siegel Jacobi theta function of a coset of the rescaled lattice $\sqrt{r}\Lambda\subset \Lambda_\IR$.
As shown in Section \ref{verticalKthree} the above series are  independent of the choice of $d$. 

As pointed out in \cite{Mock_theta}, note that such an expression does not hold in general 
for arbitrary D4-D2-D0 BPS states on Calabi-Yau threefolds. The fact that it holds here relies on the  invariance of vertical BPS indices under endomorphisms of the charge lattice 
of the form \eqref{eq:chargemdrmy}, which is proven in 
Section \ref{verticalsheaves}. 
Moreover, a similar decomposition 
of the thermal partition function occurs in 
\cite[Sect. 2.2]{Denef:2007vg} with one significant difference. In the 
cases studied in loc. cit. the generic supersymmetric D4-D2-D0 configuration consists of a single D4-brane wrapping a smooth very ample divisor in $X$ bound to some D2 and anti-D0 branes. 
In the present case, a generic stable configuration consists of a 
rank $r$ stable vector bundle supported on a smooth K3 fiber, 
and cannot be deformed to a single D4-brane wrapping a 
smooth surface. This is reflected in the presence of the cosets 
$\Lambda^\vee/ r\Lambda$ with $r\geq 1$ in the right hand side of \eqref{eq:strgenfctC}, as opposed to the analogous formulas 
in \cite[Sect. 2.2]{Denef:2007vg}. This will turn out to 
have some important arithmetic consequences. 

Mathematically the BPS indices $\Omega(\gamma)$ are generalized Donaldson-Thomas invariants \cite{RT,wallcrossing, genDTI} counting semistable vertical sheaves on $X$ suppported on K3 
fibers. More precisely, the formalism of \cite{wallcrossing, genDTI} 
yields rational invariants $DT(\gamma) \in \IQ$ which are conjecturally related 
to the integral ones $\Omega(\gamma)$ by the  multicover 
formula
\[
DT(\gamma) = \sum_{\substack{ k\in \IZ, \, 
k\geq 1\\ \gamma = k\gamma'\\ }}
{1\over k^2} \Omega(\gamma').
\]
Some aspects of generalized Donaldson-Thomas invariants
are reviewed in Section \ref{verticalsheaves}.  

The importance of rational invariants in this context was
emphasized by Manschot 
in \cite{Wall_trees}, where it is shown that partition functions 
of rational as opposed to integral invariants are expected to 
exhibit modular properties. This is indeed the case for the 
explicit formulas derived in this paper, as explained below. 

The partition function of rational invariants is obtained by 
summing over  multicovers. In order to write an explicit formula, note that for any 
$r, r' \in \IZ$, $r, r'\geq 1$, such that $r= k r'$ with $k \in \IZ$ 
there is an injective morphism 
\[ 
f_{r',k} : \Lambda^\vee/ r' \Lambda \to \Lambda^\vee/ r\Lambda, \qquad f_{r', k}([d]_{r'}) = [kd]_r \ {\rm for\ all}\ d \in \Lambda^\vee.
\]
Here $[d]_s\in \Lambda^\vee/ s\Lambda$ denotes the equivalence 
class of $d \in \Lambda^\vee$ mod $s\Lambda$ for any $s\in \IZ$, 
$s\geq 1$. 
Then the rank $r\geq 1$ partition function for rational invariants reads 
\be\label{eq:DTgenfct}
Z_{DT}(X, r; \tau, {\bar \tau},B,C) = \sum_{\delta \in \Lambda^\vee/ r\Lambda} Z_{DT}(X, r, \delta; \tau)\Theta^*_{r, \delta}(\tau, 
{\bar \tau}; C,B)
\ee
where 
\[
Z_{DT}(X, r, \delta; \tau)=
\sum_{\substack{k \in \IZ,\ k\geq 1\\ (r, \delta)=
(k r', f_{r',k}(\delta'))}}
{1\over k^2} Z_{BPS}(X, r', \delta'; k\tau). 
\]

\subsection{The main formula}\label{mainf}
The main result of this paper is an explicit conjectural formula 
for the partition function \eqref{eq:DTgenfct}. 
This formula will be written in terms of a certain vector valued 
modular form ${\widetilde \Phi}$ associated to an irreducible 
nodal K3 pencil $X \to C$ 
 via the construction of \cite{GW_NL, NL_YZ}. As 
explained in detail in Section \ref{setup}, this form 
encodes the Noether-Lefschetz numbers of an associated 
smooth pencil ${\wX}$ over a double cover of $C$. 
The form ${\widetilde \Phi}$ takes values in the Weil 
representation determined by the lattice $\Lambda$ 
equipped with the intersection form, and has weight 
$11 - {\rm \rk}(\Lambda)/2$. Leaving the technical details 
for section \ref{NLmodular}, note that the latter is a 
representation of the metaplectic cover of $SL(2, \IZ)$ 
on the vector space $V=\IC[\Lambda^\vee/\Lambda]$. 
The components of ${\widetilde\Phi}$ with respect to the 
natural basis of $(e_g)$, $g\in \Lambda^\vee/\Lambda$ 
will be denoted by ${\widetilde \Phi}_g$. 
Note also that for any $l\geq 1$ there is a $\IQ/\IZ$-valued quadratic form $\theta_l$ on 
$\Lambda^\vee/l\Lambda$ by 
\[
\theta_l([d]_l) = {d^2 \over 2l} \quad {\rm mod}\ \IZ.
\]
Then the main formula reads
\be\label{eq:mainformulaA} 
Z_{DT}(X,r,\delta; \tau) = {1\over 2r^2} 
\sum_{\substack{ k,l\in \IZ,\ k,l\geq 1\\ kl=r,\ \delta = f_{l,k}(\eta)}} 
\sum_{s=0}^{l-1} l
e^{-2\pi i s \theta_l(\eta)}\big(\Delta^{-1}{\widetilde \Phi}_{[\eta]_1}\big)\left({k\tau +s\over l}\right)
\ee
where $\Delta(\tau) = \eta(\tau)^{24}$.
Note that in the right hand side $\eta \in \Lambda^\vee/ l \Lambda$ 
is uniquely determined by $\delta$ since the morphism 
$f_{l,k}: \Lambda^\vee/ l \Lambda\to \Lambda^\vee/ r \Lambda$ 
is injective. Moreover $[\eta]_1\in\Lambda^\vee/\Lambda$ denotes the image 
of $\eta$ under the natural projection $\Lambda^\vee/l \Lambda\twoheadrightarrow \Lambda^\vee/ \Lambda$. 

Special cases of the above  formula have been rigorously 
proven
by Gholampour and Sheshmani  \cite{DT_twodim} for K3 fibrations 
without singular fibers and primitive 
charge vectors as well as for rank $r=1$ invariants on 
irreducible nodal K3 pencils. 
Other special cases for arbitrary $r\geq 1$ were obtained in
\cite{vertical} for 
elliptic K3 pencils in Weierstrass form using the Fourier-Mukai transform. 

A string theoretic derivation of equation \eqref{eq:mainformulaA} is given in Section \ref{adiabatic} using adiabatic IIA/heterotic duality. The main idea is that for the purpose of vertical BPS index the K3 pencil can be regarded as a family of 
world sheet conformal field theories for a $T^4$ compactification of the 
$E_8\times E_8$ heterotic string. Such an identification is not canonical because 
the restriction of the Calabi-Yau threefold metric to a K3 surface 
is not in general hyper-K\"ahler. Nevertheless it can be used 
as a very efficient tool since the BPS index of D4-D2-D0 supported on a K3 fiber is invariant under metric perturbations. 
The conclusion is that counting D4-D2-D0 bound states supported 
on nodal fibers is equivalent to counting Dabholkar-Harvey states \cite{DH}
in a smooth heterotic conformal field theory. Then one can use the general formulas for the degeneracies of such states
obtained in \cite{exact_deg}. 

At the same time, a mathematical derivation of the same formula is given 
in Section \ref{recursion} using previous results of 
Gholampour, Sheshmani and Toda \cite{G_S_Toda}.  
Using recursive methods, Section \ref{recursion} gives a rigorous proof of 
equation \eqref{eq:mainformulaA} provided that one 
assumes the Gromov-Witten/stable pair correspondence 
\cite{stabpairsI} to hold. This proof is based on the explicit results 
obtained by Maulik and Pandharipande \cite{GW_NL} for vertical 
Gopakumar-Vafa invariants of K3 pencils, as well as a remarkable 
combinatorial identity for the Fourier coefficients of a 
certain meromorphic Jacobi form of negative index.

\subsection{Modularity results} 
The rank $r$ partition function is expected to be a  non-holomorphic modular form of weight $(-3/2,1/2)$, as predicted in \cite{M5_genus,Farey_tail,Denef:2007vg} 
on physics grounds. In order to show this for the explicit results derived in this paper, note that formula \eqref{eq:mainformulaA} yields the following expression 
\be\label{eq:mainformulaB} 
\bal
& Z_{DT}(X,r; \tau, {\bar \tau}, B,C ) = \\
& {1\over 2r^2} 
\sum_{\substack{k,l\in \IZ_{\geq 1}\\ kl=r}} 
\sum_{\eta \in \Lambda^\vee/l\Lambda} 
\sum_{s=0}^{l-1}
l \big(\Delta^{-1}{\widetilde \Phi}_{[\eta]_1}\big)\left({k\tau+s\over l}\right) 
\Theta^*_{1,[\eta]_1}\left({k\tau+s\over l}, {k{\bar \tau}+s\over l}; kC+sB, lB\right). \\
\eal
\ee
This is proven in Section \ref{multicoversect}. 
Since $Z_{DT}(X,1; \tau,{\bar \tau},B,C)$ is a non-holomorphic modular function of weight $(-3/2,1/2)$ in $(\tau, {\bar \tau})$, the above formula is an order 
$r$ Hecke transform of the rank one result. 
Recall that the partition function of 
Vafa-Witten theory with gauge group $U(r)$ was previously shown in 
\cite{E_strings} to be given precisely by an order $r$ Hecke transform 
of its $U(1)$ counterpart. The above formula generalizes this construction to vertical D4-D2-D0 partition functions 
for K3 fibered Calabi-Yau threefolds. 

As further discussed in 
Section \ref{modularsect}, an important consequence of this result  is that 
for  fixed $r\geq 1$ the 
collection $\big(Z_{DT}(X, r,\delta; \tau)\big)$, $\delta \in 
\Lambda^\vee/ r \Lambda$, forms a vector valued modular form for a certain representation of the metaplectic group 
$Mp(2, \IZ)$ of weight $(-1-{\rm rk}(\Lambda)/2)$. The representation in question is the dual  Weil representation 
of the rescaled lattice $\sqrt{r}\Lambda \subset \Lambda_\IR$. 
Since the above argument is rather indirect, a second direct 
proof of this statement is given in Section \ref{secondproof} 
for the skeptical reader. 

It is also worth noting that these results yields a new transformation 
acting on vector valued modular forms which is somewhat similar to a construction carried out in 
\cite[Thm. 3.1]{Constr_vvmf}. 
Given the present context, this construction
could be regarded as a 
generalization of the Hecke transform to vector 
valued modular forms in the Weil representation. An interesting open 
question is whether this construction is related to the existing one in the mathematical literature \cite{Hecke_vvmf}. These issues are currently under 
investigation in \cite{Hecke_in_progress}.

\subsection*{{\it Acknowledgments.}}  We are very grateful to
Terry Gannon for collaboration at the incipient 
stage of this project. D.E-D. owes special thanks to 
 Steve Miller and Greg Moore for very helpful discussions and suggestions. In particular Section 2 grew out of illuminating discussions with Greg Moore during a string theory group meeting at Rutgers. 
A. S. would like to thank Yukinobu Toda for helpful
discussions. We would also like to thank the referee for 
a careful reading of the manuscript and very insightful 
feedback. 
V.B and T. C. acknowledge the support of the Natural Sciences
and Engineering Research Council of Canada. 
The work of D.-E. D. was partially supported by NSF grant 
DMS-1501612. C.F.D. would like to acknowledge support from the Natural Sciences and
Engineering Research Council of Canada and the Visiting Campobassi
Professorship of the University of Maryland.
The work of C.Q. was supported in part through
fellowships from the Pacific Institute for the Mathematical Sciences and
the Natural Sciences and Engineering Research Council of Canada. A.S. would also like to thank Kavli IPMU for their kind and wonderful
hospitality. His work at IPMU was supported by the World Premier International
Research Center Initiative (WPI Initiative), MEXT, Japan.

\section{Vertical D4-D2-D0 bound states on K3
 fibrations}\label{verticalKthree}

This section will review some basic facts on vertical supersymmetric 
D4-D2-D0 bound states on K3 fibered Calabi-Yau threefolds. The 
presentation will follow closely the treatment in Sections 2.1 and 2.2 of 
\cite{Denef:2007vg}, adapting the construction of loc. cit. to K3 fibers 
rather than very ample divisors. 
Below 
$X$ will be a smooth projective Calabi-Yau threefold equipped with a projection map $\pi: X \to \IP^1$ and a section $\sigma: \IP^1 \to X$. 
The Calabi-Yau condition implies that the generic fibers of $\pi$ will be smooth  K3 surfaces. Singular fibers will be in general present, 
but throughout this paper it will be assumed that all such fibers 
have only isolated simple nodal points. It will be also assumed  
for simplicity that the integral cohomology of 
$X$ is torsion free and that $h^{1,0}(X) =0$. These assumptions are easily satisfied for sufficiently generic models. 

\subsection{The vertical charge lattice} 

For a type IIA compactification on $X$, the lattice of 
electric and magnetic charges of four dimensional BPS particles 
is isomorphic to the  degree zero K-theory of $X$. 
In the absence of torsion, the lattice of charges can be identified 
with the even integral homology of $X$ using the Chern character 
and Poincar\'e duality. 
In particular the charges of 
D4-D2-D0 bound states supported on the fibers of $\pi$ take values 
in the sublattice 
\[
\IZ\langle D\rangle \oplus H_2(X,\IZ)^{\pi} \oplus H_0(X,\IZ) \subset H_{\sf even}(X, \IZ), 
\]
where $D \in H_4(X,\IZ)$ is the K3 fiber class and $H_{2}(X,\IZ)^{\pi}\subset H_2(X,\IZ)$ is the kernel of the 
pushforward map $\pi_*: H_{2}(X,\IZ) \to H_2(\IP^1,\IZ)$.

Since one assumes the existence of a section, there are direct sum decompositions 
\[ 
H_4(X, \IZ)\simeq \IZ\langle D \rangle \oplus \Lambda, \qquad 
H_2(X, \IZ) \simeq \IZ\langle C \rangle \oplus H_2(X, \IZ)^\pi
\] 
where $C,D$ are the section class and the K3 fiber class respectively.
Moreover $\Lambda \subset H_4(X,\IZ)$ is the sublattice 
generated by horizontal 4-cycles relative to the map $\pi$, i.e.
4-cycles which project surjectively to $\IP^1$. Poincar\'e duality 
yields a nondegenerate bilinear pairing 
\[
H_2(X, \IZ)^\pi \times \Lambda \to \IZ
\]
which identifies $H_2(X,\IZ)^\pi$ with the dual lattice $\Lambda^\vee$. 
In conclusion, the charges of vertical 
D4-D2-D0 bound states take values in a lattice of the form
in 
\[
\Gamma=\IZ\oplus \Lambda^\vee \oplus \IZ. 
\]
A typical charge vector will be denoted by $\gamma=(r, d, n)$ 
where $r \geq 1$ is the D4 brane multiplicity, $d$ is the effective 
D2 charge and $n$ the anti-D0 brane charge, not including 
gravitational corrections. 

Moreover, for any smooth fiber $X_p$, $p\in \IP^1$, one has a natural 
restriction map 
\[
H^2(X,\IZ) \to H^2(X_p, \IZ)
\]
whose image is contained in the sublattice of $(1,1)$ classes, 
$H^{1,1}(X_p) \cap H^2(X,\IZ)$. Using Poincar\'e duality, this yields a 
natural restriction map 
\be\label{eq:Lembedding}
\Lambda \to H^{1,1}(X_p) \cap H^2(X,\IZ)
\ee
which is furthermore injective for any smooth fiber $X_p$. 
For sufficiently generic models this map is in fact an isomorphism 
for any sufficiently generic smooth fiber $X_p$. However, it fails to 
be surjective at special points $p\in \IP^1$, where the rank of 
$H^{1,1}(X_p) \cap H^2(X,\IZ)$ jumps. This jumping phenomenon is studied in 
detail in the mathematics literature on Noether-Lefschetz loci 
\cite{GW_NL,NL_YZ}. The results used in this paper are 
reviewed in  detail in Section \ref{setup}. 
For the purpose of this 
section, note that the map \eqref{eq:Lembedding} will be assumed to be 
a primitive lattice embedding for all smooth fibers $X_p$. Moreover,
note that the intersection pairing on $H^2(X_p, \IZ)$ determines by restriction a nondegenerate symmetric bilinear form on $\Lambda$. 
Throughout this paper this form will be assumed to be independent 
of the point $p\in \IP^1$. These conditions are part of the definition 
of a lattice polarized family of K3 surfaces, which is reviewed in detail 
in Section \ref{setup}. In addition one also requires the signature of 
the induced form on $\Lambda$ to be $(1, {\rm rk}(\Lambda)-1)$. 

 In conclusion, $\Lambda$ will be endowed from this point on with  a nondegenerate symmetric bilinear form which agrees with the intersection product on each smooth fiber. This form will 
be denoted by 
\[ 
(\alpha, \alpha') \mapsto \alpha\cdot \alpha'
\]
for any $\alpha, \alpha'\in \Lambda$. By construction, this pairing is integral and even, but not unimodular. Therefore it determines an 
embedding $\Lambda \hookrightarrow \Lambda^\vee$ which will not be in general surjective. The quotient 
$\Lambda^\vee / \Lambda$ 
is a finite abelian group which will play a central part in this paper. 
Using this embedding $\Lambda$ will be identified in the following with 
a sublattice of $\Lambda^\vee$. 
Furthermore, the symmetric nondegenerate bilinear form on $\Lambda$ 
determines a $\IQ$-valued symmetric nondegenerate bilinear form 
on $\Lambda_\IQ = \Lambda \otimes_\IZ \IQ$, hence also on $\Lambda^\vee$, which is contained in $\Lambda_\IQ$. 
This form will be similarly written as 
\[
(d,d') \mapsto d\cdot d' 
\]
for any $d, d'\in \Lambda^\vee$. Note that 
$d\cdot \alpha =d(\alpha)\in \IZ$ 
for any $d\in \Lambda^\vee$, $\alpha \in\Lambda$. 

The main goal of this paper is to derive explicit formulas 
for the supersymmetric indices of vertical D4-D2-D0 BPS configurations 
of fixed charge $\gamma$, for an arbitrary  fixed K\"ahler class $\omega$ on $X$.

\subsection{BPS indices and partition functions}\label{BPSindices} 

For each charge vector $\gamma$ there is a Hilbert space of  
four dimensional massive  BPS particles $\CH_\omega(X,\gamma)$ which 
carries an action of the three dimensional rotation group 
$SO(3) \subset SO(1,3)$. On general 
symmetry grounds this space is of the form 
\[ 
\CH_\omega(X, \gamma) \simeq 2\left((0)+(1/2)\right) \otimes 
\CH^{\sf int}_\omega(X,\gamma)
\]
where $\CH^{\sf int}_\omega(X,\gamma)$ is obtained by quantizing the 
internal degrees of freedom of these particles. 
The BPS index $\Omega_\omega(\gamma)$ is then defined as the Witten index 
\be\label{eq:fourDindex}
\Omega_\omega(\gamma) = {\rm Tr}_{\CH_\omega^{\sf int}(X,\gamma)} (-1)^{2J_3}
\ee
where $J_3\subset SO(3)$ is the Cartan generator of the four dimensional little group. 

Note that a simple application of 
the Kontsevich-Soibelman wallcrossing formula \cite{wallcrossing}
shows that for vertical configurations the indices $\Omega_\omega(\gamma)$ are independent of the K\"ahler class $\omega$. This follows from the observation that $(\gamma, \gamma')=0$ for any two vertical charge vectors with respect to the natural symplectic pairing on $H^{\sf even}(X,\IZ)$. Therefore 
the subscript $\omega$ will be omitted in the following.

One more important observation is that the BPS degeneracies 
$\Omega(r,d,n)$ are invariant under transformations of the form 
\be\label{eq:chargemdrmy}
(r,d,n) \mapsto
\big( r, d+ r \alpha, n - d\cdot \alpha - {1\over 2} r \alpha^2\big),
\ee
for any element $\alpha \in \Lambda$. This is proven 
at the end of Section \ref{verticalsheaves} using the 
mathematical construction of BPS degeneracies as Donaldson-Thomas invariants. 

As explained in Sections 2.1 and 2.2 of 
\cite{Denef:2007vg}, there is a natural 
construction for generating functions of these indices derived 
from the natural DBI action on D4-branes. 
A central element in this construction is the jumping phenomenon associated to Noether-Lefschetz loci for K3 fibrations, which was 
briefly explained above. 
As opposed to the current case, the treatment of \cite{Denef:2007vg} is focused on a rank one D4-brane wrapping a smooth very ample 
divisor in $X$. However, their considerrations can be 
easily adapted to the situation at hand. 

Suppose first the D4-brane rank is $r=1$. In this case a 
generic supersymmetric 
D4-D2-D0 configuration of charge $\gamma = (1, d, n)$ will 
be a bound state of one D4 brane and $k$ anti-D0 branes supported 
on an arbitrary  smooth fiber $X_p$. One also has an
$U(1)$ flux $\beta\in H^2(X_p,\IZ)$ 
on the D4-brane such that 
\be\label{eq:Fd}
 \beta \cdot \alpha  = d\cdot \alpha
\ee
for any $\alpha \in \Lambda$. By supersymmetry, $F$ has to be a 
$(1,1)$ class. Note also that
\be\label{eq:dzerocharge} 
n = k -{\beta^2\over 2}
\ee
by an easy application of the Grothendieck-Riemann-Roch theorem. 

Recall that under the current assumptions $\Lambda$ is a  sublattice of $H^{1,1}(X_p)\cap H^2(X_p,\IZ)$, which is in turn a sublattice of $H^2(X_p,\IZ)$. Let $\Lambda^\perp\subset 
H^2(X_p,\IZ)$ denote the sublattice consisting of all elements which 
are 
orthogonal to $\Lambda$ with respect to the natural intersection 
pairing. Then note that there is a direct sum decomposition
\be\label{eq:Qdecomp}
H^2(X_p, \IQ) \simeq \Lambda_\IQ\oplus \Lambda^\perp_\IQ. 
\ee
Let $\beta  = \beta^\parallel + \beta^\perp$ be the corresponding decomposition 
of $\beta$. Using the natural inclusion $\Lambda^\vee \subset \Lambda_\IQ$ equation \eqref{eq:Fd} yields an identification $\beta^\parallel = d$.

The construction of the string theoretic generating function for the BPS degeneracies will follow closely Sections 2.1 and 2.2 of 
\cite{Denef:2007vg}. The generating function will be a thermal partition function working in an euclidean four dimensional background 
of the form $\IR^3 \times S^1$, where the time direction is periodic, 
with period $T$. The background B-field will be an element 
$-B \in \Lambda$ while 
the bacground RR fields will be set to 
\[
C_3=-Cdt/T, \qquad C_1 = C_0dt/T
\]
where $C\in \Lambda_\IR$, $C_0\in \IR$, and $0\leq t <T$ is the Euclidean time coordinate. 
As in equation (2.9) of loc. cit. the contribution of all such supersymmetric configurations supported on a fixed smooth fiber $X_p$ to the partition function 
will be given by 
\[ 
{\rm Tr}_{\CH^{\sf int}(X_p, 1;\beta, k)} (-1)^{2J_3} e^{-TH +2\pi i C_0 \big(k - (\beta+B)^2/2-\chi(X_p)/24\big) 
+2\pi i (\beta+B)\cdot C} 
\]
where $H$ is the Hamiltonian acting on the internal Hilbert space 
$\CH^{\sf int}(X_p, 1;\beta, k)$. For any cohomology class 
$\alpha \in H^2(X_p,\IR)$ let $\alpha_+, \alpha_-\in H^2(X_p, \IR)$ denote the self-dual, respectively anti-self-dual parts 
of $\alpha$ with respect to the restriction $\omega|_{X_p}$ 
of the K\"ahler form. 
Then, evaluating this trace as in equations 
(2.10)-(2.14) of loc. cit. results in an expression of the form 
\[ 
c(X_p, 1,\beta, k) e^{2\pi i\tau \big(k - (\beta_- + B_-)^2/2 -\chi(X_p)/24\big) - 2\pi i {\bar \tau} (\beta_+ + B_+)^2/2+2\pi i (\beta+B)\cdot C} 
\]
where 
\[
\tau = C_0 + {iT\over g_s }, \qquad 
c(X_p, 1,\beta, k) = {\rm Tr}_{\CH^{\sf int}(X_p, 1;\beta, k)} (-1)^{2J_3}.
\]

Next recall that by assumption $\omega|_{X_p} \in \Lambda_\IR$, hence
$\beta_+ = \beta^\parallel_+$. 
This yields
\[
k - \beta_-^2/2 = n + \beta_+^2/2 = n + \big(\beta^\parallel_+\big)^2/2 = 
n + \big(\beta^\parallel\big)^2/2 - \big(\beta^\parallel_-)^2/2,
\]
where $\beta^\parallel =d$. Moreover, since $B\in  \Lambda_\IR$, note that $B_-, B_+\in \Lambda_\IR$, hence 
\[
\beta_-\cdot B_- = \beta^\parallel_-\cdot B_-.
\]
Similarly, 
\[
(\beta+B)\cdot C = (\beta^\parallel+B)\cdot C. 
\]
Since $\chi(X_p)=24$, one obtains 
\[ 
e^{2\pi i\tau \big(k - (\beta_- + B_-)^2/2 -\chi(X_p)/24\big) - 2\pi i {\bar \tau} (\beta_+ + B_+)^2/2} = 
e^{2\pi i \tau (n + d^2/2 -1)} e^{-2 \pi i \tau
(d_- + B_-)^2/2 -2\pi i {\bar \tau} (d_+ + B_+)^2/2}  
\]
which depends only on $(d,n)$ and $B$. 
Then, summing over $(\beta, k)$, one obtains 
\be\label{eq:sumbetakA}
\bal
& \sum_{\beta,k} c(X_p, 1, \beta, k) 
e^{2\pi i\tau \big(k - (\beta + B)_-^2/2 -\chi(X_p)/24\big) - 2\pi i {\bar \tau} (\beta+B)_+^2/2+2\pi i (\beta+B)\cdot C} =\\
&
\sum_{d,n} \Omega(X_p,1,d,n) e^{2\pi i \tau (n + d^2/2 -1)} e^{-2 \pi i \tau (d+ B)_-^2/2 -2\pi i {\bar \tau} (d+ B)_+^2/2+2\pi i (d+B)\cdot C},\\
\eal
\ee
where 
\[
\Omega(X_p, 1,d,n) = \sum_{\beta, \beta^\parallel=d} c(X_p, 1,\beta, n + \beta^2/2).
\]

Moreover, note that $\Omega(X_p, 1,d,n)$ and  $n+d^2/2$ 
are invariant under transformations of the form \eqref{eq:chargemdrmy}
with $r=1$. Therefore for any $d\in \Lambda^\vee$, the   sum 
\[
\sum_{n} \Omega(X_p, 1,d,n) e^{2\pi i \tau (n + d^2/2 -1)}
\] 
depends only on the equivalence class $\delta = [d]_1 \in \Lambda^\vee/\Lambda$ of $d$ mod $\Lambda$. 
Hence it will be denoted by $Z(X_p,1,\delta; \tau)$. 
At the same time, for any fixed $d\in \Lambda^\vee$, $B,C\in \Lambda_\IR$ note that 
\[ 
\sum_{\alpha \in \Lambda} e^{-2 \pi i \tau 
(d+ B+\alpha)_-^2/2 -2\pi i {\bar \tau} (d+
B+\alpha)_+^2/2+2\pi i (d+B+\alpha)\cdot C} = 
e^{-\pi i B\cdot C} \Theta^*_\delta(\tau, {\bar \tau}; C,B)
\]
where $\Theta_\delta(\tau, {\bar \tau}; C,B)$ is a Siegel Jacobi theta function, which also depends only on the equivalence class $\delta$ of $d$ mod $\Lambda$. 
In conclusion for any smooth fiber $X_p$, the sum 
\eqref{eq:sumbetakA}
can be written as 
\[ 
\sum_{\delta \in \Lambda^\vee/\Lambda} Z(X_p,1,\delta; \tau) \Theta^*_\delta(\tau, 
{\bar \tau}; C,B)
\]
up to the constant phase factor $e^{-\pi i B\cdot C}$, which can be omitted by a suitable choice of normalization of the 
partition function. 

In order to compute the full degeneracies $\Omega(\gamma)$ one has to 
integrate over all possible locations of the fiber $X_p$, including 
singular fibers. The mathematical framework for such a computation 
is the theory developed by Behrend \cite{micro} which allows one 
to write $\Omega(\gamma)$ as the integral of a certain constructible function 
on the moduli space of supersymmetric D4-D2-D0 configurations. 
Leaving the details for Sections \ref{verticalsheaves} and \ref{primitivesect}, it suffices to note here that the effect of integrating over all fibers is to convert
$\Omega(X_p, 1,d,n)$ into $\Omega(1,d,n)$ leaving the general form 
of the partition function unchanged. In conclusion the partition function for rank $r=1$ 
invariants will take  the form 
\be\label{eq:strgenfctB} 
Z_{BPS}(X,1; \tau, {\bar \tau}, B,C) = \sum_{\delta \in \Lambda^\vee/\Lambda} Z_{BPS}(X,1,\delta; \tau) \Theta^*_\delta(\tau, {\bar \tau}; C,B)
\ee 
where 
\[
Z_{BPS}(X,1,\delta;\tau)= \sum_{n} \Omega(1,d,n) e^{2\pi i \tau (n + d^2/2 -1)}
\] 
and $d\in \Lambda^\vee$ is an arbitrary fixed representative of $\delta$ for each $\delta \in \Lambda^\vee/\Lambda$.
Note that $Z_{BPS}(X,1,\delta;\tau)$ is a power series in 
$q^{1/2m}=e^{\pi i \tau/m}$ for any $\delta \in \Lambda^\vee/\Lambda$, where $m$ is the absolute value of the  determinant of the intersection form on $\Lambda$.

The next goal is to generalize the above construction to higher rank $r\geq 1$. This will be carried out in this section for primitive charge 
vectors $\gamma=(r,d,n)$. The general case can then be obtained 
by summation using the multicover formula for Donaldson-Thomas  
invariants, as shown in Section \ref{genfctsect}. For primitive charge vectors the generic supersymmetric D4-D2-D0 configuration is 
a D4-brane supported on a smooth fiber $X_p$ equipped with an 
$U(r)$ gauge field $A$. The topological invariants of $A$ are the first 
Chern class
\[
\beta = {\rm Tr}(F)
\]
and the second Chern class, or  instanton number, 
\[
k = {\beta^2\over 2} - {{\rm Tr}(F^2)\over 2},
\]
where $F$ is the field strength of $A$. Note again that 
\[
n = k - {\beta^2\over 2}\qquad {\rm and}\qquad 
\beta \cdot \alpha = d\cdot \alpha
\]
for any $\alpha \in \Lambda$. Moreover,
$d =\beta^\parallel$
with respect to the decomposition \eqref{eq:Qdecomp}.

By supersymmetry the $U(r)$ gauge field must satisfy the Donaldson-Uhlenbeck-Yau equations. 
Therefore the field strength $F$ of $A$ must be a $(1,1)$ 
form on $X_p$ such that 
\be\label{eq:DUY}
\omega_p \wedge F = \lambda \omega_p^2 
\ee
where $\omega_p = \omega|_{X_p}$ is the restriction of the ambient 
K\"ahler class, and $\lambda = (\omega_p \cdot \beta) / \omega_p^2$
is constant. Now let $A_0= A- {\rm Tr}(A)/r$ be the traceless part of $A$. This is a $PU(r) \simeq SU(r)/ {\bf \mu}_r$ gauge field on $X_p$ 
with field strength 
\[
F_0 = F - {\beta\over r} {I_r}. 
\]
Equation \eqref{eq:DUY} implies that 
\[
\omega_p \wedge F_0 = 0, 
\]
that is 
 $A_0$ is an antiself-dual field configuration. Hence one has 
\[ 
F_- = F_{0} + \beta_-, \qquad F_+ = \beta_+ I_r.
\]
Let $k_0 = - {\rm Tr}(F_0^2)/2$, which is a rational number in 
$(1/ r)\IZ \subset \IQ$. Since $F_0$ is traceless, it follows that 
\[
n = k_0 - {\beta^2\over 2r}.
\]

The contribution of all such supersymmetric configurations 
to the thermal partition function will be given by 
\[ 
{\rm Tr}_{\CH^{\sf int}(X_p, 1;\beta, k)} (-1)^{2J_3} e^{-TH +2\pi i C_0 \big(k - (\beta+rB)^2/2-\chi(X_p)/24\big) 
+2\pi i (\beta+rB)\cdot C} 
\]

Evaluating this trace by analogy with equations 
(2.10)-(2.14) in \cite{Denef:2007vg} yields in this case
\[ 
c(X_p, r, \beta, k) e^{2\pi i\tau \big(k_0 - (\beta+rB)_-^2/2r -r\chi(X_p)/24\big) - 2\pi i {\bar \tau} (\beta+rB)_+^2/2r+2\pi i (\beta+rB)\cdot C}.
\]
By analogy with the rank one case, this expression is further equal to 
\[ 
c(X_p, r, \beta, k) e^{2\pi i\tau (n + d^2/2r -r) - 2\pi i \tau 
(d+rB)_-^2/2r-2\pi i {\bar \tau} (d+rB)_+^2/2r+2\pi i (d+
rB)\cdot C}.
\]
Again, the BPS index $\Omega(X_p,\gamma)$ 
is obtained by summing over all pairs $(\beta, k)$ 
with $\beta^\parallel =d$ and $k = n + \beta^2/2$:
\[
\Omega(X_p,\gamma) = \sum_{\beta, \beta^\parallel =d} 
c(X_p,r,\beta, n+\beta^2/2).
\]
Moreover, $\Omega(X_p,\gamma)$ 
and $n+ d^2/2r$ are invariant under  transformations of the form \eqref{eq:chargemdrmy}. Therefore for any $d\in \Lambda^\vee$, the   sum 
\[
\sum_{n} \Omega(X_p, \gamma) e^{2\pi i \tau (n + d^2/2r -r)}
\] 
depends only on the equivalence class $\delta = [d]_r\in \Lambda^\vee/r\Lambda$ of $d$ mod $r\Lambda$. 
Hence it will be denoted by $Z(X_p, r, \delta; \tau)$. 
At the same time, for any $d\in \Lambda^\vee$, 
the sum 
\[ 
\sum_{\alpha \in \Lambda} e^{-2 \pi i \tau (d+rB+r\alpha)_-^2/2r -2\pi i {\bar \tau} (d+rB+r\alpha)_+^2/2r+2\pi i (d+rB+ r\alpha)\cdot C}
\]
also depends only on the 
equivalence class of $d$ mod $r\Lambda$. As shown in Section \ref{modularsect}, this sum is in fact equal to 
$e^{-\pi i r B \cdot C} 
\Theta^*_{r, \delta}(\tau, {\bar \tau}; C,B)$, where 
the complex conjugate for a Siegel Jacobi theta function 
for a coset of the rescaled lattice $\sqrt{r}\Lambda\subset \Lambda_\IR$. 
Choosing again a suitable normalization, one obtains a final expression for the rank $r$ partition function of the form 
\be\label{eq:strgenfctC} 
Z_{BPS} (X,r; \tau, {\bar \tau}, B,C ) = \sum_{\delta \in \Lambda^\vee/r\Lambda} Z_{BPS}(X,r,\delta; \tau) \Theta^*_{ r,\delta} (\tau, {\bar \tau}; C,B)
\ee 
where 
\[
Z_{BPS}(X,r, \delta;\tau)= \sum_{n} \Omega(\gamma) e^{2\pi i \tau (n + d^2/2r -1)}
\] 
and $d\in \Lambda^\vee$ is a fixed arbitrary representative for each equivalence class $\delta 
\in \Lambda^\vee/r\Lambda$. 
Note that $Z_{BPS}(X,r, \delta;\tau)$ is a power series in 
$q^{1/2mr}=e^{\pi i \tau/mr}$ for any $\delta \in \Lambda^\vee/\Lambda$, where $m$ is the absolute value of the determinant of the intersection form on $\Lambda$.

\subsection{Mathematical approach via Donaldson-Thomas invariants}\label{verticalsheaves}

Mathematically, supersymmetric D4-D2-D0 bound states on $X$ 
are Bridgeland stable 
objects in the derived category of $X$. In this paper 
it will be assumed that for fixed numerical invariants 
Bridgeland stability reduces at large radius to 
 Gieseker-stability for coherent sheaves. 
Therefore a supersymmetric D4-D2-D0 configuration 
will be a Gieseker semistable torsion coherent sheaf $E$ 
with respect to a certain polarization $\omega$ 
on $X$. For vertical D4-D2-D0 configurations $E$ will be also required to be set theoretically supported  on a 
finite union of K3 fibers. Using the isomorphism $H_2(X,\IZ)^\pi\simeq \Lambda^\vee$ found in the previous section, such a sheaf has  numerical 
invariants $\gamma=(r,d,n) \in \IZ_{\geq 1}\times \Lambda^\vee \times \IZ$ 
where 
\be\label{eq:numinvE} 
\ch_1(E) = rD, \qquad 
 \ch_2(E) =d,  \qquad \ch_3(E) = -n \ch_3(\CO_x). 
\ee

For completeness recall Gieseker 
stability for pure dimension two sheaves on $X$.
Given a real ample class $\omega$ on $X$, for any such nonzero sheaf $E$ let 
\[
\mu_{\omega}(E) = {\omega \cdot \ch_2(E)\over 
\omega^2\cdot \ch_1(E)/2}, \qquad
\nu_{\omega}(E) = {\chi(E)\over \omega^2\cdot  \ch_1(E) /2}.
\]
Then Gieseker (semi)stability  with respect to
$\omega$ is defined by
the conditions
\be\label{eq:twistedstabA}
\mu_{\omega}(E') \ (\leq) \ \mu_{\omega}(E)
\ee
for any proper nonzero subsheaf $0\subset E' \subset E$, and
\be\label{eq:twistedstabB}
\nu_{\omega}(E') \ (\leq) \ \nu_{\omega}(E)
\ee
if the slope inequality \eqref{eq:twistedstabA} is saturated. 

For any $\gamma =(r,d,n)$ let 
$M_\omega(\gamma)$ denote the coarse moduli 
space of $\omega$-semistable sheaves $E$ with 
discrete invariants \eqref{eq:numinvE}. 
In the absence of semistable objects, the Donaldson-Thomas invariants $DT_\omega(\gamma)$ are defined in terms of virtual cycles; a special case of  \cite{RT}. This is for example the 
case if $\gamma$ is primitive. Moreover in this 
case, there exists a constructible function $\nu: M_\omega(\gamma) \to \IZ$ constructed by Behrend \cite{micro} such that 
\be\label{eq:weightedchi}
DT_\omega (\gamma) = \chi(M_\omega(\gamma), \nu).
\ee
By definition, given a constructible function $\phi: S \to \IZ$ 
on any scheme of finite type, the weighted Euler characteristic 
$\chi(S, \phi)$ is defined by
\[
\chi(S, \phi) = \sum_{n\in \IZ} n \chi(\phi^{-1}(n)),
\]
where $\chi({}\ )$ denotes the topological Euler characteristic. 

 The value of $\nu$ at a point 
${\mathfrak m}\in M_\omega(\gamma)$ is determined by the local 
scheme structure of the moduli space near ${\mathfrak m}$.
According to \cite{vancycles}, the moduli space admits a local presentation 
as a critical locus of a polynomial function $W$ 
defined on the space of infinitesimal deformations of the corresponding 
sheaf $E$ on $X$. Then 
\be\label{eq:nuvalue}
\nu({\mathfrak m}) = (-1)^{d}(1-\chi(MF_{\mathfrak m}(W))
\ee
where $d$ is the dimension of the space of infinitesimal deformations 
and $MF_{\mathfrak m}(W)$ is the Milnor fiber of $W$ at ${\mathfrak m}$. Note that if the moduli space 
is smooth at ${\mathfrak m}$, 
\be\label{eq:smoothnuvalue}
\nu({\mathfrak m}) = (-1)^{\rm dim(T_{\mathfrak m}M_\omega(\gamma))}.
\ee 
In particular, if the moduli space is a smooth projective variety, the constructible 
function $\nu$ takes the constant value 
$(-1)^{{\rm dim}(M_\omega(\gamma))}$, hence 
\[
\chi(M_\omega(\gamma), \nu) = (-1)^{{\rm dim}(\CM_\omega(\gamma))}
\chi(M_\omega(\gamma)).
\]

From a physical perspective, the infinitesimal deformations 
of $E$ are associated to complex chiral superfields in the low energy 
effective action of the corresponding D4-D2-D0 configuration, and 
$W$ is a superpotential interaction. The moduli space is locally 
identified with the critical locus of $W$. For an isolated critical point, 
Milnor's 
result \cite{sing_points} shows that 
$\nu({\mathfrak m})$ is the same as the dimension of the 
chiral ring of $W$. Formula \eqref{eq:nuvalue} provides a suitable generalization for non-isolated vacua.

In the presence of 
semistable objects, one has to employ the formalism of Kontsevich and Soibelman 
\cite{wallcrossing} or Joyce and Song \cite{genDTI} to construct 
generalized Donaldson-Thomas invariants $DT_\omega(\gamma)\in \IQ$.
For sufficiently generic 
$\omega$ there are conjectural integral invariants 
$\Omega_\omega(\gamma)\in \IZ$ related to the rational ones by 
the multicover formula 
\be\label{eq:multicover}
DT_\omega(\gamma) = \sum_{\substack{ k\in \IZ, \, 
k\geq 1\\ \gamma = k \gamma'\\ }}
{1\over k^2} \Omega_\omega(\gamma').
\ee
According to \cite{DM_crossing,DG}  the integral invariants $\Omega_\omega(\gamma)$ are mathematical avatars of  
the BPS indices defined in \eqref{eq:fourDindex}. This justifies 
using the same notation in both cases. 
Moreover, as noted in the previous 
section, the wallcrossing formulas of 
\cite{wallcrossing, genDTI} show that these invariants are in fact independent of the K\"ahler class, therefore 
the subscript $\omega$ can be omitted. 

Finally, to conclude this section, the following is a detailed proof of invariance of Donaldson-Thomas invariants under 
the automorphisms \eqref{eq:chargemdrmy} of the charge 
lattice. 

First 
note that the transformations \eqref{eq:chargemdrmy} are obtained by taking a tensor 
product by a line bundle $L$ on $X$.
More precisely for any vertical sheaf $E$ one has the Chern 
class relations 
\be\label{eq:tensorchern}
\bal 
\ch_1(E&\otimes_X L) = \ch_1(E) ,\qquad \ch_2(E\otimes_X L) = \ch_2(E) + c_1(L) \cdot \ch_1(E), \\
& \ch_3(E\otimes_X L) = \ch_3(E)+c_1(L)\cdot \ch_2(E) + {1\over 2} c_1(L)^2 \cdot 
\ch_1(E). \\
\eal
\ee
For $c_1(L)=\alpha$, the numerical invariants of $E$ change according to 
equation \eqref{eq:chargemdrmy}. 
Below it will be shown that taking a tensor product as above yields an isomorphism of moduli spaces for sufficiently generic K\"ahler classes. Since the vertical Donaldson-Thomas invariants do not change under wallcrossing, this implies the invariance statement needed 
in Section \ref{verticalKthree}. 

Suppose $F$ is a vertical pure dimension two sheaf  with numerical 
invariants $\gamma=(r,d,n)$.
Then $F\otimes_X L$ is also a 
vertical  pure dimension two sheaf on $X$ with the same 
support as $F$ and numerical 
invariants as in \eqref{eq:chargemdrmy} i.e. 
\[
{\tilde \gamma} = \big( r, d+ r\alpha, n - d\cdot \alpha - {1\over 2} r \alpha^2 \big).
\]
Let $\omega = t D + \eta$ be an arbitrary K\"ahler class on $X$, where $\eta \in \Lambda_\IR$ is a relatively ample class.  Since $D$ is orthogonal to all vertical curve classes with respect to the 
intersection product on $X$, and $D^3=0$ one can easily check that 
\[
\mu_\omega(F) = {2d\cdot \eta\over r \eta^2}, 
\qquad 
\nu_\omega(F) = {4r-2n \over r \eta^2}.
\]
Note that $\chi(F)=2r-n$ by Riemann-Roch. This yields 
\[
\mu_\omega(F\otimes_X L) = \mu_{\omega}(F) + 
{\alpha \cdot \eta\over \eta^2}, \qquad 
\nu_\omega(F\otimes_X L) = \nu_\omega(F) + {2d\cdot \alpha \over r \eta^2} + {\alpha^2\over \eta^2}
\]

Now let $E$ be a vertical sheaf as above and $0\subset E'\subset E$ be a nontrivial proper subsheaf. Then $E'$ 
has to be vertical as well, hence it will have numerical 
invariants $\gamma'=(r',d',n')$. Using the above formulas
it follows that 
\be\label{eq:tensormu}
\mu_\omega(E\otimes_X L) - \mu_\omega(E'\otimes_X L) =\mu_\omega(E) - \mu_\omega(E')
\ee
and 
\be\label{eq:tensornu}
\nu_\omega(E\otimes_X L) - \nu_\omega(E'\otimes_X L) = 
\nu_\omega(E) - \nu_\omega(E') + {2\over \eta^2} \left({d\over r} - {d'\over r'}\right)\cdot \alpha.
\ee
Suppose $E$ is $\omega$-stable. This implies 
\[
\mu_\omega(E) - \mu_\omega(E') > 0 
\]
or 
\[
\mu_\omega(E) - \mu_\omega(E') =0\quad {\rm and}\quad 
\nu_\omega(E) - \nu_\omega(E') > 0. 
\]
In the first case, equation \eqref{eq:tensormu} implies that 
\[
\mu_\omega(E\otimes_X L) - \mu_\omega(E'\otimes_X L)
>0.
\]
In the second case, note that 
\[
\mu_\omega(E) - \mu_\omega(E') = {2\over \eta^2} \left({d\over r} - {d'\over r'}\right)\cdot \eta. 
\]
For sufficiently generic $\eta \in \Lambda_\IR$, equality of the slopes implies $d/r - d'/r' =0$. For example, if $\eta$ is a linear combination of lattice generators with sufficiently generic irrational coefficients. Then equation \eqref{eq:tensornu} further implies 
\[
\nu_\omega(E\otimes_X L) - \nu_\omega(E'\otimes_X L) = 
\nu_\omega(E) - \nu_\omega(E') >0.
\]
To cover all cases, suppose $E$ is strictly $\omega$-semistable and let $0\subset E'\subset E$ be a proper nontrivial 
subsheaf. If $E'$ does not saturate the stability condition, 
the proof is identical to the one given above. Hence suppose 
that 
\[
\mu_\omega(E') = \mu_\omega(E), \qquad 
\nu_\omega(E') = \nu_\omega(E).
\]
Then, under the same genericity assumption, the first 
equality implies again that $d/r = d'/r'$, which  yields 
\[
\mu_\omega(E'\otimes_X L) = \mu_\omega(E\otimes_X L), \qquad 
\nu_\omega(E'\otimes_X L) = \nu_\omega(E\otimes_X L).
\]
The map $E'\mapsto E'\otimes_X L$ is a bijection between the proper nontrivial subsheaves of $E$ and those of $E\otimes_X L$. Moreover, one can run 
the above argument in reverse taking a tensor product by 
$L^{-1}$. Therefore, for sufficiently generic $\eta\in \Lambda_\IR$ it follows that $E$ is $\omega$-(semi)stable 
if and only if $E\otimes_X L$ is $\omega$-(semi)stable. 

The plan for the rest of the paper is to provide two derivations
for the expression \eqref{eq:mainformulaA} encoding all the above 
invariants. The first is a string theoretic derivation based on 
adiabatic IIA/heterotic duality while the second is based on the 
mathematical results of \cite{G_S_Toda}. 
Both derivations rely heavily on a detailed understanding 
of lattice polarizations and Noether-Lefschetz loci, which is the subject of the next section.

\section{Lattice polarization and Noether-Lefschetz loci}\label{setup}

This section is a review of lattice polarized K3 fibrations and 
Noether-Lefshetz numbers mainly following \cite{GW_NL,NL_YZ}. 
The presentation will be fairly technical by neccessity, since 
it lies the groundwork  for the following sections.

Let $\pi: X \to \IP^1$ be a 
K3 fibered smooth projective Calabi-Yau threefold with a section $\sigma : \IP^1 \to X$ satisfying the following 
conditions:
\begin{itemize}
\item[$a)$] All K3 fibers are irreducible, reduced. The generic fiber is smooth and there are finitely many singular fibers, each of them with exactly one simple node. 
In order to simplify the presentation, 
 it will also be assumed that the number of 
singular fibers is even, although this is not an essential 
assumption. All the following considerations extend with minor modifications to the fibrations with odd numbers of singular fibers. 
\end{itemize}

Let ${S^\pi}\subset \IP^1$ be 
the set of critical values of $\pi$. Under the above assumptions, ${S^\pi}$ is a finite set consisting of an even 
number of points. Let $f: \Sigma\to \IP^1$ be a smooth 
generic double cover with branch locus ${S^\pi}$. Hence 
$\Sigma$ is a hyperelliptic curve of genus 
\[
{g}(\Sigma) = |{S^\pi}|/2 -1.
\] 
Note that there is a unique ramification point of  $f$ mapping to $\sigma \in {S^\pi}$. Abusing notation, it will be denoted by $\sigma$ as 
well, the distinction being clear from the context. 
The set of ramification points of $f$ will be denoted 
by $R^f \subset \Sigma$. Note also that if $|S^\pi|$ is odd one has to choose an extra point $\sigma_\infty\in \IP^1$ parameterizing a 
generic smooth fiber, and consider a double cover with branch locus 
$S^\pi \cup \{\sigma_\infty\}$. All the following considerations 
will go through with minor modifications. 

Let $X'=X\times_{\IP^1} \Sigma$. Then  $X'$ is a singular 
threefold with finitely many ordinary double points corresponding to the nodal points in the fibers of $\pi$. 
Under the current assumption there is one nodal 
point $x'_\sigma\in X'$ for each $\sigma \in {S^\pi}$. 
Let $\wX\to X'$ be a small crepant resolution of singularities. 
Let ${\tilde \pi}: \wX \to \Sigma$ and 
${\tilde p}: \wX \to X$ be the natural projections. 
The exceptional locus consists of finitely many disjoint 
$(-1,-1)$ curves $\wC_\sigma$ on $\wX$, in one-to-one correspondence with points $\sigma \in {S^\pi}$. 
These are projective lines on $\wX$ 
with normal bundles isomorphic to $\CO(-1)\oplus \CO(-1)$. 
Each such curve is at the same time a $(-2)$-curve 
on the fiber $\wX_{\sigma}={\tilde \pi}^{-1}({\sigma})$, which is an embedded resolution of the nodal 
surface  $X_\sigma$. 

Let $U$ denote the
 Lorentzian rank two lattice generated by 
two null vectors vectors $u,u^*$ with 
$u\cdot u^*=1$. 
Let $\Lambda_{K3} \simeq U^{\oplus 3}\oplus \Lambda_{E_8}(-1)^{\oplus 2}$ be the 
middle homology lattice of a smooth generic K3 surface, 
where the $\Lambda_{E_8}(-1)$ denotes the $E_8$ root lattice 
equipped with a bilinear pairing given by the negative of the Cartan 
form. Let $\Lambda \subset \Lambda_{K3}$ be a sublattice of 
rank $1\leq \ell \leq 20$ and signature $(1, \ell-1)$, and 
let $(v_1, \ldots, v_\ell)$ be an integral basis of $\Lambda$. 
Let also $\Lambda^\vee$ be the dual lattice and 
$({\check v}^i)$, $1\leq i \leq \ell$ be the dual basis with respect 
to $(v_i)$.   

The pencil $\pi: X \to \IP^1$ will be assumed to satisfy the following 
additional conditions, which are easily satisfied for generic complete intersections in toric varieties. 
\begin{itemize}
\item[$(b)$] There exists a collection of divisor classes
 $H_1, \ldots, H_\ell\in {\rm Pic}(X)$, 
$m\geq 1$ such that 
the data $\big(\wX \to \Sigma,\ {\tilde p}^*H_1, \ldots, 
{\tilde p}^*H_\ell\big)$ is a family of $\Lambda$-polarized 
K3 surfaces as defined in \cite[Sect. 0.2.1]{NL_YZ}. 
In particular for any closed point $s\in \Sigma$ there is a 
primitive embedding $\Lambda \hookrightarrow {\rm Pic}(\wX_s)$ 
mapping $v_i$ to $\wH_{i,s} = {\tilde p}^*H_i|_{\wX_s}$ for all $1\leq i \leq \ell$. One also requires the 
existence of an element $\lambda \in \Lambda$ which 
is mapped to a quasi-polarization of $\wX_s$ for 
each $s\in \Sigma$. 

A stronger condition will be 
assumed here, namely that $ {\rm Pic}(X)\simeq H^2(X,\IZ)$ is freely generated by 
$H_1, \ldots, H_\ell$ and the K3 fiber class $D$. 
Moreover, there is a relatively ample class on $X$ over $\IP^1$ 
which 
restricts to $\lambda$ on each smooth fiber of $\pi$.

\item[$(c)$] For each point $\sigma \in {S^\pi}$ there is 
an orthogonal decomposition
\be\label{eq:picspecfibers}
{\rm Pic}(\wX_{\sigma}) \simeq \Lambda \oplus \IZ\langle \wC_\sigma\rangle
\ee 
with respect to the intersection  product. 

\item[$(d)$] For any sufficiently generic point $s \in \Sigma\setminus R^f$ the primitive embedding 
$\Lambda \hookrightarrow {\rm Pic}(\wX_s)$ is an isomorphism. 
\end{itemize}

\subsection{Noether-Lefschetz numbers}\label{NLdef} 
Next recall the definition of Noether-Lefschetz numbers for 
the family ${\tilde \pi}:\wX \to \Sigma$. 
Let $h\in \IZ$ and $d=\sum_{i=1}^\ell d_i {\check v}^i \in \Lambda^\vee$. Since the bases $(v_i)$, $({\check v}^i)$ are fixed, 
$d$ will be often  written as $d=(d_i)_{1\leq i\leq \ell}$. 
Informally the Noether Lefschetz number 
${\widetilde {NL}}_{h,d}\subset \Sigma$ is the number of 
points $s\in \Sigma$, counted with 
multiplicity, such that there exists a divisor 
class $\beta\in {\rm Pic}(\wX_s)$ satisfying
\be\label{eq:NLA} 
\beta^2 = 2h-2, \qquad \beta \cdot \wH_{i,s} = d_i, \quad 
1\leq i \leq \ell.
\ee

A rigorous definition of Noether-Lefschetz numbers
involves excess intersection theory, as shown in \cite[Sect. 1.4]{GW_NL} and \cite[Sect. 0.2.2, 0.2.3]{NL_YZ}. 
Following \cite[Sect. 1.4]{GW_NL},  consider the local system $\wcV=R^2{\tilde \pi}_*\IZ$ on $\Sigma$ and let 
$h : \wcH \to \Sigma$ be the ${\tilde\pi}$-relative 
moduli space of Hodge structures of type $(1,20,1)$ on 
$\wcV\otimes_\IZ \IC$. 
For any pair $(h,d)\in \IZ \times \IZ^{\ell}$ there exists a countable union of divisors
${\mathcal D}_{h,d}\subset \wcH$ 
parameterizing Hodge structures 
which contain a class $\beta \in \wcV_s$ 
satisfying 
conditions \eqref{eq:NLA}. 
One also has a canonical section 
$\phi: \Sigma \to \wcH$ such that $\phi(s) = [H^0(\wX_s, \IC)]\in \wcH_s$ for any $s\in \Sigma$. 
Then 
\be\label{eq:NLB}
{\widetilde {NL}}_{h,d} = \int_\Sigma \phi^*{\mathcal D}_{h,d}. 
\ee
According to \cite[Prop. 1]{GW_NL} the right hand side 
of equation \eqref{eq:NLB} is finite although  ${\mathcal D}_{h,d}$ may have infinitely 
many components. The proof of \cite[Prop. 1]{GW_NL}
shows that the image $\phi(\Sigma)$ intersects only finitely many of them.

\subsection{Local systems and jump loci}\label{jumpsect}
Given a pair $(h,d)\in \IZ\times \Lambda^\vee$ and a
point $s\in \Sigma$ let $\CB^{\tilde \pi}_s(h,d)$ denote the 
set of classes $\beta \in {\rm Pic}(\wX_s)$ satisfying 
conditions \eqref{eq:NLA}. This is a finite set by 
\cite[Prop. 1]{GW_NL}. The union 
$\CB^{\tilde \pi}(h,d) =\cup_{s\in \Sigma} 
\CB^{\tilde \pi}_s(h,d)\subset \wcV$ decomposes as 
\be\label{eq:NLC} 
\CB^{\tilde \pi}(h,d) = \wCB^{\sf iso}(h,d) \cup \wCB^{\infty}(h,d)
\ee
where $\wCB^{\sf iso}(h,d)$ projects to a finite subset 
of $\Sigma$, while
$\wCB^{\infty}(h,d)\subset \wcV$ is a local 
system over $\Sigma$. For any pair $(h,d)$ let 
$J^{\tilde \pi}_{h,d}\subset \Sigma$ be the projection of $\wCB^{\sf iso}(h,d)$ to $\Sigma$. This finite subset of $\Sigma$ will be called 
the jump locus of type $(h,d)$. 
The Noether-Lefschetz numbers decompose accordingly as 
\be\label{eq:NLD}
{\widetilde {NL}}_{h,d} = NL^{\sf iso}_{h,d}+ NL^{\infty}_{h,d}.
\ee
The first term in the right hand side of \eqref{eq:NLD} 
is a finite sum of the form 
\be\label{eq:isoNL}
NL^{\sf iso}_{h,d} = \sum_{s\in J^{\tilde \pi}_{h,d}}\ \sum_{\beta 
\in \wCB_s^{\sf iso}(h,d)} {\tilde \mu}(h,d,\beta)
\ee
where ${\tilde \mu}(h,d,\beta)\in \IZ$ is the intersection multiplicity 
of the section $\phi(\Sigma)$ with ${\mathcal D}(h,d)$ at 
the closed point corresponding to $\beta$. 

By definition, the
second term in the right hand side of \eqref{eq:NLD} 
is computed as follows. 
Note that there is a line bundle $\CK = R^0{\tilde \pi}_*\omega_{\tilde \pi}$ on $\Sigma$, where 
$\omega_{\tilde \pi}$ is the relative dualizing sheaf. 
Then 
\be\label{eq:NLE}
NL^{\infty}_{h,d} = -\int_{\wCB^{\infty}(h,d)} c_1(\CK). 
\ee
Using the natural inclusion $\Lambda\hookrightarrow \Lambda^\vee$ determined by the intersection form, 
condition $(d)$ in this section implies that 
\be\label{eq:NLF}
\wCB_{s}^\infty(h,d) = \left\{ \begin{array}{ll} 
 \{\alpha\}\subset \Lambda, & {\rm if}\  
d=\alpha\ {\rm for\ some}\ \alpha \in \Lambda\ {\rm and}\ h = 1+\alpha^2/2,\\
& \\
 \emptyset, & {\rm otherwise}. \\
\end{array}\right.
\ee
for any $s\in \Sigma$. 
Therefore $\CB^\infty(h,d)$ is either empty or a rank one local 
system on $\Sigma$. 
Since $X$ is $K$-trivial, one then obtains 
\be\label{eq:NLE}
NL^{\infty}_{h,d} = \left\{ \begin{array}{ll} 
 -4,  & {\rm if}\  
d=\alpha\ {\rm for\ some}\ \alpha \in \Lambda\ {\rm and}\ h = 1+\alpha^2/2,\\
& \\
 0, & {\rm otherwise}. \\
\end{array}\right.
\ee

For future reference consider the following example. 
Let  ${\sigma}\in \Sigma$ be a ramification point of $f$. 
Using the isomorphism 
\eqref{eq:picspecfibers}, 
 any  class $\beta \in 
{\rm Pic}(\wX_{\sigma})$ is written as 
\[ 
\beta = \alpha + l {\wC_\sigma} 
\]
with $l \in \IZ$ and $\alpha \in \Lambda$. 
This implies that $\beta\cdot \wH_{i,{\sigma}} = \alpha\cdot \wH_{i,{\sigma}}$ 
for $1\leq i \leq \ell$ and $\beta^2 = \alpha^2 
-2l^2$. 
Therefore, under the current assumptions, for any ramification point ${\sigma}\in R^f$, the component $\wCB^{\sf iso}_{\sigma}(h,d)$ 
is empty unless 
\[
(h,d)=(1+\alpha^2/2-l^2, \alpha)
\]
for some $\alpha \in \Lambda$, 
$l\in \IZ\setminus \{0\}$, in which case 
\[ 
\wCB^{\sf iso}_{\sigma}(1+\alpha^2/2-l^2,\alpha)\simeq \{ \alpha -lC_\sigma, 
\alpha + l C_\sigma\}. 
\]
Furthermore, \cite[Lemma 2]{NL_YZ},
 implies that 
\be\label{eq:multspecfiberA}
{\tilde \mu}(1-l^2, 0, \pm lC_\sigma) = 2
\ee
for any $\sigma\in R^f$, $l\in \IZ \setminus \{0\}$. 
More generally, by analogy with loc. cit., using condition (d) 
it can also be proved that
\be\label{eq:multspecfiberB}
{\tilde \mu}(1+\alpha^2/2-l^2, \alpha\pm lC_\sigma) = 2
\ee
for any $\sigma\in R^f$, $\alpha\in \Lambda$, $l\in \IZ \setminus \{0\}$. 

One can similarly define local systems and jump loci 
for the restriction of the family $\pi: X \to \IP^1$ to the 
open subset $U^\pi = \IP^1\setminus {S^\pi}$. For any $p\in U^\pi$ 
and any $(h,d)\in \IZ\times \Lambda^\vee$ 
let ${\mathcal B}^\pi_p(h,d)$ be the subset of classes 
$\beta \in {\rm Pic}(X_p)$ such that 
\be\label{eq:NLG}
\beta^2 = 2h-d, \qquad \beta\cdot {H_i}|_{X_p} = d_i, \quad 
1\leq i \leq \ell. 
\ee
By construction, $\CB^\pi_p(h,d) \simeq \CB^{\tilde \pi}_{s_1}(m,h,d) \simeq 
\CB^{\tilde \pi}_{s_2}(m,h,d)$ for any $p\in U^\pi$ and any $(h,d)$, 
where $f^{-1}(p)=\{s_1, s_2\} \subset \Sigma$. In particular  
all $\CB^\pi_p(h,d)$ are finite and the union 
$\CB^\pi(h,d) = \cup_{p\in U^\pi}\CB^\pi_p(h,d)$ 
decomposes again as 
\[
\CB^\pi(h,d) = \CB^{\sf iso}(h,d) \cup \CB^{\infty}(h,d)
\]
by analogy with  \eqref{eq:NLC}.
 Clearly, $\CB^{\sf iso}_p(h,d)
\simeq {\widetilde \CB}^{\sf iso}_{s_1}(h,d)\simeq 
{\widetilde \CB}^{\sf iso}_{s_2}(h,d)$ and 
$\CB^{ \infty}_p(h,d)
\simeq {\widetilde \CB}^{\infty}_{s_1}(h,d)\simeq {\widetilde \CB}^{\infty}_{s_2}(h,d)$ for any $p\in U^\pi$. Again, 
the jump locus $J^\pi_{h,d} \subset U^\pi$ is the projection of $\CB^{\sf iso}(h,d)$. Obviously, 
${J^{\tilde \pi}}_{h,d}\setminus R^f = f^{-1}(J^\pi_{h,d})$. Moreover, 
one can again define the intersection multiplicity $\mu(h,d,\beta)$ 
for any $\beta \in \CB^{\sf iso}(h,d)$. This will coincide 
with the multiplicity of the corresponding classes 
 $\beta_i\in {\widetilde \CB}^{\sf iso}_{s_i}(h,d)$, $1\leq i\leq 2$. Then equation \eqref{eq:multspecfiberB} yields
\be\label{eq:NLF} 
NL^{\sf iso}_{h,d} = 
\left\{ \begin{array}{ll} 
2 \sum_{p\in J^\pi_{h, d}} \sum_{\beta \in \CB^{\sf iso}_p(h,d)} \mu(h,d,\beta) + 2|S^\pi|, & {\rm if}\  
d=\alpha,\ h = 1+\alpha^2/2-l^2\\
&  {\rm for\ some}\ \alpha \in \Lambda,\ l\in \IZ\setminus\{0\},\\
2 \sum_{p\in J^\pi_{h, d}} \sum_{\beta \in \CB^{\sf iso}_p(h,d)} \mu(h,d,\beta) & {\rm otherwise}, \\
\end{array}\right.
\ee
which will be used in the computation of vertical 
D4-D2-D0 degeneracies in Section \ref{singfibers}. 

\subsection{Noether-Lefschetz numbers and modular forms}\label{NLmodular}  

According to \cite{Borcherds,KM} any  smooth lattice polarized K3 pencil determines a vector valued modular form which 
encodes all its Noether-Lefschetz numbers. This is briefly reviewed in 
\cite[Sect. 0.2.4]{NL_YZ}. 

Let $m= |{\rm det}(M)|$, 
where $M_{ij} = v_i \cdot v_j$ is the intersection matrix of the basis 
of $\Lambda$. Let $G_1=\Lambda^\vee/\Lambda$, where the injection 
$ \Lambda\hookrightarrow \Lambda^\vee$ is determined by the intersection form.
As explained in Appendix \ref{Weilreps} there is a canonical representation
$\rho_\Lambda: Mp(2,\IZ)\to {\rm End}(\IC[G_1])$ constructed by 
Weil \cite{Weil_rep},  where 
$Mp(2,\IZ)$ is the metaplectic double cover of $SL(2,\IZ)$. 

Using the isomorphism \eqref{eq:Weilrepisom}, 
the main result of \cite{Borcherds} yields the following 
modularity statement for Noether-Lefschetz numbers. 
For each pair $(h, d) \in {\IZ} \times\Lambda^\vee$, $d=(d_1, \ldots, d_\ell)$, let 
\[
\Delta(h,d) =(-1)^\ell {\rm det}\left(\begin{array}{cc} 
M & d^t \\ d & 2h-2 \end{array}\right). 
\]
Note that 
\[
{\Delta(h,d)\over 2 m} = 1+{d^2\over 2} - h.
\]
Then there is a vector valued modular form 
\[
{\widetilde \Phi}(q) = \sum_{\delta \in G_1} 
{\widetilde \Phi}_\delta(q)e_\delta \in  \IC[[q^{1/2m}]]\otimes \IC[G_1]
\]
of weight $w= {(22-\ell)/2}$ and type $\rho_\Lambda$ such that 
\be\label{eq:NLmodularA}
{\widetilde {NL}}_{h,d} = {\widetilde \Phi}_\delta \left[\Delta(h,d)/2m\right].
\ee
where $[d]_1= \pm \delta$. Here $\Psi[s]$, $s\in (1/ 2m)\IZ\subset \IQ$, $s\geq 0$ 
are the Fourier coefficients of the series $\Psi(q)\in \IC[[q^{1/2m}]]$, that is
$$\Psi(q) = \sum_{\substack{s\in (1/2m)\IZ, \\  s\geq 0}}
\Psi[s] q^s.$$
As immediate consequence, this implies 
\be\label{eq:algHodgeineq}
h\leq {d^2\over 2} +1 
\ee
for any class $\beta\in \CB^{\tilde \pi}(h,d)$, which can be proved directly using the algebraic Hodge theorem. 

Note that in its original form \cite[Thm 4.5]{Borcherds} 
implies the existence of such a vector valued modular form 
with values in $\rho_{\Lambda^\perp}^*$, where $\Lambda^\perp\subset \Lambda_{K3}$ is the sublattice 
consisting of all elements $u \in \Lambda_{K3}$, 
 $u \cdot \Lambda=0$. The above statement follows from the isomorphism \eqref{eq:Weilrepisom}. 

\section{Vertical BPS indices from adiabatic IIA/heterotic duality}\label{adiabatic}  

This section consists of a string theoretic derivation of the main formula 
\eqref{eq:mainformulaA} from adiabatic IIA/heterotic duality
for K3 fibrations. 

\subsection{Primitive charge vectors}\label{primitivesect}
Using the notation of Section \ref{verticalsheaves}, recall 
that the topological invariants of vertical two dimensional 
sheaves are given  by triples $\gamma=(r,d,n)\in 
\IZ\times \Lambda^\vee \times \IZ$ where 
$\Lambda$ is the polarizing lattice of the K3 pencil
$\pi: X \to \IP^1$. The dual lattice $\Lambda^\vee$ 
is naturally identified with the sublattice $H_2(X, \IZ)^\pi\subset H_2(X,\IZ)$ parameterizing vertical 
curve classes. 
In this section, $\gamma=(r,d,n)$ will be assumed to be  primitive. 
In this case all 
semistable vertical sheaves with invariants $\gamma$ are stable. 
According to \cite[Lemma 3.1]{DT_twodim}, any such sheaf $E$ must be the extension 
by  zero of a stable torsion sheaf $F$ on a reduced 
fiber $X_p$ of $\pi$. For any $p\in \IP^1$ let 
 $M_p(\gamma)$ denote the closed 
subspace of the coarse moduli space parameterizing isomorphism 
classes of stable sheaves $E$ supported on $X_p$. 
Recall that $U^\pi=\IP^1\setminus {S^\pi}$ is the open subset parameterizing smooth fibers. Let 
$M_{U^\pi}(\gamma)$ denote 
the open subset of the moduli space parameterizing stable sheaves 
$E$ supported on $X_p$ with $p\in U^\pi$. 
Then using equation \eqref{eq:weightedchi} one has 
\be\label{eq:omegasumA}
\Omega(\gamma) = \chi(M_{U^\pi}(\gamma), \nu) + \sum_{\sigma\in {S^\pi}} 
\chi(M_{\sigma}(\gamma), \nu). 
\ee

The first term in the right hand side of equation \eqref{eq:omegasumA} 
can be explicitly evaluated using the results of \cite{DT_twodim}. 
To explain this in some detail, let $X_p$ be a smooth fiber of $\pi$ and $\iota_p:X_p \hookrightarrow X$ denote the natural embedding. 
Then any stable sheaf $E$ suported on $X_p$ is the extension 
by zero, $E\simeq \iota_{p*}(F)$, of an $\omega|_{X_p}$-stable 
sheaf $F$ on $X_p$. The numerical invariants of $F$ are related 
to those of $E$ by the Grothendieck-Riemann-Roch formula:
\be\label{eq:GRRa}
\rk(F) =r, \qquad \iota_*\beta =d, \qquad 
k-{\beta^2\over 2} =n,
\ee
where 
$\beta = c_1(F)$ and 
$k=\int_{X_p} c_2(F)$. 
This implies that $M_p(\gamma)$ has disjoint 
components $M_{p,\beta}(\gamma)$ in one-to-one correspondence 
with classes $\beta \in {\rm Pic}(X)$ such that $\iota_*\beta = d$. 
Using the definition and the main properties of Noether-Lefschetz loci 
reviewed in Sections \ref{NLdef}, respectively \ref{jumpsect} and \ref{NLmodular}, the set of all such classes is a union 
\[
\bigcup_{\substack{d\in \Lambda^\vee,\ h \in \IZ\\ 
h \leq 1+d^2/2 }} \CB^{\pi}_p(h,d). 
\]
where $h = 1+ \beta^2/2$. 

For each $\beta \in \CB^\pi_p(h,d)$, $k\in \IZ$, let $M(X_p,r,\beta,k)$ 
be the moduli space of $\omega|_{X_p}$-stable torsion free sheaves on $X_p$ 
with numerical invariants $(r,\beta,k)$, where $k = n + \beta^2/2=n+ h-1$.
This is smooth and projective, 
of dimension
\be\label{eq:moddimA}
{\rm dim}\, M(X_p,r,\beta,k)= 2\left(rk - (r-1)\beta^2/2 - r^2 +1\right) = 2(rn -r^2 +h).
\ee
In fact, according to \cite{Symplectic_structure} and  \cite[Sect. 6]{huylehn}, 
for primitive invariants $(r,\beta,k)$, the moduli space 
$M(X_p, r,\beta, k)$ is a smooth deformation of a Hilbert scheme of points 
$H^{{\rm dim}(M(X_p, r,\beta, k))/2}(S)$ on a smooth algebraic 
K3 surface $S$. 
 
One can easily construct a closed embedding 
$M(X_p, r,\beta, k) \hookrightarrow 
M_{p,\beta}(\gamma)$ which yields an isomorphism between the two sets of closed points. This implies that the reduced scheme $M_{p,\beta}(\gamma)^{\sf red}$ is isomorphic to $M(X_p, r,\beta, k)$, 
hence $M_{p,\beta}(\gamma)$ has 
the same dimension as $M(X_p, r,\beta, k)$. 
In particular $M_{p,\beta}$ is nonempty if and only if 
\be\label{eq:hboundA}
h \geq r(r-n).
\ee 
Therefore the disjoint components of $M_p(\gamma)$ are in one-to-one correspondence 
with elements of 
\[
\bigcup_{\substack{d\in \Lambda^\vee,\ h \in \IZ\\ 
r(r-n)\leq h \leq 1+d^2/2 }} \CB^{\pi}_p(h,d),
\]
which is a finite set for fixed $\gamma=(r,d,n)$, possibly 
empty. Note however that in general 
$M_{p,\beta}(\gamma)$ will not be isomorphic to $M(X_p, r,\beta, k)$ as a scheme since its structure sheaf can in principle contain nilpotent elements. 
The different scheme structure 
of $M_{p,\beta}(\gamma)$ will lead to nontrivial 
values of the Behrend function, as explained below equation 
\eqref{eq:omegasumB}. 

To summarize, one has a decomposition 
\be\label{eq:chiMpA}
\chi(M_{p}(\gamma),\nu) = 
\sum_{\substack{h\in \IZ\\  r(r-n)\leq h \leq 1+d^2/2}} \sum_{\beta \in \CB^\pi_p(h,d)}  
\chi(M_{p,\beta}(\gamma),\nu),
\ee
where the sum in the right hand side
is finite.
Next recall that for any $p\in U^\pi=\IP^1\setminus {S^\pi}$ the set 
$\CB^\pi_{p}(h,d)$ decomposes as  $\CB^\pi_p(h,d)=\CB^\infty_p(h,d)\cup \CB^{\sf iso}_p(h,d)$.
Note that $M_{p, \beta}(\gamma)$ is an isolated  closed component of the moduli space
for each $\beta \in \CB^{\sf iso}_p(h,d)$. 
Using the results of \cite{micro}, its contribution to the right hand side of \eqref{eq:chiMpA} follows 
from \cite[Thm. 3.18]{DT_twodim}, 
\be\label{eq:chiMpB} 
\chi(M_{p,\beta}(\gamma), \nu) = \mu(h,d,\beta) \chi(M(X_p,r,\beta,k)) = \mu(h,d,\beta) c(r(n-r)+h)),
\ee
where 
\[ 
c(r(n-r)+h) = \left\{\begin{array}{ll} 
\chi(H^{r(n-r)+h}(S)), & {\rm if}\ h\geq r(r-n), \\ 
0 & {\rm otherwise}.\\ \end{array}\right. 
\]
In the above formula $H^k(S)$ denotes the Hilbert scheme of 
$k$ points on a smooth generic algebraic K3 surface $S$. 
The coefficients $c(r(n-r)+h)$ are determined by  G{\"o}ettsche's formula \cite{Betti_Hilbert} applied to K3 surfaces, 
\be\label{eq:Hilbgen}
q^{-1}\sum_{k=0}^\infty \chi(H^k(S)) q^k = {1\over \eta(q)^{24}}.
\ee
The coefficient $\mu(h,d,\beta)$ is the same as the contribution of the isolated class $\beta$ 
to the Noether-Lefschetz number in \eqref{eq:isoNL}. 

Furthermore, according to 
condition $(e)$  in Section \ref{jumpsect},
\[
\CB^\infty_p(h,d)= \left\{ \begin{array}{ll} 
 \CB_p^\infty(\alpha)=\{\alpha\}\subset \Lambda, & {\rm if}\ (h,d) = (h(\alpha),\alpha)\ {\rm for\ some}\ \alpha \in \Lambda, \\
& \\
 \emptyset, & {\rm otherwise}, \\
\end{array}\right.
\]
where $h(\alpha) =1 +\alpha^2/2$. Therefore, if $d \neq \alpha$ for some $\alpha \in \Lambda$, the moduli space 
$M_{U^\pi}(\gamma)$ will be a finite union of isolated closed 
components whose contributions are given by \eqref{eq:chiMpB}. 
If $d=\alpha$ for some $\alpha \in \Lambda$, according to  \cite[Lemma 3.7]{DT_twodim}, there is a smooth connected 
component $M^\infty_{U^\pi}(\gamma)$ of $M_{U^\pi}(\gamma)$ 
whose set of closed points coincides with the union 
$\cup_{p\in{ U^\pi}} M_{p,\alpha}(\gamma)$. Moreover for each $p\in U$, there is an isomorphism $M_{p,\alpha}(\gamma) \simeq M(X_p, r,\alpha, k)$, 
with $k= n +\alpha^2/2$. 
Since $M(X_p, r,\alpha, k)$ 
is smooth and projective of dimension \eqref{eq:moddimA}, this implies that 
\be\label{eq:chiMpC}
\chi(M_{p,\alpha}(\gamma), \nu) = - \chi(M_{p,\alpha}(\gamma)) 
= -c(r(n-r)+h(\alpha))
\ee
and 
\be\label{eq:chiMU}
\chi(M_U^\infty(r,\alpha, n), \nu) = -\chi(U^\pi)c(r(n-r)+h(\alpha)).
\ee
Therefore, for any $\gamma=(r,d, n)$, the contribution of 
$M_{U^\pi}(\gamma)$ to the Donaldson-Thomas invariant $\Omega(\gamma)$ is 
\be\label{eq:omegasumB} 
\bal
\chi(M_{U^\pi}(\gamma), \nu) =  & - \chi(U^\pi) c(r(n-r)  +d^2/2+1) \delta_{d, \Lambda} \\
& + 
\sum_{\substack{h\in \IZ\\  r(r-n)\leq h \leq d^2/2+1}} 
\sum_{\beta \in \CB^{\sf iso}(h,\alpha)} \mu(h,d,\beta) c(r(n-r)+h),\\
\eal
\ee
where 
\[ 
\delta_{d,\Lambda} = \left\{\begin{array}{ll} 
1,& {\rm if}\ d\in \Lambda, \\ 
0,& {\rm otherwise.}\\ \end{array}\right.
\]

For completeness note that the weights $\mu(h,d,\beta)$ in 
equation \eqref{eq:chiMpB} have a clear physical interpretation. 
This was first observed in a similar context in \cite[App. G]{Denef:2007vg}. 
Namely, one can easily check that any vertical stable D4-D2-D0 
configuration has exactly one normal infinitesimal deformation 
corresponding to translations along the base of the K3 fibration. 
More precisely, given a stable vertical sheaf $E= \iota_{p*}(F)$ 
supported on a 
reduced K3 fiber $X_p$ one can easily check that the space ${\rm Ext}^1_X(E,E)$ of infinitesimal deformations splits as 
\[ 
{\rm Ext}^1_X(E,E) \simeq {\rm Ext}_{X_p}^1(F,F) \oplus {\rm Ext}^0_{X_p}(F,F). 
\] 
The first summand parameterizes infinitesimal deformations 
of $F$ as a sheaf on $X_p$, while the second parameterizes 
normal deformations in the Calabi-Yau threefold $X$. 
Moreover, stability implies that ${\rm Ext}^0_{X_p}(F,F)\simeq \IC$ 
is one dimensional. This means that the low energy effective 
action of the corresponding  D4-D2-D0 configuration will contain 
a complex  chiral fields $\CX_1, \ldots, \CX_{d-1}$ associated to tangent fluctuations and an additional chiral field 
$\Phi$ associated to normal fluctuations to the fiber. 
Here $d= {\rm dim}\, {\rm Ext}^1_X(E,E)\geq 1$. 
Since the moduli space $M(X_p, r, \beta, k)$ is smooth, the tangent  deformations parameterized by 
$\CX_1, \ldots, \CX_{d-1}$ are unobstructed. However, 
if $\beta = c_1(F)$ is an isolated curve class on $X_p$,
the normal deformations of $E$ will be obstructed. This will 
be encoded in a superpotential interaction 
$W(\CX_1, \ldots, \CX_d, \Phi)$ such that the critical 
scheme defined by $dW=0$ is locally isomorphic to a 
nilpotent extension of the moduli space $M(X_p,r,\beta,k)$. 
Then using equation \eqref{eq:nuvalue}, the 
value of the Behrend function $\nu([E])$ at the point $[E]$ will be determined by the resulting nilpotent extension. 
In principle, $\nu([E])$ may jump as $[E]$ moves in the moduli space. 
However,  it is natural 
to conjecture it is constant along $M_{p, \beta}(\gamma)$ 
and takes value $\nu([E])=\mu(h,d,\beta)$ at all points. This is certainly in agreement with equation \eqref{eq:chiMpB}. While a rigorous proof 
would be quite difficult, intuitively one expects this to be the case 
since the only obstructions to the normal deformations of $E$ 
come from obstructions to normal deformations of the curve class 
$\beta$, which are independent of $E$.

\subsection{Singular fibers and adiabatic IIA/heterotic duality}\label{singfibers}

In order to finish the computation 
one has to evaluate the contributions of the singular fibers
 $X_\sigma$, $\sigma \in \Sigma$ to the right hand 
side of equation \eqref{eq:omegasumA}. 
The presence of singularities 
makes a direct geometric approach difficult. However one can gain 
important insight using fiberwise heterotic/IIA duality for
the K3 fibration $\pi: X\to \IP^1$. 
Since the Donaldson-Thomas 
invariants are independent of the K\"ahler class $\omega$, the latter 
can be chosen such that the volume of the section of $\pi$ 
is much larger than that of the K3 fibers. In this regime, it is natural 
to define a constructible function $\mu: \IP^1 \to \IZ$, 
\be\label{eq:mufctA} 
\mu(p) = \chi(M_p(\gamma), \nu). 
\ee 
Clearly, the value of $\mu$ at $p$ represents the contribution 
of the fiber $X_p$ to the Donaldson-Thomas invariant. More concretely, one 
can write 
\be\label{eq:muomegaA}
\bal
\Omega(\gamma)  = \chi(\IP^1, \mu).
\eal 
\ee
The main idea emerging from 
heterotic/IIA 
duality is that the value
of $\mu$ at a point $p\in \IP^1$ must be related to degeneracies of perturbative BPS states for  a $T^4$ compactification of the $E_8\times 
E_8$ heterotic string.   A concrete conjecture will be formulated below. 

First recall that six dimensional heterotic/IIA duality 
identifies a $T^4$ compactification of the $E_8\times E_8$ heterotic 
string to a K3 compactification of the IIA string. The heterotic Narain lattice $\Gamma_{4,20}$ is isomorphic to a direct sum $U\oplus 
\Lambda_{K3}$, where $\Lambda_{K3} \simeq U^{\oplus 3} \oplus \Lambda_{E_8}(-1)^{\oplus 2}$ is the middle homology lattice of a smooth generic K3 surface. This identification singles out a topologically nontrivial  circle $S^1_A\subset T^4$  
corresponding to the first $U$ summand. 

The conformal field theory moduli space $\CM_{\sf het}$ 
of the $E_8\times E_8$
heterotic string on $T^4$ is a quotient of the form 
\[ 
\CM_{\sf het} = {\rm Aut}(\Gamma_{4,20})\backslash 
{\widetilde \CM}_{\sf het}
\]
where ${\widetilde \CM}_{\sf het}=O^+(4,20)/SO(4)\times O(20)$
and  ${\rm Aut}(\Gamma_{4,20})$ is the automorphism group of the 
Narain lattice, acting naturally on the coset space.
The latter is isomorphic to the grassmannian of space-like 4-planes $\Pi\subset \Gamma_{4,20}\otimes_\IZ\IR$, hence 
it is a smooth complex manifold. The quotient by the $T$-duality group 
will have orbifold singularities. 

According to \cite[Thm. 6]{K3_string}, a certain open subspace of 
this moduli space is precisely identified 
with the moduli space of conformally invariant nonlinear sigma 
models with target space K3, including metric and $B$-field moduli. 
Hence $\CM_{\sf het}$ is in fact a compactifiction of the sigma 
model moduli space. As explained for example in 
\cite[Sect. 4.3]{K3_string} certain points in $\CM_{\sf het}$
correspond to nonperturbative IIA compactification on 
K3 surfaces with ${\sf ADE}$ quotient singularities. Such points are associated to six dimensional gauge symmetry enhancement. 
More precisely, for a generic point in the moduli space, the six dimensional gauge group is $U(1)^{24}$. Let 
${\widetilde {\mathcal D}} \subset 
{\widetilde {\CM}}_{\sf het}$ be the locus where the spacelike 4-plane 
$\Pi$ is orthogonal to some vector $\beta \in \Gamma_{4,20}$, $\beta^2=-2$. Then the abelian gauge group is enhanced to 
$SU(2)\times U(1)^{23}$ at generic points on 
${\mathcal D}= O(\Gamma_{4,20})\backslash {\widetilde {\mathcal D}}
\subset \CM_{\sf het}$.  
For the purposes of the present discussion, it should be emphasized that the points on ${\mathcal D}$ parameterize smooth well behaved 
heterotic conformal field theories, although the corresponding  K3 surfaces in IIA theory develop $A_1$ singularities.

The duality also leads to a precise 
indentification of the Hilbert spaces of six dimensional 
BPS states in the two string theories, as explained for example in \cite{Alg_BPS,prec_counting}. 
As shown in \cite{exact_deg}, \cite[Sect. 6.2]{prec_counting}, D4-D2-D0 BPS states with charge vector $\gamma=(r,\beta, k)$ supported on $S$ are in one-to-one correspondence with certain perturbative heterotic string states with momentum $k$ and 
winding number $r$ on the circle $S^1_A$. These states are obtained 
by tensoring the ground state of the right moving superconformal sector with a level $N$ state of the bosonic left moving sector, where $N$ is determined by level matching:
\be\label{eq:level}
N = r(k-r)-(r-1)\beta^2/2 +1. 
\ee
These are the Dabholkar and Harvey states considered in 
\cite{DH} in relation to black hole entropy. A general formula 
for the degeneracies of such states is derived in 
\cite[Sect 3]{exact_deg}. 

Assuming $S$ to be algebraic, note
 that $N$ is half the dimension of the moduli space of 
stable torsion free sheaves on $S$. 
Since the left moving sector consists of 24 bosons, it follows that 
the degeneracy of these states is the $N$-th coefficient $c_N$ 
in the 
expansion of $q/\eta(q)^{24}$, in agreement with Goettsche's 
formula \eqref{eq:Hilbgen}. This follows from  
\cite[Sect. 3]{exact_deg} as well as \cite[Sect. 6]{prec_counting}.

Finally, suppose $S$ is a singular algebraic K3 surface with a single 
node corresponding to a generic point in ${{\wCD}}$, and let $\wS$ be its minimal crepant resolution. Let $\beta_C \subset 
H^2(\wS,\IZ)$ denote the Poincar\'e dual of the exceptional $(-2)$-curve $C\subset \wS$. Note that $\beta_C$ is identified with a root 
vector of one of the $E_8$ sublattices of $H_2(\wS, \IZ)$. 
 As explained above, the dual heterotic conformal field theory 
associated to $S$ is 
still smooth, except that the six dimensional gauge group of the 
 corresponding six dimensional vacuum is 
enhanced to $SU(2)$. The extra massless $W$-bosons correspond to 
heterotic vertex operators associated to the root $\beta_C$. 
In particular the six dimensional theory exhibits a gauge symmetry 
which maps $\beta_C \mapsto -\beta_C$. This is the action 
of the generator of the Weyl group of the enhanced $SU(2)$ gauge 
group. 

The degeneracies of DH states in the conformal theory associated 
to the nodal surface $S$ are exactly the same as those computed 
in the conformal field theory associated to the blow-up $\wS$. 
This is manifest from the counting algorithm, which is independent 
of deformations of conformal field theory as long as the theory remains smooth. However, since the reflection $\beta_C \mapsto -\beta_C$ is a 
gauge symmetry, any two states DH related by this reflection are physically identical, so such a pair should be counted only once 
in the six dimensional BPS spectrum. 

Returning to the family of $\Lambda$-polarized algebraic K3 surfaces  $\pi : X \to \IP^1$, note that this 
family cannot be canonically identified with a family of heterotic 
conformal field theories since the restriction of the Calabi-Yau 
threefold metric to a K3 fiber need not be hyper-K\"ahler.  
However, the BPS index for D4-D2-D0 states supported on a fiber is independent of metric perturbations, 
hence one can still derive a precise conjecture for 
the constructible function $\mu: \IP^1 \to \IZ$ by counting perturbative heterotic string states. 

Using equations \eqref{eq:chiMpA},   \eqref{eq:chiMpB} and  \eqref{eq:chiMpC}, the contribution of the fiber $X_p$, $p\in U^\pi$, 
 to the four dimensional 
BPS index of charge $\gamma=(r,d, n)$ is given by 
\be\label{eq:smoothmu}
\mu(p)= - 
c(r(n-r)+d^2/2+1)\delta_{d,\Lambda}  + \sum_{\substack{
h\in \IZ, \\  r(r-n)\leq h \leq d^2/2+1\\}} \sum_{\beta \in \CB^{\sf iso}(h,d)}
 \mu(h,d,\beta) 
c(r(n-r)+h).
\ee
The sum in the right
hand side represents the contribution of all charge vectors 
$\beta\in H^2(X_p, \IZ)$ which yield the same charge vector $\alpha$ with respect to the four-dimensional abelian gauge group i.e. $\iota_*\beta =\alpha$. Aside from the weights $\mu(h,d, \beta)$, 
the contribution of each class $\beta$ is given by the corresponding 
degeneracy of heterotic DH states. 
As explained below \eqref{eq:omegasumB}, 
for any class 
$\beta \in \CB^{\sf iso}_{p}(h,d)$, the weight $\mu(h,d,\beta)$ 
represents the vacuum multiplicity of the corresponding BPS D4-D2-D0 
configurations supported on $X_p$. 

Moreover,
recall that the 
components $M_{p,\alpha}(\gamma)$ of the moduli space of D4-D2-D0 branes on $X_p$ 
fit in the smooth family $M_{U^\pi}^\infty(\gamma)$ over $U^\pi$. Therefore, according to formula \eqref{eq:smoothnuvalue}, 
the contribution of degeneracies of states with $\beta=\alpha$ 
to the four dimensional index 
should be weighted by $(-1)$, which encodes 
their four dimensional spin quantum number.

Employing heterotic/IIA duality as above,  it follows that the contribution $\mu(\sigma)$, 
$\sigma \in {S^\pi}$, of a singular fiber can be inferred 
from counting D4-D2-D0 bound states supported on its 
blow-up $\wX_\sigma$. The main point is that, choosing 
appropriate hyper-K\"ahler metrics on $X_\sigma, \wX_\sigma$, 
one obtains smooth heterotic conformal field theories related 
by a smooth deformation. Hence the DH degeneracies as well 
as the four dimensional spin quantum number will be the same 
in the two theories. The only difference is the gauge symmetry 
$\beta_C \mapsto -\beta_C$ in the six dimensional vacuum 
associated to the nodal surface, which implies that 
DH states with charge $\alpha + l \beta_C$ are physically indistinguishable from DH states with charge $\alpha - l \beta_C$. Working under the genericity assumptions 
formulated in Section \ref{setup}, the multiplicity of all
curve classes supported on $\wX_\sigma$ is given by equation 
\eqref{eq:multspecfiberB}. Therefore, collecting the facts, one is led to the following conjectural expression 
\be\label{eq:singmu} 
\mu(\sigma) = -c(r(n-r)+d^2/2+1)\delta_{d,\Lambda} + \sum_{\substack{ 
l\in \IZ,\
l \geq 1 \\ l^2 \leq r(n-r)+d^2/2+1\\ } } 
c(r(n-r) +d^2/2+1 - l^2)\delta_{d,\Lambda}
\ee  
for the contribution of a singular nodal fiber to the BPS index. 
Using equations \eqref{eq:smoothmu} and \eqref{eq:singmu} in 
 equation \eqref{eq:muomegaA}, one then obtains 
\[
\bal 
\Omega(\gamma) 
= & -\chi(\IP^1) c(r(n-r)+d^2/2+1)\delta_{d,\Lambda} +
\sum_{\substack{h\in \IZ\\  r(r-n)\leq h < d^2/2+1}} 
\sum_{\beta \in \CB^{\sf iso}(h,d)} \mu(h,d,\beta) c(r(n-r)+h)\\
& + |{S^\pi}| \sum_{\substack{
l\in \IZ,\ l \geq 1 \\ l^2 \leq r(n-r)+d^2/2+1\\ } } 
c(r(n-r) + d^2/2+1 - l^2) \delta_{d,\Lambda}.\\
\eal 
\]
Finally,  using equations \eqref{eq:NLE},  \eqref{eq:NLF} the above formula 
can be rewritten as 
\be\label{eq:omegaNLa} 
\Omega(\gamma) ={1\over 2} \sum_{\substack{ h\in \IZ\\  r(r-n)\leq h \leq d^2/2+1}} c(r(n-r)+h) {\widetilde {NL}}_{h,d}
\ee
where ${\widetilde {NL}}_{h,d}$ are the Noether-Lefschetz numbers of the 
threefold $\wX$ constructed in Section \ref{setup}.

\subsection{Generating functions for primitive charge vectors }\label{genfctsect} 

Suppose the pair $(r,\delta) \in \IZ_{\geq 1}\times \Lambda^\vee/r\Lambda$ is primitive. This means there
is no integer $k \in \IZ$, $k\geq 1$, such that $k|r$ 
and $\delta = [kd']_r$ with $d'\in \Lambda^\vee$. 
Then any charge vector $\gamma=(r,d,n)$ with 
$\delta = [d]_r$ will 
be primitive. Recall that the rank $r$ partition function of vertical 
D4-D2-D0 invariants is an expression of the form \eqref{eq:strgenfctC}, where 
\[
Z_{BPS}(X,r,\delta; \tau) = \sum_{n\in \IZ} \Omega(r,d, n) 
q^{n+d^2/2r-r}. 
\]
In the above formula $d\in \Lambda^\vee$ is a fixed arbitrary representative of $\delta\in \Lambda^\vee/r\Lambda$. For any $r\geq 1$ let $G_r = \Lambda^\vee/ r\Lambda$ and let $[d]_r\in G_r$ denote the 
equivalence class of $d\in \Lambda^\vee$. 
Let $h(d)\in \IZ$ be defined by 
\[ 
{d^2\over 2} +1 = {h(d)\over 2m}
\]
where $m = |{\rm det}(M)|$. 
Recall that the Noether-Lefschetz numbers ${\widetilde {NL}}_{h,d}$ are identified in equation \eqref{eq:NLmodularA} with the Fourier coefficients 
of a vector valued modular form ${\widetilde \Phi}(q)$ with 
values in the Weil representation of $\Lambda$. 
Then equations \eqref{eq:NLmodularA} and  \eqref{eq:omegaNLa} yield
\[ 
Z_{BPS}(X,r,\delta;\tau) = {1\over 2} \sum_{\substack{ n\in \IZ \\ 
r(r-n) \leq h(d)/2m}} q^{n+d^2/2r-r} 
\sum_{r(r-n)\leq h \leq h(d)/2m} 
c(r(n-r)+h){\widetilde \Phi}_{[d]_1}[h(d)/2m-h],
\]
Let 
\[
{l\over 2m} = r(n-r) + {h(d)\over 2m}, \qquad k = r(n-r)+h
\]
with $k,l\in \IZ$. In particular 
\[ 
n - r= {l-h(d)\over 2rm} \in \IZ. 
\]
Then the right hand side of the above equation can be 
written as 
\[
\bal 
& {1\over 2}q^{d^2/2r}\sum_{\substack{
l\in \IZ,\ l\geq 0,\\ 
(l-h(d))/2rm \in \IZ}} q^{(l-h(d))/2rm}
\sum_{\substack{k\in \IZ\\ 0\leq k \leq l/2m}} c({k}) {\widetilde \Phi}_{[d]_1}[l/2m -k]=\\
& {1\over 4rm} q^{d^2/2r}\sum_{l \in \IZ,\ l\geq 0} q^{(l-h(d))/2rm}
\sum_{s=0}^{2rm-1} e^{2\pi i (l-h(d)) s/2rm} 
\sum_{\substack{k\in \IZ\\ 0\leq k \leq l/2m}} c({k}) {\widetilde \Phi}_{[d]_1}[l/2m -k].\\
\eal
\]
Let $\Delta(q) =  \eta(q)^{24}$, which is a modular form of weight $(-12)$. Note that 
$\Delta^{-1}(q)= q^{-1} \sum_{k\geq 0} c(k) q^k$
by Goettsche's formula \eqref{eq:Hilbgen}.  Then one 
has a series identity 
\[
\sum_{l \in \IZ,\ l\geq 0} q^{l/2rm}
e^{2\pi i ls/2rm} 
\sum_{\substack{k\in \IZ\\ 0\leq k \leq l/2m}} c({k}) {\widetilde \Phi}_{[d]_1}[l/2m -k] = q^{1/r} e^{2\pi is/r} \big(\Delta^{-1} 
{\widetilde \Phi}_{[d]_1}\big)\left({\tau +s\over r}\right).
\]
Therefore 
\be\label{eq:omegaNLc}
\bal
Z_{BPS}(X,r,\delta; \tau) & = 
{1\over 4rm}q^{d^2/2r} q^{(1 -h(d)/2m)/r} 
\sum_{s=0}^{2rm-1} e^{2\pi i s(1 - h(d)/2m)/r} 
\big(\Delta^{-1}{\widetilde \Phi}_{[d]_1}\big)\left({\tau +s\over r}\right)\\
& = {1\over 4rm} 
\sum_{s=0}^{2rm-1} e^{-\pi i sd^2/r} 
\big(\Delta^{-1}{\widetilde \Phi}_{[d]_1}\big)\left({\tau +s\over r}\right).\\
\eal
\ee
Since 
$\Delta^{-1}{\widetilde \Phi}(q)$ is a vector valued modular form 
of 
weight $(-1-\ell/2)$ with values in the Weil representation 
$\rho_\Lambda$, one has 
\[
\big(\Delta^{-1}{\widetilde \Phi}_{[d]_1}\big)\left({\tau +s+kr\over r}\right) = e^{\pi i k d^2} \big(\Delta^{-1}{\widetilde \Phi}_{[d]_1}\big)\left({\tau +s\over r}\right). 
\]
At the same time 
\[
e^{-\pi i (s+kr) d^2/r} = e^{-\pi i k d^2} e^{-\pi i sd^2/r}. 
\]
Therefore each term in the right hand side of equation \eqref{eq:omegaNLc} is invariant under $s\mapsto s+ kr$, $k\in \IZ$. 
Then equation \eqref{eq:omegaNLc} reduces to  
\be\label{eq:omegaNLcc} 
Z_{BPS}(X,r,\delta;\tau) = {1\over 2r} 
\sum_{s=0}^{r-1} e^{-\pi i sd^2/r} 
\big(\Delta^{-1}{\widetilde \Phi}_{[d]_1}\big)\left({\tau +s\over r}\right).
\ee
In particular, for $r=1$ one obtains $Z_{BPS}(X,1,\delta;\tau) = 
\big(\Delta^{-1}{\widetilde \Phi}_{\delta}\big)\left(\tau\right)$ for 
any $\delta \in \Lambda^\vee/ \Lambda$, 
in agreement with the results of \cite{DT_twodim}. 
As required by physical arguments \cite{M5_genus,Farey_tail,Denef:2007vg}, 
the collection $(Z_{BPS}(X,1,\delta;\tau))_{\delta \in \Lambda^\vee/\Lambda}$ 
determines a weight $(-1-\ell/2)$ vector-valued modular form 
with values in the Weil representation. 

\subsection{Non-primitive charge vectors and multicover contributions}\label{multicoversect}

To conclude this section suppose $\gamma$ is not primitive. 
Then in general there will exist strictly semistable objects in the 
moduli space of stable sheaves, making the theory of Donaldson-Thomas 
invariants more difficult. In particular one has the rational 
invariants $DT(\gamma)$ which are related to the integral ones 
$\Omega(\gamma)$ by the multicover formula \eqref{eq:multicover},
\[
DT(r,d,n) = \sum_{\substack{ k \in \IZ,\ k\geq 1\\(r,d,n) = k (r',d',n')}} 
{1\over k^2} \Omega(r',d',n').
\]
For any pair $(r,\delta)$, $\delta \in \Lambda^\vee/ r\Lambda$,  let 
\[
Z_{DT}(X,r,\delta; \tau )= \sum_{n\in \IZ} DT(r,d,n) q^{n+d^2/2r-r},
\]
where $d\in \Lambda^\vee$ is an arbitrary representative of $\delta$. 
Again, the right hand side of the above equation depends only 
on the equivalence class $\delta$ of $d$ mod $r\Lambda$ 
since the rational Donaldson-Thomas invariants are invariant 
under transformations \eqref{eq:chargemdrmy} as well. 
In order to evaluate this series, first note that for any pair $(k, r')$ 
with $kr'=r$ there is an injective morphism  
\[ 
f_{r',k} : \Lambda^\vee/ r'\Lambda \to \Lambda^\vee/ r \Lambda, \qquad 
f_{r',k}([d]_{r'}) = [kd]_r, \quad {\rm for\ all}\ d \in \Lambda^\vee. 
\]
Then, using the above multicover formula, the generating 
functional is written as 
\be\label{eq:refeq}
\bal 
Z_{DT}(X,r,\delta; \tau )
 & = \sum_{\substack{ k\in \IZ,\ k\geq 1\\ (r,d)=k(r',d')}} 
{1\over k^2} \sum_{n'\in \IZ} q^{k(n'+{d'}^2/2r'-r')}\Omega(r',d',n') \\
& = 
\sum_{\substack{ k\in \IZ, k\geq 1\\ r = kr', \ \delta = f_{r',k}(\delta')} }
{1\over k^2}Z_{BPS}(X,r',\delta'; k\tau).
\eal
\ee
In the right hand side, $\delta' \in \Lambda^\vee/ r'\Lambda$ is 
uniquely determined by $(k, \delta)$ since $f_{r',k}$ is injective. 

For the next step
one needs a generalization of the 
conjectural formula \eqref{eq:omegaNLc} to all pairs
$(r,\delta)$, not just primitive ones.  On physics grounds, the natural conjecture at this point is that \eqref{eq:omegaNLc} is in fact valid for all such pairs,  including non-primitive ones. The main physical argument 
for this conjecture is based on modularity constraints. Physics arguments based on S-duality \cite{Denef:2007vg} or M5-brane 
elliptic genus \cite{M5_genus,Farey_tail} 
imply that 
the collection of partition functions 
$Z_{DT}(X,r,\delta; \tau )$, 
$\delta \in \Lambda^\vee/r\Lambda$  must be a meromorphic
vector valued modular form of weight $(-1 - \ell/2)$. 
This vector valued modular form
must take values in a finite dimensional unitary representation of the metaplectic cover of $SL(2, \IZ)$   
on the $\IC$-linear span $\IC[\Lambda^\vee/r\Lambda]$. 

Granting this statement, it follows that 
the whole generating function $Z_{DT}(X,r,\delta; \tau )$
is completely determined by the Donaldson-Thomas invariants for primitive charges. The main point  is that although $(r,\delta)$ is non-primitive, for any representative $d \in \Lambda^\vee$ of $\delta$, there are infinitely many values of $n\in \IZ$ such that 
$\gamma =(r,d,n)$ is primitive. By the arguments of the previous 
section, the conjectural formula 
\eqref{eq:omegaNLa} will apply to all such values. Then
the generating function $Z_{BPS}(X,r,\delta; \tau)$ 
will be given 
by equation \eqref{eq:omegaNLc} for all pairs $(r, \delta)$ since the 
vector space of weight $(-1-\ell/2)$ vector valued modular forms is finite dimensional. As shown below this leads to 
the final expression  \eqref{eq:omegaNLd} for the partition 
function of rational invariants. In Section \ref{modularsect},
it will be shown that these generating functions are indeed the coefficients of a weight $(-1-\ell/2)$ vector valued modular form with values in the Weil representation associated to the lattice ${\sqrt{r} \Lambda} \subset \Lambda_\IR$. Moreover, 
the vector space of such vector valued modular forms 
is indeed finite dimensional. Further confimation 
of this conjecture will be obtained 
in Section \ref{Vert_DT_GV} employing a more 
mathematical approach. 

Granting formula \eqref{eq:omegaNLc} for all charge 
vectors, 
one obtains 
\[
Z_{DT}(r,\delta; \tau) = \sum_{\substack{ k\in \IZ,\ k\geq 1\\ 
r=kr', \ \delta= f_{r',k}(\delta')}} 
{1\over k^2} {1\over 2r'} \sum_{s=0}^{r'-1} 
e^{-\pi i s(d')^2/r'}\big(\Delta^{-1}{\widetilde \Phi}_{[d']_1}\big)\left({k\tau +s\over r'}\right).
\]
Now note that for any $l\in \IZ$, $l\geq 1$, the $\IQ$-valued symmetric bilinear pairing 
of $\Lambda^\vee$ induces a $\IQ/\IZ$-valued symmetric bilinear pairing $({}\ ,\ )_l$ on $\Lambda^\vee/l\Lambda$ given by 
\[ 
([d]_l, [d']_l)_l = {d\cdot d'\over l} \quad {\rm mod}\ \IZ. 
\]
Moreover, since the intersection pairing on $\Lambda$ is even, there is a well defined $\IQ/\IZ$-valued 
quadratic form $\theta_l : \Lambda^\vee/l\Lambda\to 
\IQ/\IZ$, 
\[
\theta_l(\eta) ={ (\eta, \eta)_l\over 2}.
\]
Then the  formula can be further rewritten as 
\be\label{eq:omegaNLd}
Z_{DT}(X,r,\delta; \tau) = {1\over 2r^2} 
\sum_{\substack{ k\in \IZ,\ k\geq 1\\ 
r=kl, \ \delta= f_{l,k}(\eta)}} 
\sum_{s=0}^{l-1} l
e^{-2\pi i s \theta_l(\eta)}\big(\Delta^{-1}{\widetilde \Phi}_{[\eta]_1}\big)\left({k\tau +s\over l}\right),
\ee
where $[\eta]_1\in \Lambda^\vee/ \Lambda$ is the equivalence 
class of $\eta \in \Lambda^\vee/ l \Lambda$ mod $\Lambda$. 
This is precisely equation \eqref{eq:mainformulaA}.

Finally, the total rank $r\geq 1$ generating function for Donaldson-Thomas invariants is obtained by summing over 
all $d\in \Lambda^\vee$. As shown in Section \ref{verticalKthree}, 
this yields a sum of the form 
\[ 
Z_{DT}(X,r; \tau, {\bar \tau}, B,C)=\sum_{\delta \in \Lambda^\vee/\Lambda} Z_{DT}(X,r,\delta; \tau)
\Theta^*_{r,\delta}(\tau, {\bar \tau}; C,B),
\]
where
\[
\Theta^*_{r, \delta}(\tau, {\bar \tau})= \sum_{\alpha \in \Lambda} e^{-2 \pi i \tau (d+rB+r\alpha)_-^2/2r -2\pi i {\bar \tau} (d+rB+r\alpha)_+^2/2r+2\pi i (d+rB+r\alpha)\cdot C}.
\] 
In the right hand side of the above expression $d\in \Lambda^\vee$ is an arbitrary representative of $\delta$. 
Using the multicover formula for Donaldson-Thomas 
one obtains 
\[ 
\bal 
Z_{DT}(X,r; \tau, {\bar \tau}, B,C )
 = \sum_{\delta \in \Lambda^\vee/r\Lambda}\
\sum_{\substack{ k\in \IZ,\ k\geq 1\\ 
r=kr', \ \delta= f_{r',k}(\delta')}} 
{1\over k^2}Z_{BPS}(X,r',{\delta'}; k\tau) \Theta^*_{r,\delta}(\tau, {\bar \tau}; C,B).
\eal
\]
Moreover
\[ 
\Theta^*_{r, \delta}(\tau, {\bar \tau}; B,C) = \Theta^*_{r', \delta'}(k\tau, k{\bar \tau}; kC,B).
\]
Then, using equation \eqref{eq:omegaNLcc}, one then obtains 
by straightforward computations 
\be\label{eq:rankrDTo}
\bal
 & Z_{DT}(X,r; \tau ) =  \\
&
{1\over 2r^2} 
\sum_{\substack{k,l\in \IZ,\ k,l\geq 1\\ kl=r}} 
\sum_{\eta \in \Lambda^\vee/l\Lambda} 
\sum_{s=0}^{l-1} l
\big(\Delta^{-1}{\widetilde \Phi}_{[\eta]_1}\big)\left({k\tau+s\over l}\right) 
\Theta^*_{l,\eta}\left({k\tau+s}, {k{\bar \tau}+s}; kC,B\right).\\
\eal
\ee
Next recall that there is an exact sequence of finite abelian groups 
\[ 
0 \to \Lambda/l\Lambda \to \Lambda^\vee/l\Lambda \to 
\Lambda^\vee/\Lambda \to 0.
\]
Given an element $\eta\in \Lambda^\vee/l\Lambda$ one can first 
sum over all classes of the form $\eta+\gamma$ with 
$\gamma\in \Lambda/l\Lambda$ in the right hand side 
of \eqref{eq:rankrDTo}. By a simple computation, this sum turns out to be 
\[
\sum_{\gamma\in \Lambda/l\Lambda} 
\Theta^*_{l,\eta+\gamma}\left({k\tau+s}, {k{\bar \tau}+s}; B, kC\right)=\Theta_{1,[\eta]_1}^*\left({k\tau+s\over l}, 
{k{\bar\tau}+s\over l}; kC+sB,lB\right),
\]
which 
depends only on the equivalence class $[\eta]_1\in \Lambda^\vee/\Lambda$. Therefore formula \eqref{eq:rankrDTo}
for the partition function can be rewritten as 
\be\label{eq:rankrDTa}
\bal
& Z_{DT}(X,r; \tau ) =  \\
& {1\over 2r^2} 
\sum_{\substack{k,l\in \IZ,\ k,l\geq 1\\ kl=r}} 
\sum_{\rho \in \Lambda^\vee/\Lambda} 
\sum_{s=0}^{l-1} l
\big(\Delta^{-1}{\widetilde \Phi}_{\rho}\big)\left({k\tau+s\over l}\right) 
\Theta^*_{1,\rho}\left({k\tau+s\over l}, {k{\bar \tau}+s\over l}; kC+sB,lB\right).\\
\eal
\ee
For $B=0$ and $C=0$ one can immediately recognize this formula as an order $r$ Hecke 
transform of the rank $1$ result, as stated in 
Section \ref{mainf}. For nonzero $B,C$ this formula 
is a Hecke transform for Jacobi forms, as discussed in more 
detail in \cite{Hecke_in_progress}. 
For completeness,  a brief definition of 
Hecke operators for modular forms is given below, following for example 
 \cite[Ch.4. Part.2]{Nt_ph}.

Let $\CM_n$ be the set of  $2\times 2$ matrices with 
entries in $\IZ$, of determinant $n$, and note that there is a finite set of orbits $\Gamma_1\backslash \CM_n$ under 
left multiplication by $\Gamma_1= PSL(2,\IZ)$. Then  
the order $n$ Hecke operator \cite[Ch.4. Part.2]{Nt_ph} is 
an endomorphism of the space of the space of holomorphic modular forms of fixed weight $w$ defined by 
\[
T_nf(\tau) = n^{w-1} \sum_{\gamma \in \Gamma_1\backslash \CM_n} (c\tau+d)^{-w} f(\gamma \cdot \tau)
\]
where $\left(\begin{array}{cc} a & b \\ c & d\end{array}\right)\in \CM_n$ is a representative of $\gamma$.  The right hand side does not depend on the choice of representative. Moreover, one can prove that 
the above operator can be written as 
\be\label{eq:Hecketrans}
T_nf(\tau) = n^{w-1} \sum_{\substack{a,d, \in \IZ, \\a,d>0,\ ad=n}} \sum_{b=0}^{d-1} d^{-w} f\left({a\tau+b\over d}\right).
\ee
This is a consequence of Theorem 1 in 
\cite[Ch.4. Part.2]{Nt_ph}. The same construction applies analogously to non-holomorphic modular forms and 
Jacobi forms.

\section{Recursive derivation from stable pair invariants}\label{recursion}

The goal of this section is to provide an alternative derivation 
for the main formula \eqref{eq:mainformulaA} based on the  mathematical results of \cite{G_S_Toda}. Using wallcrossing techniques, vertical Donaldson-Thomas invariants for nodal 
K3 pencils are related in
loc. cit. to the stable pair invariants defined by Pandharipande and Thomas in \cite{stabpairsI}. It will be shown in this section that this 
relation determines the Donaldson-Thomas invariants in terms 
of stable pair ones by a recursive algorithm. Moreover, assuming Gromov-Witten/stable pair correspondence and the multicover 
formula \eqref{eq:multicover}, this algorithm yields a proof of 
the main formula \eqref{eq:mainformulaA}. Using the results of Maulik and Pandharipande on Gopakumar-Vafa 
invariants of K3 pencils, the recursive algorithm reduces the proof of the main formula to a remarkable combinatorial identity which is 
proven in Section \ref{combid}. 

\subsection{Background}
Let $X$ be a nonsingular projective threefold, $\upsilon \in H_2(X,\IZ)$ a curve class, and $n\in \IZ$. 
According to \cite{stabpairsI}, 
the moduli space of stable pairs $\pP_n(X,\upsilon)$ parametrizes the pairs $\CO_{X}\xrightarrow{s} \cf$ where $\cf$ is a pure 1-dimensional sheaf on $X$ with $\ch_{2}(\cf)=\upsilon$ and $\chi(\cf)=n$ such that the cokernel of $s$ is 0-dimensional. It was shown by the authors that $\pP_n(X,\upsilon)$ is a locally complete moduli space of complexes in the derived category, which enabled them to construct a perfect obstruction theory on $\pP_n(X,\upsilon)$. In the case where the virtual dimension of the moduli space is zero, the stable pair invariants 
$P_{n,\upsilon}$ are defined by taking the degree of the virtual cycle obtained from this obstruction theory. 

Now suppose $X$ has a $K3$ fibration structure 
$\pi : X \to C$ 
over a smooth projective curve $C$ with at most nodal fibers. 
The sublattice of 
vertical classes will be denoted by $H_2(X,\IZ)^\pi= {\rm Ker}(\pi_*)\subset H_2(X,\IZ)$. Since the canonical class of $X$ is vertical, it is easy to show that the virtual dimension of the moduli space of stable pairs is zero for any $\upsilon\in H_2(X,\IZ)^\pi$ and any $n \in \IZ$.  Such stable pairs, will be called ``\textit{vertical stable pairs}".  The vertical stable pair series of $X$ is then defined by 
\[ 
PT^{\sf vert} (X)=\sum_{\substack {n\in \IZ,\\ 
\upsilon \in H_2(X,\IZ)^\pi}}P_{n,\upsilon}\; q^nt^\upsilon.
\]
Note that in the above formula one has to sum only over 
nonzero, effective curve classes. This condition will be written as $\upsilon >0$.

For $X$ a smooth Calabi-Yau threefold, one of the main results of \cite{G_S_Toda} relates the 
vertical stable invariants of $X$  to the 
vertical DT invariants $DT(r,\upsilon,n)$ introduced in Section 
\ref{verticalsheaves}. Recall that the latter are virtual counting invariants for semistable sheaves $E$ on $X$ 
with numerical invariants 
\[
{\ch_0}(E)=0, \qquad \ch_1(E) = r D, \qquad 
\ch_2(E) = \upsilon, \qquad \ch_3(E) = - n \ch_3(\CO_x),
\]
where $x\in X$ is an arbitrary closed point. 
Therefore they are related to the invariants 
used in \cite[Thm. 2]{G_S_Toda} by $DT(r,\upsilon, n)= 
J(r, \upsilon, -n)$. Assuming a certain technical conjecture, 
\cite[Conjecture 2.3]{G_S_Toda}, 
the authors of \cite{G_S_Toda} proved the following
identity \cite[Thm. 2]{G_S_Toda}
using  wall-crossing techniques:
\be\label{eq:GST_formula}
\begin{aligned} 
\PT^{\sf vert}(X)=\prod_{r\ge 0, \upsilon>0, n\leq 0} &
\exp\left((-1)^{n-1}DT(r, \upsilon, n)q^{-n} t^{\upsilon} \right)^{-n+2r} \\
&\cdot \prod_{r> 0, \upsilon >0, n<0}
\exp\left((-1)^{n-1}DT(r, \upsilon, n) q^{n}t^{\upsilon} 
\right)^{-n+2r}.\\
\eal
\ee

In order to make contact with the set up of Sections
\ref{setup} and 
\ref{verticalsheaves} suppose the K3 fibration $\pi: X \to C$ satisfies conditions $(a)-(d)$ in Section \ref{setup}. 
Then Poincar\'e duality yields an isomorphism 
$H_2(X,\IZ)^\pi\simeq \Lambda^\vee$, where $\Lambda$ 
is the polarizing lattice for the pencil $\wX$. 
Recall that $\Lambda$ is a rank $\ell$ sublattice of the middle cohomology lattice $\Lambda_{K3}$ of a smooth generic K3 surface.
Moreover, the  natural intersection form on $\Lambda_{K3}$ restricts to a nondegenerate symmetric bilinear form on $\Lambda$ of signature $(1, \ell-1)$. 
Therefore the vertical classes $\upsilon$ will be identified with dual lattice vectors $d\in \Lambda^\vee$. 
Using the given bilinear form, $\Lambda$ is identified 
with a sublattice of $\Lambda^\vee$. Furthermore the 
integral bilinear form on $\Lambda$ extends to a $\IQ$-valued bilinear form on $\Lambda^\vee$, which will be denoted by 
$(d,d') \mapsto d\cdot d'$. By construction, $\alpha\cdot d = d\cdot \alpha = d(\alpha) \in \IZ$ for any any $\alpha \in \Lambda$, $d \in \Lambda^\vee$. 
Finally note that for any such pair $(\alpha, d)$, 
the lattice product $d\cdot \alpha$ is the same 
as the intersection product of $\alpha$ and $d$ viewed 
as homology classes on $X$ as explained above. Therefore 
no notational distinction between such cases will be 
made in this section. 

The next subsection will provide a recursive method for computing the invariants $DT(r,d,n)$ on the right hand side of \eqref{eq:GST_formula} directly from the partition function $\PT(X)$ of $P_{n,d}$ invariants on the left hand side.

\subsection{A recursive algorithm for $DT(r,d,n)$}

The recursive algorithm for the invariants
$\{DT(r,d,n)\}$ is based on the following two key properties. 

\begin{enumerate}
\item[1.] $DT(r,d, n)\neq 0$ implies either $r=0$, or 
\be\label{eq:Bogomolov_I}
d^2 +2rn \ge 0.
\ee 
\item[2.] For any $(r,d,n)$ there is an 
identity 
\[
DT(r, d, n)= DT(r, d + r\alpha, n - r\alpha^2/2 - r d\cdot \alpha),
\]
\end{enumerate}

\noindent
where $\alpha$ is a divisor class of $X$ restricted to the fiber of $\pi$. 

The first property follows from the Bogomolov inequality. 
In more detail, the proof proceeds in several steps, as follows. Suppose first $E$ is a stable sheaf on $X$
supported on a smooth $K3$ fiber $\iota_p:X_p\hookrightarrow X$, with invariants $(r,d,n)$. Then $E$ is the extension by zero of a stable 
sheaf $F$ on $X_p$ with invariants 
\[
{\rm rk}(F)=r, \qquad c_1(F) =\beta, \qquad c_2(F) =k
\]
such that 
\[ 
\iota_{p*}(\beta) =d, \qquad k- {\beta^2\over 2} = n. 
\]
If $r>0$, $F$ must be torsion free and the
Bogomolov inequality reads 
\[
k - {r-1\over 2r} \beta^2 \geq 0,
\]
which is equivalent to
\[ 
\beta^2 + 2rn \geq 0.
\]
As explained in Section \ref{NLmodular}, the algebraic Hodge theorem yields  a second inequality, 
\[
\beta^2 \leq d^2.
\]
This proves inequality \eqref{eq:Bogomolov_I}. 

Next suppose $E$ is a stable sheaf scheme theoretically supported on a nodal fiber  $\iota_\sigma: X_\sigma \to X$. In that case inequality \eqref{eq:Bogomolov_I} is proven by analogy with Lemmas 4.3 and 4.4 in \cite{vertical}. 
Finally, suppose $E$ is semistable. Then inequality
\eqref{eq:Bogomolov_I} is determined by from the above 
results for stable shaves using a Jordan-H\"older filtration 
by analogy with Lemma 4.5 in \cite{vertical}. 

As noted in Section \ref{verticalsheaves}, the second property follows from the observation that the tensor product by the line bundle $\CO_X(\alpha)$ yields 
an isomorphism of moduli stacks of semistable sheaves for 
any $\alpha \in \Lambda \subset {\rm Pic}(X)$. 

In order to construct the recursive algorithm for 
$DT(r,d,n)$, note that equation \eqref{eq:GST_formula}
yields 
\begin{align}\label{eq:log}
&
\log(\PT(X))=\sum_{r\ge 0, d>0, n\leq 0} (-1)^{n-1}(-n+2r) DT(r, d, n)q^{-n} t^{d}\notag\\
&
+\sum_{r>0, d>0, n<0} (-1)^{n-1}(-n+2r) DT(r, d, n)q^{n}t^{d}. 
\end{align}
Then one proceeds inductively by $r\in \IZ$, $r\geq 0$. 

\textbf{Step 1. $r=0$.} According to property (2) above, 
$DT(0,d,n)= DT(0, d, n - d\cdot \alpha)$ for any divisor 
$H\in \Lambda \subset {\rm Pic}(X)$. 
Now note that if $H$ is sufficiently relatively ample with respect to the projection map $\pi: X \to \IP^1$, the coefficient of $q^{-n+d\cdot \alpha} t^{d}$ on the right hand side of \eqref{eq:log} does not contain any $r>0$ terms. Indeed if otherwise it does, then inequality \eqref{eq:Bogomolov_I} 
implies 
$$ d^2 + 2r (n - d\cdot \alpha) \geq 0.$$ 
Since $d>0$, this leads to a contradiction for sufficiently relatively ample $\alpha$, keeping $(d,n)$ fixed. 
Therefore the invariant $DT(0, d, n)$ is determined by the  coefficient of $q^{-n+d \cdot \alpha}t^{d}$ for $\alpha\gg 0$ in the left hand side of \eqref{eq:log}.\\ 

\textbf{Step 2. Induction on $r>0$ and computation of 
$DT(r,d, n)$ invariants}: Now apply the induction on $r$. Suppose that $r>0$ and every $DT(r', d, n)$ is expressed with respect to the stable pair invariants for $r'<r$.  By 
property  $(2)$ above, 
$$DT(r, d, n)=DT(r, d+r\alpha, n-r\alpha^2/2 - d\cdot \alpha)$$ 
for any $\alpha\in \Lambda$. Let $\gamma=(r,d,n)$ and 
$N(\gamma,\alpha) = -n+r\alpha^2/2 + d\cdot \alpha$.
Again, the key observation is that if $\alpha$ is sufficiently
relatively ample,  the coefficient of  
$q^{N(\gamma, \alpha) }t^{d+r\alpha}$ in the right hand side of 
\eqref{eq:log} does  not contain any invariants 
$$DT(R, d+r\alpha, -N(\gamma,\alpha))$$ with $R>r$. Indeed if it does otherwise, then property (2) 
implies that 
\[
(d+r\alpha)^2 + 2R\big(n - r\alpha^2/2 - d\cdot \alpha) \geq 0
\]
 Keeping $(r,d,n)$ fixed, this leads to a contradiction for any $R>r$, if  $\alpha$ is sufficiently
relatively ample. Therefore for such divisors $\alpha$, the sum  \eqref{eq:log} 
\[ 
(-1)^{N(\gamma, \alpha)-1} \sum_{s=0}^r (N(\gamma,\alpha)+2s) 
DT(s, d+r\alpha, -N(\gamma,\alpha)) 
\]
is identified with the coefficient of $q^{N(\gamma, \alpha) }t^{d+r\alpha}$ in the left hand side of \eqref{eq:log}.
Using the induction, one can then 
express $DT(r,d,n)$ in terms of stable pair invariants.

\subsection{Vertical Donaldson-Thomas invariants from Gopakumar-Vafa invariants}\label{Vert_DT_GV}
The goal of this section is to show that 
the conjectural formula \eqref{eq:omegaNLa} follows from the above 
recursive algorithm assuming the 
 Gromov-Witten/stable pair correspondence to hold  in the current setup. 
 If this is the case, the vertical stable pair partition function can be written in terms of vertical Gopakumar-Vafa invariants, which have been
computed in \cite{GW_NL}: 
\be\label{eq:vertGVa} 
 \log(\PT(X))= 
\sum_{g\geq 0} \sum_{\substack{d\in \Lambda^\vee\\ d>0\\}} \sum_{k=1}^\infty {(-1)^{g-1}\over k} n_{g,d}(X) 
\left({(1-(-q)^k)^2 \over (-q)^{k}}\right)^{g-1}
t^{kd} 
\ee
where $n_{g,d}(X)$ are the vertical Gopakumar-Vafa invariants of $X$. 
According to \cite[Thm. 1]{GW_NL}, 
\be\label{eq:GVK3}
n_{g,d}(X) = {1\over 2} \sum_{h=g}^\infty r_{g,h} 
{\widetilde {NL}}_{h,d}
\ee
where $r_{g,h}$ are the local Gopakumar-Vafa invariants of a smooth 
K3 surface and ${\widetilde {NL}}_{h,d}$ 
are the Noether-Lefschetz numbers of the K3 pencil ${\tilde \pi}: 
\wX \to \Sigma$ constructed in Section \ref{setup}. 
An explicit conjecture for the invariants $r_{g,h}$ was formulated by Katz, Klemm and Vafa \cite{KKV}, and proven by 
Pandharipande and Thomas \cite{proofKKV}. This reads 
\be\label{eq:BPSK3} 
\sum_{g \geq 0} \sum_{h \geq 0} 
(-1)^g r_{g,h} \left({(1-y)^2\over y}\right)^{2g} q^h = 
\prod_{n\geq 1} {1\over (1-q^n)^{20} (1-yq^n)^2 (1-y^{-1}q^n)^2}. 
\ee
The above formula implies in particular that $r_{g,h}=0$ for 
$g>h$. 

According the inductive step in the previous section, 
for a given triple $\gamma=(r, d, n)$, one has to choose  
a sufficiently relatively ample divisor $\alpha\in \Lambda \subset {\rm Pic}(X)$. Then the linear combination
\be\label{eq:DTsideA}
(-1)^{N(\gamma, \alpha)-1} \sum_{s=0}^r (N(\gamma,\alpha)+2s) 
DT(s, d+r\alpha, -N(\gamma, \alpha)) 
\ee
is identified with the 
the coefficient $L_{N(\gamma,\alpha),d+r\alpha}$ 
of $q^{N(\gamma, \alpha )} t^{d+r\alpha}$ in 
\eqref{eq:vertGVa}, and 
$N(\gamma, \alpha) = -n + r\alpha^2/2 + d\cdot \alpha$. 
Note that $L_{N(\gamma,\alpha),d+r\alpha}$ can be written as 
\[ 
L_{N(\gamma,\alpha),d+r\alpha} = L_{N(\gamma,\alpha),d+r\alpha}^{0} + L_{N, d+r\alpha}^{\geq 2} 
\] 
where $L_{N(\gamma,\alpha),d+r\alpha}^{0}$ is the contribution 
of genus $g=0$ terms while $L_{N, d+r\alpha}^{\geq 2}$ 
encodes the terms with $g\geq 2$. Genus $g=1$ terms are 
obviously absent for sufficiently large $N(\gamma, \alpha)>0$. 
Expanding the right hand side of \eqref{eq:vertGVa} 
in powers of $q$ yields 
\be\label{eq:genuszeroA}
L_{N(\gamma,\alpha),d+r\alpha}^{0}  = (-1)^{N(\gamma,\alpha)-1} N(\gamma,\alpha)
\sum_{\substack{k\in \IZ,\ k\geq 1\\ 
k | (d+r\alpha, N(\gamma, \alpha))}} {1\over k^2} n_{0, (d+r\alpha)/k} 
\ee
and 
\be\label{eq:highergenusA}
L_{N(\gamma,\alpha), d+r\alpha}^{\geq 2} = (-1)^{N(\gamma,\alpha)} \sum_{\substack{k\in \IZ,\ k\geq 1\\ 
k | (d+r\alpha, N(\gamma, \alpha))}} \sum_{g \geq 1 + 
N(\gamma,\alpha)/k} 
{(-1)^{N(\gamma,\alpha)/k}\over k} n_{g,(d+r\alpha)/k} \binom{2g-2}{g-1-N(\gamma,\alpha)/k}. 
\ee
Note that for $r=0$, the expression \eqref{eq:DTsideA} 
reduces to 
\[ 
(-1)^{N(\gamma,\alpha)-1}N(\gamma, \alpha) DT(0, d+r\alpha, -N(\gamma, \alpha))
\]
Using the multicover formula for Donaldson-Thomas invariants, this is further equal to 
\[ 
(-1)^{N(\gamma,\alpha)-1} N(\gamma,\alpha)
\sum_{\substack{k\in \IZ,\ k\geq 1\\ 
k | (d+r\alpha, N(\gamma, \alpha))}} {1\over k^2} \Omega(0, (d+r\alpha)/k, -N(\gamma, \alpha)/k). 
\]
Moreover, as conjectured in \cite{GVgenuszero}, for any
$\upsilon \in H^2(X, \IZ)$ and any 
$N\in \IZ$, one has $\Omega(0, \upsilon, N) = n_{0,\upsilon}$. Therefore $L_{N(\gamma,\alpha),d+r\alpha}^{0}$ in equation 
\eqref{eq:genuszeroA} equals the $s=0$ term  in \eqref{eq:DTsideA}. 

Next, using \eqref{eq:GVK3}, the higher genus 
contributions $L_{N(\gamma,\alpha), d+r\alpha}^{\geq 2}$ can be written in terms 
of the local invariants $r_{g,h}$ and Noether-Lefschetz numbers. Note that ${\widetilde {NL}}_{h,d}=0$ for 
$h > d^2 /2 +1$ and $r_{g,h}=0$ for $g>h$. 
This implies that 
\be\label{eq:genusrangeA}
{N(\gamma, \alpha)\over k} +1 \leq g \leq h \leq {(d+r\alpha)^2\over 2k^2} +1 
\ee
for any $r_{g,h}$ occuring by substitution in the right hand 
side of equation \eqref{eq:highergenusA}. 
Then equations \eqref{eq:GVK3}, \eqref{eq:highergenusA}, 
yield
\be\label{eq:highergenusC}
\bal
& L_{N(\gamma,\alpha), d+r\alpha}^{\geq 2} =
 \sum_{\substack{k\in \IZ,\ k\geq 1\\ 
k | (d+r\alpha, N(\gamma, \alpha))}} {(-1)^{(k+1)N(\gamma, \alpha)/k}\over 2k}\\
& 
\sum_{\substack{h \in \IZ\\ 
N(\gamma, \alpha)/k +1 \leq h \leq (d+r\alpha)^2/2k^2 +1}} {\widetilde {NL}}_{h,(d+r\alpha)/k}\
\sum_{g= 1 + N(\gamma,\alpha)/k}^h  r_{g,h}\binom{2g-2}{g-1-N(\gamma,\alpha)/k}.
\\
\eal 
\ee
At the same time, using the multicover formula for 
Donaldson-Thomas invariants, the $s\geq 1$ part of the sum in \eqref{eq:DTsideA} becomes
\be\label{eq:DTsideB} 
\bal
& (-1)^{N(\gamma, \alpha)-1} \sum_{s=0}^r (N(\gamma,\alpha)+2s)\ \sum_{\substack{ k \in \IZ,\ k\geq 1\\ 
k |(s, d+r\alpha, N(\gamma, \alpha))}}\ 
{1\over k^2} 
\Omega(s/k, (d+r\alpha)/k, -N(\gamma, \alpha)/k)= \\
& (-1)^{N(\gamma, \alpha)-1} \sum_{\substack{k\in \IZ,\ k\geq 1\\ 
k | (d+r\alpha, N(\gamma, \alpha))}} 
\sum_{\substack{ u\in \IZ\\ 1 \leq u \leq r/k}}
{1\over k}\left({N(\gamma, \alpha)\over k} + 2u\right)
\Omega(u, (d+r\alpha)/k, -N(\gamma, \alpha)/k).
\\
\eal 
\ee
Now recall the conjectural formula \eqref{eq:omegaNLa} 
for the vertical D4-D2-D0 degeneracies. For any triple 
$\gamma=(r,d,n)$, $r\geq 1$, one has 
\be\label{eq:conjformula}
\Omega(\gamma) ={1\over 2} \sum_{\substack{ h\in \IZ\\  r(r-n)\leq h \leq d^2/2+1}} c(r(n-r)+h) {\widetilde {NL}}_{h,d}
\ee
where 
\[ 
\sum_{n\geq 0} c(n) q^n = {1\over \prod_{n\geq 1} (1-q^n)^{24}}. 
\]
Note that by comparison with equation \eqref{eq:GVK3}, it follows that 
$c(n) = r_{0,n}$ for any $n \geq 0$. 
Using equation \eqref{eq:conjformula}, one obtains 
\be\label{eq:DTsideC}
\bal 
\Omega(u,(d+r\alpha)/k, - N(\gamma, \alpha)/k) ={1\over 2} \sum_{\substack{ h\in \IZ\\ u(u+N(\gamma,\alpha)/k) 
\leq h \leq (d+r\alpha)^2/2k^2+1}} c(h -u(u+N(\gamma,\alpha)/k)
) {\widetilde {NL}}_{h,(d+r\alpha)/k}
\eal
\ee
Next note that for $u\geq 1$ the inequality 
\be\label{eq:uineq}
u\left(u+ {N(r,\alpha)\over k}\right) \leq {(d+r\alpha)^2\over 2k^2}
+1
\ee
is equivalent to
\[
 u \leq -{N(\gamma, \alpha)\over 2k} + {1\over 2k} 
\left(N(\gamma, \alpha)^2 + 2(d+r\alpha)^2 + 4k^2\right)^{1/2}.
\]
Moreover note that one must have $k\leq r$ for all nonzero 
terms in the right hand side of \eqref{eq:DTsideB} since $u\geq 1$. Keeping in mind that 
$N(\gamma, \alpha) = - n + r\alpha^2/2 + d\cdot \alpha$, 
a straightforward series expansion shows that for fixed $\gamma=(r,d,n)$ 
there exists a sufficiently relatively ample $\alpha$ such that 
inequality \eqref{eq:uineq} is equivalent to $u \leq r/k$. The latter is in turn precisely 
the upper bound on $u$ in the sum \eqref{eq:DTsideB}. 

At the same time, since $u\geq 1$, for any nonzero term in the right hand side of equation \eqref{eq:DTsideC} one must clearly have 
\[ 
{N(\gamma, \alpha) \over k} + 1 \leq h \leq 
{(d+r\alpha)^2\over 2k^2} +1. 
\]
Therefore 
expression \eqref{eq:DTsideB} can be further written as 
\be\label{eq:DTsideD} 
\bal 
& (-1)^{N(\gamma, \alpha)-1} \sum_{\substack{k\in \IZ,\ k\geq 1\\ 
k | (d+r\alpha, N(\gamma, \alpha))}} {1\over 2k}
\sum_{\substack{h \in \IZ\\ 
N(\gamma, \alpha)/k +1 \leq h \leq (d+r\alpha)^2/2k^2 +1}} {\widetilde {NL}}_{h,(d+r\alpha)/k}\\
& 
\sum_{\substack{ u\in \IZ,\ u\geq 1\\ u(u+N(\gamma,\alpha)/k) \leq h}}
\left({N(\gamma, \alpha)\over k} + 2u\right)r_{0, h -u(u+N(\gamma,\alpha)/k)} .
\\
\eal 
\ee
In order to facilitate comparison with \eqref{eq:highergenusC}, note 
that the  latter reads
\[
\bal
& L_{N(\gamma,\alpha), d+r\alpha}^{\geq 2} =
\sum_{\substack{k\in \IZ,\ k\geq 1\\ 
k | (d+r\alpha, N(\gamma, \alpha))}} {(-1)^{(k+1)N(\gamma, \alpha)/k}\over 2k}\\
& 
\sum_{\substack{h \in \IZ\\ 
N(\gamma, \alpha)/k +1 \leq h \leq (d+r\alpha)^2/2k^2 +1}} {\widetilde {NL}}_{h,(d+r\alpha)/k}\
\sum_{g= 1 + N(\gamma,\alpha)/k}^h  r_{g,h}\binom{2g-2}{g-1-N(\gamma,\alpha)/k}.
\\
\eal 
\]
Therefore it follows that the two expressions are in agreement provided 
that the following identity holds for any $h,n\in \IZ$,  $n\geq 1$, 
$h \geq n+1$:  
\be\label{eq:conjidA} 
\sum_{g=n+1}^h r_{g,h} \binom{2g-2}{g-n-1} = 
(-1)^{n-1} \sum_{\substack{ s\in \IZ,\ s\geq 1\\ s(s+n) \leq h}}
(n + 2s)r_{0, h -s(s+n)} .
\ee 
This is rigorously proven in the next subsection. 

In conclusion, formula \eqref{eq:omegaNLa} 
derived in Section \ref{singfibers} on physics grounds 
indeed follows from the wallcrossing indentity \eqref{eq:GST_formula}. The above mathematical derivation 
holds for all charge vectors, not just primitive ones, confirming the string theoretic conjecture made in Section \ref{multicoversect} below equation \eqref{eq:refeq}.

\subsection{Proof of the combinatorial identity}\label{combid} 
Identity \eqref{eq:conjidA} is equivalent to 
\be\label{eq:conjidB}
\sum_{s=1}^\infty (n+2s) r_{0, h-s(s+n)} = (-1)^{n-1} \sum_{g=n+1}^h \begin{pmatrix} 2g-2 \\ g-n-1 \end{pmatrix} r_{g, h}.
\ee
for any $n \in \IZ$, $0\leq n \leq |h|$, provided that, by convention, 
$r_{0,k}=0$ for $k <0$. 

In order to prove this, first recall the relevant statements of \cite{GW_NL}. The $r_{g, h}$ are given by
\[
\sum_{g\geq 0} \sum_{h\geq 0} (-1)^g r_{g, h} \left(y^{\frac{1}{2}}-y^{-\frac{1}{2}}\right)^{2g} q^h = \prod_{n\geq 0} \frac{1}{\left(1-q^n\right)^{20} \left(1-yq^n\right)^{2}\left(1-y^{-1}q^n\right)^{2}}
\]
where the right-hand side is a meromorphic function on $\mathbb H\times \mathbb C$ and this identity only holds if $|qy|<1$ and $|y^{-1}q|<1$. Suppose in addition $|y|<1$. 
Then this formula has two consequences, first $r_{g, h}=0$ if $g>h$ and also in the specialization $y=1$ it reduces to
\[
\sum_{h\geq 0} r_{0, h}q^h = \prod_{n\geq 1} \frac{1}{\left(1-q^n\right)^{24}}.
\]

Secondly we recall the meromorphic index minus two Jacobi form $\Phi_2(y, q)$ of \cite{BCR}, actually its rescaled version, rescaled by $\eta(q)^{-18}$ is
\[
\frac{\Phi_2(y, q)}{\eta(q)^{18}} =  -\frac{1}{q\left(y^{\frac{1}{2}}-y^{-\frac{1}{2}}\right)^{2}} \prod_{n\geq 0} \frac{1}{\left(1-q^n\right)^{20} \left(1-yq^n\right)^{2}\left(1-y^{-1}q^n\right)^{2}}.
\]
The main result of that article are Fourier expansions of meromorphic negative index Jacobi forms and in that case Theorem 1.3 of \cite{BCR} yields
\[
\frac{\Phi_2(y, q)}{\eta(q)^{18}} =  -\frac{1}{\eta(q)^{24}} \sum_{n\in\mathbb Z} \frac{(2n+1)yq^{n(n+1)}}{\left(1-yq^n\right)}+\frac{y^2q^{n(n+2)}}{\left(1-yq^n\right)^2}.
\]
Define the Fourier coefficients $a_{n, h}$ as
\[
\frac{\Phi_2(y, q)}{\eta(q)^{18}} =  \sum_{\substack{n\in \mathbb Z\\ h\geq -1}} a_{n, h} y^nq^h,
\]
where $|qy|<1$ and $|y^{-1}q|<1$ and also $|y|<1$. 
These coefficients can be computed in two fashions and the comparison will yield the theorem. 
First, using the formula of \cite{GW_NL} we get
\begin{equation}\nonumber
\begin{split}
\frac{\Phi_2(y, q)}{\eta(q)^{18}} &= -\frac{1}{q\left(y^{\frac{1}{2}}-y^{-\frac{1}{2}}\right)^{2}} \prod_{n\geq 0} \frac{1}{\left(1-q^n\right)^{20} \left(1-yq^n\right)^{2}\left(1-y^{-1}q^n\right)^{2}} \\
&= -\frac{1}{q\left(y^{\frac{1}{2}}-y^{-\frac{1}{2}}\right)^{2}}  \sum_{h\geq 0}\sum_{g= 0}^h (-1)^g r_{g, h} \left(y^{\frac{1}{2}}-y^{-\frac{1}{2}}\right)^{2g} q^h\\
&= -\sum_{h\geq -1}\sum_{g= 0}^{h+1} (-1)^g r_{g, h+1} \left(y^{\frac{1}{2}}-y^{-\frac{1}{2}}\right)^{2(g-1)} q^h \\
&= -\sum_{h\geq -1} r_{0, h+1} \left(y^{\frac{1}{2}}-y^{-\frac{1}{2}}\right)^{-2} q^h-\sum_{h\geq -1}\sum_{g=1}^{h+1} (-1)^g r_{g, h+1} \left(y^{\frac{1}{2}}-y^{-\frac{1}{2}}\right)^{2(g-1)} q^h\\
&= -\sum_{h\geq -1}\sum_{n=0}^\infty r_{0, h+1} ny^n q^h - 
\sum_{h\geq -1}\sum_{g=1}^{h+1} (-1)^{g+k} r_{g, h+1} \begin{pmatrix} 2(g-1) \\ k \end{pmatrix} y^{g-k-1} q^h.
\end{split}
\end{equation}
In the last equality we have expanded $\left(y^{\frac{1}{2}}-y^{-\frac{1}{2}}\right)^{-2}$ in the domain $|y|<1$ and used the binomial formula for $\left(y^{\frac{1}{2}}-y^{-\frac{1}{2}}\right)^{2(g-1)}$. It follows that
\begin{equation}\label{eqLFourierone}
-a_{n, h} = nr_{0, h+1} + (-1)^{n+1} \sum_{g=1}^{h+1} (-1)^{g+k} r_{g, h+1} \begin{pmatrix} 2(g-1) \\ g-n-1 \end{pmatrix}.
\end{equation}
Second, we will compute these coefficients using Theorem 1.3 of \cite{BCR} together with the expression of $\eta(q)^{-24}$ in terms of the $r_{0, h}$, namely
\begin{equation}\nonumber
\begin{split}
\frac{\Phi_2(y, q)}{\eta(q)^{18}} &=  -\frac{1}{\eta(q)^{24}} \sum_{n\in\mathbb Z} \frac{(2n+1)yq^{n(n+1)}}{\left(1-yq^n\right)}+\frac{y^2q^{n(n+2)}}{\left(1-yq^n\right)^2} \\
&= -\sum_{h\geq 0} r_{0, h}q^h \sum_{n\in\mathbb Z} \frac{(2n+1)yq^{n(n+1)}}{\left(1-yq^n\right)}+\frac{y^2q^{n(n+2)}}{\left(1-yq^n\right)^2} \\
&= -\sum_{h\geq 0} r_{0, h}q^h \sum_{n\in\mathbb Z}\sum_{m=0}^\infty (2n+1)y^{m+1}q^{n(n+m+1)}+(m+1)y^{m+2}q^{n(n+m+2)} \\
\end{split}
\end{equation}
and hence the Fourier coefficients are
\begin{equation}\label{eqLFouriertwo}
-a_{n, h} = \sum_{s=0}^\infty (2s+n) r_{0, h+1-s(s+n)} =
 nr_{0, h+1} +\sum_{s=1}^\infty (2s+n) r_{0, h+1-s(s+n)} .
\end{equation}
So that comparing $a_{n, h-1}$ in \eqref{eqLFourierone} and \eqref{eqLFouriertwo} yields identity \eqref{eq:conjidB}.

\section{Modularity of partition functions}\label{modularsect} 

The goal of this section is to summarize the main modularity results obtained in this paper.  

\subsection{The statement}
In order to fix ideas and notation, the following is a brief self-contained presentation of the construction carried in the previous sections.

\begin{itemize} 
\item Let $\Lambda$ be a rank $\ell\geq 1$ lattice equipped with an integral even 
nondegenerate symmetric bilinear form of signature $(1, \ell-1)$ written as $(\alpha, \alpha') \mapsto 
\alpha \cdot \alpha'$. This form will be often referred to as the  intersection form on $\Lambda$ because of its geometric origin. 
Let $(v_i)_{1\leq i\leq \ell}$ be a basis in 
$\Lambda$. The matrix $(v_i\cdot v_j)$ will be denoted by $M=(M_{ij})$. Let $m = |{\rm det}(M)|$. 
\item Let $\Lambda^\vee= {\rm Hom}_\IZ(\Lambda, \IZ)$ be the dual lattice. Using the bilinear form, 
$\Lambda$ is identified with a sublattice of $\Lambda^\vee$. 
For any $r\in \IZ$, $r\geq 1$, 
the  quotient $\Lambda^\vee /r\Lambda$ is denoted by $G_r$. 
The equivalence class of an element $d\in \Lambda^\vee$ in $G_r$ 
will 
be denoted by $[d]_r$. Furthermore for any $r\geq 1$ there is a natural projection 
$G_r \twoheadrightarrow G_1$. The image of an element $\delta \in G_r$ in $G_1$ will be denoted by $[\delta]_1$. 
\item Note that $\Lambda^\vee\subset \Lambda_\IQ$ and 
 there is a $\IQ$-valued nondegenerate symmetric bilinear form  pairing on $\Lambda^\vee$ induced by the given intersection
form on $\Lambda$. In terms of the dual basis
$({\check v}^i)_{1\leq i \leq \ell}$ this pairing is given by 
$({\check v}^i, {\check v}^j) \mapsto (M^{-1})_{ij}$. Moreover, 
for any $l\in \IZ$, $l\geq 1$, the induced form  descends to a $\IQ/\IZ$-valued nondegenerate symmetric bilinear form  $({}\ ,\ )_l$ 
on $G_l$ given by 
\[
([d]_l, [d']_l)_l = {d\cdot d'\over l} \quad {\rm mod}\ \IZ. 
\]
In particular one obtains a $\IQ/\IZ$-valued quadratic form 
$\theta_l: G_l\to \IQ/\IZ$, 
\be\label{eq:quadformB}
\theta_l(\delta) = {(\delta,\delta)_l\over 2}
\ee
for all $\delta\in G_l$. 
\item As shown by Weil in  \cite{Weil_rep}, the above data 
determines a representation 
$\rho_\Lambda: Mp(2,\IZ)\to {\rm End}(\IC[G_1])$
of the metaplectic group $Mp(2, \IZ)$. The detailed construction is presented in Appendix \ref{Weilreps} for 
completeness. For the purposes of this section, note that 
for a fixed $\tau$ in the upper half plane, 
the metaplectic group consists of pairs 
$(\gamma, \sqrt{c\tau+d})$ where $\gamma= \left(\begin{matrix} a & b \\ c & d\end{matrix}\right)
\in SL(2,\IZ)$ and is generated by $\wT=(T, 1)$ and 
$\wS=(S,\sqrt{\tau})$. 
The Weil representation $\rho_\Lambda$ is given by 
\be\label{eq:metrepAB} 
\rho_\Lambda(\wT)(e_\delta) = e^{2\pi i\theta_1(\delta)}e_\delta, \qquad 
\rho_\Lambda(\wS)(e_\delta) = {e^{\pi i (\ell-2)/4}\over \sqrt{m}} 
\sum_{\delta'\in G_1} e^{-2\pi i (\delta, \delta')_1} e_{\delta'},
\ee
where $(e_\delta)_{\delta\in G_1}$ is a basis of $\IC[G_1]$. 
Note that this is a unitary symmetric representation, hence the dual representation $\rho^*_\Lambda$ is naturally isomorphic to its complex conjugate. This identification will be implicit throughout this section.

\item A scaled form of the Weil representation will 
be needed below. Namely for any $r \in \IZ$, $r\geq 1$, 
let ${\widetilde \Lambda} \subset \Lambda_\IR=\Lambda\otimes_\IZ \IR$ be the sublattice 
\[
{\tilde \Lambda} = \{ \sqrt{r} \alpha \, |\, \alpha \in \Lambda\}.
\]
Clearly, the intersection form on $\Lambda$ induces an intersection 
form on ${\widetilde \Lambda}$. 
The dual lattice ${\widetilde \Lambda}^\vee\subset 
\Lambda_\IR$ is naturally identified with the sublattice 
\[
{\widetilde \Lambda}^\vee = \{ d / \sqrt{r} \, |\, d\in \Lambda^\vee\}.
\]
Then note 
that there is a  lattice isomorphism
\[ 
f: \Lambda\, {\buildrel \sim \over \longto}\, {\tilde \Lambda}, \qquad f(\alpha) = \sqrt{r} \alpha
\]
which yields an isomorphism of finite abelian groups 
\be\label{eq:quotisom} 
\phi: G_r\, {\buildrel \sim \over \longto}\, {\widetilde \Lambda}^\vee/
{\widetilde \Lambda}, \qquad \phi([d]_r) = [d/\sqrt{r}]_1. 
\ee
Using this isomorphism, the Weil representation 
associated to ${\widetilde \Lambda}$ with the induced bilinear form yields a representation $\rho_{\Lambda,r}: Mp(2,\IZ) \to 
{\rm End}(\IC[G_r])$ given by 
\be\label{eq:metrepB}
\rho_{\Lambda,r}(\wT)(e_{\delta}) = e^{2\pi i\theta_r(\delta)}e_\delta, \qquad 
\rho_{\Lambda,r}(\wS)(e_\delta) = {e^{\pi i (\ell-2)/4}\over r^{\ell/2} \sqrt{m}} 
\sum_{\delta'\in G_r} e^{-2\pi i (\delta, \delta')_r} e_{\delta'}
\ee
where $(e_{\delta})_{\delta \in G_r}$ is a basis of $\IC[G_r]$. 

\item 
For completeness recall some basic facts on 
vector valued modular forms, also 
needed below.  Given a representation 
$\rho: Mp(2, \IZ)\to {\rm End}(V)$, with $V$ a finite dimensional complex vector space, a weight $(w_+, w_-)$ vector valued modular form of type $\rho$ is a $V$-valued function $\Phi(\tau, {\bar \tau})$ 
defined on the upper half plane such that 
\[
\Phi\left(\gamma\cdot \tau, \gamma \cdot {\bar \tau} \right) = \left(c\tau +d\right)^{w_+} \left(c\bar\tau +d\right)^{w_-} \rho\left(\gamma, \sqrt{c\tau+d} \right) \Phi(\tau, \bar\tau).
\]
for any $\gamma = \begin{pmatrix} a & b \\ c & d \end{pmatrix}
\in SL(2, \IZ)$, 
where, as usual, $\gamma\cdot \tau = \frac{a\tau +b}{c\tau+d}$. 
An important class of examples to be used in the construction below are the Siegel theta functions 
constructed in \cite[Thm. 4.1]{Borcherds2}.

\item 
For any $r\in \IZ$, $r\geq 1$ and for any $\delta \in  G_r$ there is a partition function for vertical Donaldson-Thomas 
invariants of a $\Lambda$-polarized K3 pencil given in equation \eqref{eq:rankrDTa}. This formula is written in terms of a  vector 
valued modular form $\Delta^{-1}(q){\widetilde \Phi}(q)$ 
with values in the Weil representation of $\Lambda$. 
Since the 
geometric details are not important for modularity questions, 
it will be convenient to consider the following abstract variant of this construction.
 
Let $\Psi(\tau)= \sum_{\delta \in G_1} \Psi_\delta(\tau)
e_\delta$ be a weight $w = -1 - \ell/2$ holomorphic vector valued modular form 
of type $\rho_\Lambda$. Note that for any $k, l\in \IZ$, $k,l\geq 1$, there is an injective morphism of abelian groups 
\[
f_{l,k} : G_l \to G_{kl}, \qquad f_{l,k}([d]_l) = [kd]_{kl} 
\]
for any $d\in \Lambda^\vee$. Hence any $\delta \in {\rm Im}(f_{l,k})$ determines a unique element $\eta = 
f_{l,k}^{-1}(\delta) \in G_l$. Then for any $r\in \IZ$, $r\geq 1$ and for any $\delta \in 
G_r$ let 
\be\label{eq:ZrdeltaI}
Z_{r,\delta}(\tau) = {1\over 2r^2} 
\sum_{\substack{ k,l\in \IZ,\ k,l\geq 1\\  r=kl,\ \delta = f_{l,k}(\eta)}} 
\sum_{s=0}^{l-1} l
e^{-2\pi i s \theta_l(\eta) }\Psi_{[\eta]_1}
\left({k\tau +s\over l}\right).
\ee
where  $[\eta]_1\in G_1$ 
denotes the projection of $\eta\in G_{l}$ onto $G_1$. 
\end{itemize}

In this framework, the main modularity result obtained in this 
paper reads:
\bigskip

{\it For any $r\in \IZ$, $r\geq 1$, the vector valued series 
\[ 
Z_r(\tau) = \sum_{\delta \in G_r} Z_{r, \delta}(\tau) e_\delta 
\]
is a  holomorphic vector valued modular form of weight $(-1-\ell/2)$ and type 
$\rho_{\Lambda, r}$.}
\bigskip

A quick proof of this statement follows from the identification 
of the rank $r$ partition function \eqref{eq:mainformulaB} with a Hecke transform of a Jacobi form. The above claim follows from the observation that the series $\Theta_{r,\delta}^*(\tau, {\bar \tau}, C,B)$ are in fact complex conjugates of Siegel Jacobi functions for the 
lattice ${\widetilde \Lambda}= \sqrt{r} \Lambda \subset 
\Lambda_\IR$. In more detail, 
using the isomorphism \eqref{eq:quotisom},
\[
\Theta^*_{r, \delta}(\tau, {\bar \tau},C,B) = \sum_{{\tilde \alpha} \in {\widetilde \Lambda}} e^{-2 \pi i \bar\tau ({\tilde d} + {\tilde \alpha}+{\tilde B}/2)_-^2/2 -2\pi i { \tau} ({\tilde d}+{\tilde \alpha}+{\tilde B}/2)_+^2/2+2\pi i 
({\tilde d} +{\tilde \alpha} + {\tilde B}/2)\cdot {\tilde C}},
\]
where ${\tilde d} = d/\sqrt{r}$, ${\tilde B}=\sqrt{r}B$, 
${\tilde C}=\sqrt{r}C$. 
Therefore they form a vector valued modular 
form of 
weight $\left((\ell-1)/2, 1/2\right)$ with values in the complex conjugate representation
$\rho_{\Lambda,r}^*$, and they satisfy the required linearly independence conditions. 
For the skeptical reader, a second proof of the above  statement will be provided below in a separate subsection. 

To conclude this section, note that this modularity statement
yields the finite dimensionality result needed in 
the derivation of final expression  \eqref{eq:omegaNLd}
for the partition function. In order to employ modularity arguments, 
one has to know that the vector space of all possible 
partitions functions allowed by modularity constraints 
is finite dimensional. Given the above statement, this
follows from the results of 
\cite{Vvmfspaces} on the finite dimensionality of spaces of vector 
valued modular forms.

\subsection{A second proof of modularity}\label{secondproof}
The goal of this section is to give a direct proof of the main modularity statement. The standard notation
$
e(x) = e^{2 \pi i x}
$
will be used for brevity. To prove modularity, one needs to show that
\begin{equation}\label{eq:toprove}
Z_{r, \delta}\left( - \frac{1}{\tau} \right) = \frac{\tau^{-1-\ell/2} e\left( \frac{\ell-2}{8} \right)}{\sqrt{m} r^{\ell/2}} \sum_{\delta' \in G_r} e(-(\delta,\delta')_r ) Z_{r,\delta'}(\tau).
\end{equation}

Recall that $Z_{r,\delta}(\tau)$ is given by 
\begin{equation}\label{eq:zz}
Z_{r,\delta}(\tau) = \frac{1}{2 r^2} \mathop{\sum_{k,l \in \mathbb{Z}_{\geq 1},\, k l = r}}_{\delta = f_{l,k}(\eta)}\ \sum_{s=0}^{l-1} l e(-s \theta_l(\eta) ) \Psi_{\eta_1}\left( \frac{k \tau + s}{l} \right).
\end{equation}
where for any $k,l\in \IZ$, $k,l\geq 1$
\[
f_{l,k}: \Lambda^\vee / l\Lambda { \longto} 
 \Lambda^\vee/ (kl)\Lambda
\]
is the injective morphism of lattices defined by 
$f_{l,k}([d]_l) = [kd]_{kl}$. 
In addition the surjective morphism of lattices 
\[
g_{k,l}: \Lambda^\vee/ (kl)\Lambda  {\longto} 
\Lambda^\vee / l\Lambda, \qquad g_{k,l}([d]_{kl}) = [d]_l
\]
will be also used in the proof. 
For any $\delta \in \Lambda^\vee/(kl)\Lambda$ let $\delta_l = g_{k,l}(\delta)$. 

On needs to distinguish between the $s=0$ and $s >0$ cases. For $s>0$, let $ p = {\rm gcd}(s,l)$ and $\tilde{s} = s/p$, $\tilde{l} = l/p$. 
Note that $0<{\tilde s}< {\tilde l}$. 
Then \eqref{eq:zz} can be rewritten as 
\begin{align}
Z_{r,\delta}(\tau) =& \frac{1}{2 r^2} \mathop{\sum_{k,l \in \mathbb{Z}_{\geq 1},\, k l = r}}_{\delta = f_{l,k}(\eta)} l \Psi_{\eta_1}\left( \frac{k \tau}{l} \right)\nonumber\\
& +  \frac{1}{2 r^2} \mathop{\sum_{k,\tilde{l},p \in \mathbb{Z}_{\geq 1},\, k \tilde{l} p = r}}_{\delta= f_{p\tilde{l},k}(\eta)} p \tilde{l} \mathop{\sum_{1\leq \tilde{s} \leq {\tilde l}-1}}_{(\tilde{s},\tilde{l})=1} e(-\tilde{s} \theta_{\tilde{l}}(\eta_{\tilde{ l}}) ) \Psi_{\eta_1}\left( \frac{k \tau + p \tilde{s}}{p \tilde{l}} \right).
\label{eq:back}
\end{align}

Using the above formula, one can 
calculate $Z_{r,\delta}(-1/\tau)$ as follows. 
For the $s=0$ terms note that
\begin{align}\label{eq:s0}
\Psi_{\eta_1} \left( - \frac{k}{l \tau} \right) = \frac{l^{-1-\ell/2} k^{1 + \ell/2} \tau^{-1-\ell/2} e\left(\frac{\ell-2}{8} \right)}{\sqrt{m}} \sum_{\mu \in G_1} e(-(\eta_1,\mu)_1) \Psi_\mu \left( \frac{l \tau}{k} \right).
\end{align}
since $\Psi(\tau)$ is a vector valued modular form of weight $(-1-\ell/2)$ and type $\rho_\Lambda$.
In order to evaluate the $s \geq 1$ terms, one has to compute 
\[
\Psi_{\eta_1} \left( - \frac{k}{p \tilde{l} \tau} + \frac{\tilde{s}}{\tilde{l}} \right)
\]
using the modular properties of $\Psi$. 
Let
\begin{equation}
\tau' = \frac{p}{k \tilde{l}} \tau + \frac{s'}{\tilde{l}},
\end{equation}
where $s' \in \{0,\ldots,l-1\}$ is uniquely defined by the requirement that $s' \tilde{s} =  - 1 \text{ mod } \tilde{l}$. Then, it is easy to check that
\begin{equation}
 - \frac{k}{p \tilde{l} \tau} + \frac{\tilde{s}}{\tilde{l}}  = \frac{a \tau' + b}{c \tau' + d},
\end{equation}
for the $SL(2,\mathbb{Z})$ transformation
\begin{equation}\label{eq:SLtwomatrix} 
\begin{pmatrix} a & b \\ c & d \end{pmatrix} = \begin{pmatrix} \tilde{s} & - \frac{\tilde{s} s' + 1}{\tilde{l}} \\ \tilde{l} & - s' \end{pmatrix}.
\end{equation}
Now note that Shintani's formula \cite{half_integral},
reviewed in Appendix \ref{Shintani}, 
yields an identity: 
\begin{align}\label{eq:s1}
\Psi_{\eta_1} \left( - \frac{k}{p \tilde{l} \tau} + \frac{\tilde{s}}{\tilde{l}} \right) = \frac{p^{-1-\ell/2} \tau^{-1-\ell/2} k^{1 + \ell/2} \tilde{l}^{-\ell/2} e\left(\frac{\ell-2}{8} \right)}{\sqrt{m}} \sum_{\sigma \in G_1}  
{\bar h}_{{\tilde l}, -s', \tilde{s}}(\eta_1,\sigma) \Psi_\sigma\left( \frac{p}{k \tilde{l}} \tau + \frac{s'}{\tilde{l}} \right).
\end{align}
where ${\bar h}_{{\tilde l}, -s',\tilde{s}}: G_1 \times G_1 \to \IC$ is a function 
defined as follows. Pick any lift $\mu \in G_{\tilde l}$ of $\sigma$, 
such that $\mu_1 =\sigma$. Then 
\[
{\bar h}_{{\tilde l},-s',\tilde{s}}(\eta_1, \sigma) = 
\sum_{u \in \Lambda/\tilde{l} \Lambda} e\left(- \frac{s'}{2} (\mu+u, \mu+u)_{\tilde l} - (\mu+u, \eta_{\tilde l})_{\tilde l} + \frac{\tilde{s}}{2} (\eta_{\tilde l}, \eta_{\tilde l})_{\tilde l} \right).
\]
As shown in Appendix \ref{Shintani}, the right hand side of the above equation does not depend on the chosen lift  
$\mu$. 

Combining \eqref{eq:s0} and \eqref{eq:s1}, one  gets
\begin{equation}
Z_{r, \delta} \left(- \frac{1}{\tau} \right) = \frac{\tau^{-1-\ell/2}e\left(\frac{\ell-2}{8} \right)}{2 r^2 \sqrt{m}}  \left( S_0(\tau) + S_{\geq 1}(\tau) \right),
\end{equation}
with
\begin{equation}\label{eq:S0}
S_0(\tau) =\mathop{\sum_{k,l \in \mathbb{Z}_{\geq 1},\, k l = r}}_{\delta = f_{l,k}(\eta)} 
 l^{-\ell/2} k^{1 + \ell/2} \sum_{\sigma \in G_1} e(-(\eta_1,\sigma)_1) \Psi_\sigma \left( \frac{l \tau}{k} \right),
\end{equation}
and
\[
\begin{aligned}
S_{\geq 1}(\tau) = &  \mathop{\sum_{k,\tilde{l},p \in \mathbb{Z}_{\geq 1},\, k \tilde{l} p = r}}_{\delta = f_{p\tilde{l},k}(\eta)} p^{-\ell/2} \tilde{l}^{1-\ell/2} k^{1+\ell/2}  \mathop{\sum_{1\leq \tilde{s} \leq {\tilde l}-1}}_{(\tilde{s},\tilde{l})=1} e(-\tilde{s} \theta_{\tilde{l}}(\eta_{\tilde{ l}}) )\   \sum_{\sigma\in G_1} 
{\bar h}_{{\tilde l},-s', \tilde{s}}(\eta_1,\sigma) \Psi_\sigma \left( \frac{p}{k \tilde{l}} \tau + \frac{s'}{\tilde{l}} \right).
\\
\eal \]
Now note that 
\[
e(-\tilde{s} \theta_{\tilde{l}}(\eta_{\tilde{ l}}) )
{\bar h}_{{\tilde l},-s', \tilde{s}}(\eta_1,\sigma) =  \sum_{u \in \Lambda/ {\tilde l}\Lambda}
e\left({- \frac{s'}{2} (\mu+u, \mu+u)_{\tilde l}
-(\mu+u,\eta_{\tilde l})_{\tilde l} }\right)
\]
where $\mu \in G_{\tilde l}$ is an arbitrary lift of $\sigma \in G_1$. Again, as observed in Appendix \ref{Shintani}, 
the function $h_{{\tilde l}, -s', \eta_{\tilde l}} : G_{\tilde l} \to \IC$, 
\be\label{eq:hfunction}
h_{{\tilde l}, -s', \eta_{\tilde l}}(\mu) = \sum_{u \in \Lambda/ {\tilde l}\Lambda}
e\left({- \frac{s'}{2} (\mu+u, \mu+u)_{\tilde l}
-(\mu+u,\eta_{\tilde l})_{\tilde l} }\right)
\ee
is invariant under shifts $\mu \mapsto \mu + x$, $x\in \Lambda/{\tilde l}\Lambda$. Hence it descends to a function 
${\bar h}_{{\tilde l}, -s', \eta_{\tilde l}}: G_1 \to \IC$. 
Therefore 
\be\label{eq:Sgeqone}
\bal
S_{\geq 1}(\tau) =   \mathop{\sum_{k,\tilde{l},p \in \mathbb{Z}^{\geq 1},\, k \tilde{l} p = r}}_{\delta = f_{p\tilde{l},k}(\eta)} p^{-\ell/2} \tilde{l}^{1-\ell/2} k^{1+\ell/2}  \mathop{\sum_{1\leq \tilde{s}\leq {\tilde l}-1}}_{(\tilde{s},\tilde{l})=1} 
 \sum_{\sigma \in G_1} 
{\bar h}_{{\tilde l}, -s', \eta_{\tilde l}}(\sigma) 
\Psi_\sigma 
 \left( \frac{p}{k \tilde{l}} \tau + \frac{s'}{\tilde{l}} \right).
\end{aligned}
\ee

Returning to  equation \eqref{eq:toprove}, one now has to prove that 
\begin{equation}
\frac{1}{r^{\ell/2}} \sum_{\delta' \in G_r} e(-(\delta,\delta')_r ) Z_{r,\delta'}(\tau) = \frac{1}{2 r^2}  \left( S_0(\tau) + S_{\geq 1}(\tau) \right).
\end{equation}
First, using \eqref{eq:back} the  left-hand-side can be written as 
\[
\begin{aligned}
\frac{1}{r^{\ell/2}} \sum_{\delta' \in G_r} e(-(\delta,\delta')_r ) Z_{r,\delta'}(\tau) = &\frac{1}{2 r^2 r^{\ell/2}}  \sum_{\delta' \in G_r} e(-(\delta,\delta')_r ) \bigg( \mathop{\sum_{k,l \in \mathbb{Z}_{\geq 1},\, k l = r}}_{\delta' = f_{l,k}(\eta)} l \Psi_{\eta_1}\left( \frac{k \tau}{l} \right) \\
& + \mathop{\sum_{k,\tilde{l},p \in \mathbb{Z}_{\geq 1},\, 
k \tilde{l} p = r}}_{\delta' = f_{p\tilde{l},k}(\eta)} p \tilde{l} \mathop{\sum_{1\leq \tilde{s} \leq {\tilde l}-1}}_{(\tilde{s},\tilde{l})=1} e(-\tilde{s} \theta_{\tilde{l}}(\eta_{\tilde{ l}}) ) \Psi_{\eta_1}\left( \frac{k \tau + p \tilde{s}}{p \tilde{l}} \right) \bigg) \\
= & \frac{1}{2 r^2} \left( Q_0(\tau) + Q_{\geq 1} (\tau) \right),
\end{aligned}
\]
where 
\begin{equation}
Q_0(\tau) = \frac{1}{r^{\ell/2}} \mathop{\sum_{k,l \in \mathbb{Z}_{\geq 1},\, k l = r}}_{\eta \in G_l} l e(-(\delta_l, \eta)_l)  \Psi_{\eta_1}\left( \frac{k \tau}{l} \right) ,
\end{equation}
and
\begin{equation}
Q_{\geq 1}(\tau) =  \frac{1}{r^{\ell/2}}  \sum_{\eta \in G_l} e(-(\delta_l,\eta)_l ) \mathop{\sum_{k,\tilde{l},p \in \mathbb{Z}_{\geq 1},\, k \tilde{l} p = r}} p \tilde{l} \mathop{\sum_{1\leq \tilde{s}\leq {\tilde l}_1}}_{(\tilde{s},\tilde{l})=1} e( -\tilde{s} \theta_{\tilde{l}}(\eta_{\tilde{ l}}) ) \Psi_{\eta_1}\left( \frac{k \tau + p \tilde{s}}{p \tilde{l}} \right).
\end{equation}
It will be shown below  that $Q_0(\tau) = S_0(\tau)$ and $Q_{\geq 1}(\tau) = S_{\geq 1}(\tau)$.

In order to compute $Q_0(\tau)$ note that there is an exact sequence of finite abelian groups 
\[ 
0\to \Lambda/ l\Lambda \to  G_l \to G_1 \to 0.
\]
Given any element $\sigma \in G_1$, let $\mu\in G_l$ denote 
an arbitrary lift of $\sigma$. Then one has 
\[ 
\sum_{\substack{ \eta \in G_l\\ \eta_1 =\sigma}} e(-(\delta_l, \eta)_l) = \sum_{u \in \Lambda/ l\Lambda} 
e(-(\delta_l, \mu+ u)_l) = e(-(\delta_l, \mu)_l)
\sum_{u \in \Lambda/ l\Lambda} e(-(\delta_l, u)_l). 
\]
Now note that 
\begin{equation}
\sum_{u\in \Lambda / l \Lambda}  e( - (\delta_l, u)_l) = \begin{cases}
l^{\ell} & \text{if $\delta = f_{k,l}(\xi)$ for some $\xi \in G_k$},\\
0 & \text{otherwise}.
\end{cases}
\end{equation}
Moreover if $\delta = f_{k,l}(\xi)$ for some $\xi \in G_k$, then 
$e(-(\delta_l, \mu)_l) = e(-(\xi_1, \mu_1)_1) = e(-(\xi_1, \sigma)_1)$. 
Therefore,
\begin{equation}
Q_0(\tau) = \mathop{\sum_{k,l \in \mathbb{Z}_{\geq 1},\, k l = r}}_{\delta = f_{k,l}(\xi)} k^{-\ell/2} l^{1+\ell/2} \sum_{\sigma \in G_1} e( -(\xi_1, \sigma)_1 )\Psi_{\sigma}\left( \frac{k \tau}{l} \right) .
\end{equation}
Comparing with $S_0(\tau)$ in \eqref{eq:S0}, exchanging the summation variables $l \leftrightarrow k$ and the symbols $\xi \leftrightarrow \eta$ yields 
\begin{equation}
Q_0(\tau) = S_0(\tau),
\end{equation}
as claimed above. 

The computation of $Q_{\geq 1}(\tau)$ is similar.  Recall that
\begin{equation}\label{eq:Qgeqone}
Q_{\geq 1}(\tau) =  \frac{1}{r^{\ell/2}}  \sum_{\eta \in G_l} e(-(\delta_l,\eta)_l ) \mathop{\sum_{k,\tilde{l},p \in \mathbb{Z}_{\geq 1},\, k \tilde{l} p = r}} p \tilde{l} \mathop{ \sum_{1\leq \tilde{s} \leq {\tilde l}-1}}_{(\tilde{s},\tilde{l})=1} e( -\tilde{s} \theta_{\tilde{l}}(\eta_{\tilde{ l}}) ) \Psi_{\eta_1}\left( \frac{k \tau + p \tilde{s}}{p \tilde{l}} \right).
\end{equation}
The right hand side of equation \eqref{eq:Qgeqone} 
can be written as 
\[ 
\frac{1}{r^{\ell/2}}  \mathop{\sum_{k,\tilde{l},p \in \mathbb{Z}_{\geq 1},\, k \tilde{l} p = r}} p \tilde{l}
\sum_{\eta \in G_l} e(-(\delta_l,\eta)_l ) A(\eta_{\tilde l}) 
\]
where 	
\[
A(\eta_{\tilde l}) = \mathop{\sum_{1\leq \tilde{s} \leq {\tilde l}-1}}_{(\tilde{s},\tilde{l})=1} e( -\tilde{s} \theta_{\tilde{l}}(\eta_{\tilde{ l}}) ) \Psi_{\eta_1}\left( \frac{k \tau + p \tilde{s}}{p \tilde{l}} \right)
\]
depends only on ${\tilde l}, \eta_{\tilde l}$. 
In order to compute the sum over $\eta\in G_l$ with 
fixed projection $\gamma = \eta_{\tilde l}\in G_{\tilde l}$, 
note that there is an exact sequence 
of finite abelian groups 
\[
0 \to \Lambda/ p\Lambda {\buildrel i\over \longto} G_l \to G_{\tilde l}\to 0.
\]
Given an element $v\in \Lambda/ p\Lambda$, one has 
$i(v) = [{\tilde l}\alpha]_l$ where $\alpha \in \Lambda$ is 
an arbitrary representative of $v$. The right hand side, 
$[{\tilde l}\alpha]_l\in G_l$ is clearly independent of the choice 
of $\alpha$. Then let $\gamma_0\in G_l$ denote an arbitrary lift of $\gamma$ and note that 
\begin{equation}
\sum_{\substack{\eta \in G_l\\ \eta_{\tilde l}=\gamma} }e(- (\delta_l, \eta)_l) = 
e(-(\delta_l, \gamma_0)_l)\sum_{ v \in \Lambda / p \Lambda} 
e( - (\delta_p,  v)_p)
\end{equation}
since $l=p{\tilde l}$. 
As before,
\begin{equation}
\sum_{v \in \Lambda / p \Lambda}  e( - (\delta_p, v)_p) = \begin{cases}
p^{\ell} & \text{if $\delta = f_{k{\tilde l}, p}(\xi)$ for some $\xi \in G_{k{\tilde l}}$},\\
0 & \text{otherwise}.
\end{cases}
\end{equation}
Moreover, if $\delta = f_{k{\tilde l}, p}(\xi)$ then 
\[ 
e(-(\delta_l, \gamma_0)_l)=e(-(\xi_{\tilde l}, \gamma)_{\tilde l}). 
\]
Therefore,
\begin{align}
Q_{\geq 1}(\tau) = &\mathop{\sum_{k,\tilde{l},p \in \mathbb{Z}_{\geq 1},\, k \tilde{l} p = r}}_{\delta = f_{k{\tilde l},p}(\xi)} k^{-\ell/2} p^{1+\ell/2} \tilde{l}^{1-\ell/2} 
 \mathop{\sum_{1\leq \tilde{s} \leq {\tilde l}-1}}_{(\tilde{s},\tilde{l})=1} \sum_{\gamma \in G_{\tilde{l}}}  
e( - (\xi_{\tilde{l}}, \gamma)_{\tilde{l}}-\tilde{s} \theta_{\tilde{l}}(\gamma ) ) \Psi_{\gamma_1}\left( \frac{k \tau + p \tilde{s}}{p \tilde{l}} \right).
\end{align}
Finally, using again the exact sequence 
\[ 
0 \to \Lambda/ {\tilde l} \Lambda \to G_{\tilde l} \to G_1 \to 0,
\]
note that 
\[ 
\sum_{\substack{\gamma\in G_{\tilde l}\\ \gamma_1 = \rho}} 
e( - (\xi_{\tilde{l}}, \gamma)_{\tilde{l}}-\tilde{s} \theta_{\tilde{l}}(\gamma ) ) = 
\sum_{v \in \Lambda/ {\tilde l}\Lambda} 
e\left(-\frac{\tilde{s}}{2}(\rho_0+v, \rho_0+v)_{\tilde l} - (\rho_0+v, \xi_{\tilde l})_{\tilde l}\right).
\]
where $\rho_0\in G_{\tilde l}$ is an arbitrary lift of $\rho\in G_1$.
Therefore \[
\sum_{\substack{\gamma\in G_{\tilde l}\\ \gamma_1 = \rho}} 
e( - (\xi_{\tilde{l}}, \gamma)_{\tilde{l}}-\tilde{s} \theta_{\tilde{l}}(\gamma ) ) = 
 h_{{\tilde l}, -{\tilde s}, \xi_{\tilde l}}(\rho_0) 
\]
where $h_{{\tilde l}, -{\tilde s}, \xi_{\tilde l}}: G_{\tilde l} \to \IC$ is the function defined in 
\eqref{eq:hfunction}, which descends to the  function 
${\bar h}_{{\tilde l}, -{\tilde s}, \xi_{\tilde l}} : G_1 \to \IC$. 
Hence one obtains 
\[ 
\sum_{\substack{\gamma\in G_{\tilde l}\\ \gamma_1 = \rho}} 
e( - (\xi_{\tilde{l}}, \gamma)_{\tilde{l}}-\tilde{s} \theta_{\tilde{l}}(\gamma ) ) = 
 {\bar h}_{{\tilde l}, -{\tilde s}, \xi_{\tilde l}}(\rho),
\]
independent of the choice of $\rho_0$. In conclusion the final 
expression for $Q_{\geq 1}(\tau)$ is 
\begin{align}
Q_{\geq 1}(\tau) = \mathop{\sum_{k,\tilde{l},p \in \mathbb{Z}_{\geq 1},\, k \tilde{l} p = r}}_{\delta = f_{k{\tilde l},p}(\xi)} k^{-\ell/2} p^{1+\ell/2} \tilde{l}^{1-\ell/2} \mathop{\sum_{1\leq \tilde{s}\leq {\tilde l}-1}}_{(\tilde{s},\tilde{l})=1} \sum_{\rho \in G_1} {\bar h}_{{\tilde l}, -{\tilde s}, \xi_{\tilde l}}(\rho) \Psi_{\rho}\left( \frac{k \tau + p \tilde{s}}{p \tilde{l}} \right).
\end{align}
Comparing with \eqref{eq:Sgeqone}, note that exchanging the summation variables $p \leftrightarrow k$, $s' \leftrightarrow \tilde{s}$, and the symbols $\xi \leftrightarrow \eta$ yields 
\begin{equation}
Q_{\geq 1 }(\tau) = S_{\geq 1}(\tau),
\end{equation}
concluding the proof. 

\appendix

\section{Weil representations and sublattices of the K3 lattice}\label{Weilreps} 

This section contains some basic facts on Weil representations with applications to sublattices of the 
middle cohomology lattice $H^2(S,\IZ)$ of a smooth generic K3 surface $S$. 

First recall the definition of the Weil representation 
associated to any lattice $\Upsilon$ equipped with 
an integral even nondegenerate symmetric bilinear form
$({}\ ,\ )_{\Upsilon}$ of signature $(b_+,b_-)$. 
Let $\kappa \in \IZ$ denote the absolute value of the 
determinant of the bilinear form and let $\Upsilon^\vee = 
{\rm Hom}(\Upsilon, \IZ)$ be the dual lattice. For brevity 
let $\Upsilon_\IQ = \Upsilon \otimes_\IZ \IQ$. Moreover 
the natural extension of $({}\ ,\ )_\Upsilon$ to $\Upsilon_\IQ$ by $\IQ$-linearity will be also denoted by 
$({}\ ,\ )_\Upsilon$. The distinction will be clear from the 
context. 

The given bilinear 
form determines a lattice embedding $\Upsilon\hookrightarrow \Upsilon^\vee$ such that 
$Q=\Upsilon^\vee/\Upsilon$ is a finite abelian group 
of rank $\kappa$. Note that the bilinear form 
$({}\ ,\ )_{\Upsilon}$ extends by $\IQ$-linearity to 
$\Upsilon^\vee \subset \Upsilon_\IQ$, hence 
it descends to a  $\IQ/\IZ$-valued bilinear pairing 
$(\ ,\ )_Q$ 
on $Q$. For any two equivalence classes $\delta_1, \delta_2\in  Q$ one has 
\[ 
(\delta_1, \delta_2)_Q = (\xi_1, \xi_2)_\Upsilon
\quad {\rm mod}\ \IZ,
\]
where $\xi_1,\xi_2\in \Upsilon^\vee$ are arbitrary representatives of $\delta_1, \delta_2$ respectively. 

According to \cite{Weil_rep} 
the pair $\Upsilon$, $({}\ ,\ )_{\Upsilon}$
determines a unitary symmetric representation of the metaplectic group $Mp(2,\IZ)$. 
For completeness, recall that in order to define the metaplectic group one has to pick a complex number $\tau$ with ${\rm Im}(\tau)>0$. 
Then  $Mp(2,\IZ)$ is the  double cover of $SL(2, \IZ)$ 
consisting of pairs 
\[
(\sigma, \sqrt{c\tau + d}), \qquad \sigma = \left(\begin{array}{cc} 
a & b \\ c & d \end{array}\right)\in SL(2, \IZ).
\]
This group is generated by 
\[
\wT = (T, 1), \qquad \wS = (S, \sqrt{\tau}),
\]
satisfying the relations $(\wS)^2 = (\wS\wT)^3 = (-I_2, i)$, 
which are independent of the choice of $\tau$. 

Let $\IC[\Upsilon^\vee/\Upsilon]$ be the complex 
vector space generated by the basis elements $e_{\delta}$ 
with $\delta \in Q$. 
Then the pair $\Upsilon$, $({}\ ,\ )_{\Upsilon}$ determines a representation $\rho_\Upsilon: Mp(2, \IZ) 
\to {\rm End}(\IC[\Upsilon^\vee/\Upsilon])$ 
given by 
\be\label{eq:metarepA} 
\rho_\Upsilon(\wT)(e_\delta) = e^{2\pi i(\delta,\delta)_Q/2}
e_\delta, \qquad 
\rho_\Lambda(\wS)(e_\delta) = {e^{\pi i (b_- - b_+) /4}\over \sqrt{\kappa}} 
\sum_{\delta'\in Q} e^{-2\pi i (\delta, \delta')_Q} e_{\delta'}.
\ee
This representation is unitary symmetric, which implies that the dual representation $\rho_\Upsilon^*$ is 
isomorphic to its complex conjugate.

Now let $\Lambda_{K3}$ be the middle cohomology lattice 
of a smooth generic K3 surface. This lattice has rank 22 
and is equipped with the natural intersection form 
$(\ ,\ )_{K3}$ which is an integral even nondegenerate symmetric bilinear form of determinant 1. Let $\Lambda\subset \Lambda_{K3}$ be a rank $1\leq \ell 
\leq 20$ sublattice such that the intersection form restricts to 
a nondegenerate symmetric bilinear form $(\ ,\ )_\Lambda$ on $\Lambda$ of signature $(1, \ell-1)$. Let 
$\Lambda^\perp\subset \Lambda_{K3}$ denote the sublattice consisting of all elements $u \in \Lambda_{K3}$, $u\cdot \Lambda=0$, and $(\ ,\ )_{\Lambda^\perp}$ denote the induced nondegenerate symmetric bilinear form on $\Lambda^\perp$. The later has signature $(2, 20-\ell)$. Then the lattices $\Lambda$, $\Lambda^\perp$ equipped with the induced bilinear 
forms determine the Weil representations $\rho_\Lambda$, $\rho_{\Lambda^\perp}$. Below it will 
be shown that there is an isomorphism 
\be\label{eq:Weilrepisom} 
\rho^*_{\Lambda^\perp}  \simeq \rho_\Lambda
\ee
of representations of the metaplectic group. 

Let $G = \Lambda^\vee/\Lambda$, $G_\perp = (\Lambda^\perp)^\vee/\Lambda^\perp.$
It suffices to prove that there is an isomorphism 
of finite abelian groups
$f: G_\perp {\buildrel \sim \over \longto} G$ 
such that 
\be\label{eq:zerosumpairing}
(\xi_1, \xi_2)_{G_\perp} + (f(\xi_1), f(\xi_2))_{G} =0 
\ee
in $\IQ/\IZ$ for any $\xi_1, \xi_2\in G_\perp$. 
The starting point is the observation that the natural projections 
\[
p: \big(\Lambda_{K3}\big)_\IQ\twoheadrightarrow 
\Lambda_\IQ
\qquad 
p_\perp: \big(\Lambda_{K3}\big)_\IQ\twoheadrightarrow \Lambda^\perp_\IQ
\]
with respect to the intersection form determine lattice isomorphisms 
\[ 
\phi: \Lambda_{K3}/\Lambda^\perp\, {\buildrel \sim \over \longto} \, \Lambda^\vee, \qquad 
\phi_\perp: \Lambda_{K3}/\Lambda \, {\buildrel \sim \over \longto} \, (\Lambda^\perp)^\vee.
\]
This follows easily since the intersection form on $\Lambda_{K3}$ is unimodular. Then one further obtains 
isomorphisms of finite abelian groups 
\[
{\bar \phi}: \Lambda_{K3}/(\Lambda\oplus \Lambda^\perp) \to G, \qquad 
{\bar \psi}: \Lambda_{K3}/(\Lambda\oplus \Lambda^\perp) \to G_\perp. 
\]
In particular there is an isomorphism $f={\bar \phi}\circ 
{\bar \psi}^{-1} : G_\perp\to G$. 

Next let $u_1, u_2\in \Lambda_{K3}$ be arbitrary elements. 
Note that 
\[ 
(p(u_i), p(u_i))_{\Lambda} = (p(u_i), p(u_i))_{{K3}}, 
\qquad 
(p_\perp(u_i), p_{\perp}(u_i))_{\Lambda^\perp} 
= (p_\perp(u_i), p_\perp(u_i))_{{K3}}
\]
 for $i\in \{1,2\}$.
Since 
\[
 \big(\Lambda_{K3}\big)_\IQ= \Lambda_\IQ \oplus \Lambda^\perp_\IQ. 
\]
one has $u_i= p(u_i) + p_\perp(u_i)$, for $i\in \{1,2\}$,
 which yields 
\[
(p(u_1), p(u_2))_{\Lambda} + (p_\perp(u_1), p_{\perp}(u_2))_{\Lambda^\perp} = (u_1,u_2)_{K3} \in \IZ. 
\]
This implies relation \eqref{eq:zerosumpairing}.

\section{Shintani's formula}\label{Shintani}

This section is a brief review of Shintani's explicit formula
\cite[Prop. 1.6]{half_integral} for the matrix elements of the Weil representation. As in Section \ref{modularsect}, 
consider a lattice $\Lambda$ equipped with an integral 
even nondegenerate symmetric bilinear form. Let $\ell$ denote the rank of $\Lambda$ and $m$ denote the absolute value of the  determinant of the given bilinear form.  As explained in the previous section, 
such a lattice determines a representation 
\[
\rho_\Lambda : Mp(2, \IZ) \to {\rm End}(\IC[G_1]) 
\]
where $Mp(2,\IZ)$ is the metaplectic double cover of 
$SL(2, \IZ)$ and $G_1 = \Lambda^\vee/ \Lambda$. 
The action of the generators of the metaplectic group is given in \eqref{eq:metarepA}. Shintani's result 
\cite[Prop. 1.6]{half_integral} provides an explicit formula for the matrix 
elements associated to an arbitrary element
\[
{\tilde \sigma} = (\sigma, \sqrt{c\tau+d}), \qquad 
\sigma = \left(\begin{array}{cc} a& b \\ c& d\\ \end{array} 
\right)
\]
of the metaplectic group with $c>0$. In order to 
write this formula in detail consider first the function 
\[
{ f_{c,a,d}}: \Lambda^\vee/ c\Lambda \times \Lambda^\vee/c\Lambda \times \Lambda/ c\Lambda \to 
\IC
\]
\[
{f_{c,a,d}}(\gamma, \zeta,  u) = 
e \left({a\over 2} (\gamma+u, \gamma+u)_c - (\gamma+u,\zeta)_c 
+ {d\over 2}(\zeta, \zeta)_c\right)
\]
Next define a function 
\[ 
g_{c,a,d}: \Lambda^\vee/ c\Lambda \times \Lambda^\vee/c\Lambda \to \IC 
\]
\be\label{eq:gfunct}
g_{c,a,d}(\gamma, \zeta) = \sum_{u \in \Lambda/c\Lambda} 
{ f}_{c,a,d}(\gamma, \zeta,  u). 
\ee
Then one checks as follows that the function $g_{a,c,d}$ 
satisfies the invariance conditions 
\[ 
g_{c,a,d}(\gamma+ x, \zeta) = g_{c,a,d}(\gamma, \zeta+ y) = g_{c,a,d}(\gamma, \zeta) 
\]
for any $(\gamma, \zeta) \in \Lambda^\vee/ c\Lambda \times \Lambda^\vee/c\Lambda$ and any 
$(x,y) \in \Lambda/c\Lambda\times \Lambda/c\Lambda$. 

Invariance under $\gamma \mapsto \gamma+x$ is  easily proven by changing the summation variable $u$ to 
$u-x$ in the right hand side of \eqref{eq:gfunct}. 

Invariance under $\zeta \mapsto \zeta + y$ requires 
more work. Recall that the pairing $({}\ ,\ {})_c$ takes integral 
values on $\Lambda^\vee/c\Lambda \times \Lambda/c\Lambda$ 
and restricts to an even integral form on  $\Lambda/c\Lambda \times \Lambda/c\Lambda$. Then, using 
the 
basic relation $ad-bc=1$,
one has the folllowing sequence of congruences mod $2\IZ$: 
\be\label{eq:congruences}
\bal 
& a (\gamma+u, \gamma+u)_c - 2(\gamma+u,\zeta+y)_c 
+ d(\zeta+y, \zeta+y)_c \equiv \\
& a (\gamma+u, \gamma+u)_c 
-2ad (\gamma+u, y)_c + ad^2 (y,y)_c 
-2(\gamma+u-dy,\zeta)_c 
+ d(\zeta, \zeta)_c \equiv \\
& a(\gamma+u-dy, \gamma+u-dy)_c -2(\gamma+u-dy,\zeta)_c 
+ d(\zeta, \zeta)_c. \\
\eal 
\ee
Now invariance reduces again to a shift $u\mapsto u+dy$ of the summation variable in \eqref{eq:gfunct}. 

In conclusion $g_{a,c,d}$ descends to a function
\[
{\bar g}_{c,a,d}: \Lambda^\vee/\Lambda \times \Lambda^\vee/\Lambda \to \IC
\]
Then Shintani's formula reads 
\[ 
\rho_\Lambda({\tilde \sigma})_{\delta_1, \delta_2} 
= {e\left({\ell-2\over 8}\right)\over c^{\ell/2} m^{1/2}}\, {\bar g}_{c,a,d}(\delta_1, \delta_2). 
\]

A slightly different presentation of Shintani's formula will 
be needed 
for the modularity proof in Section \ref{proof}. 
Namely let 
\[
f'_{c,a,d} : \Lambda^\vee/c\Lambda \times \Lambda^\vee/c\Lambda 
\times \Lambda/c\Lambda \to \IC
\]
be the function defined by 
\[
{f'_{c,a,d}}(\gamma, \zeta,  u) = 
e \left({a\over 2} (\gamma, \gamma)_c - (\gamma,\zeta+u)_c 
+ {d\over 2}(\zeta+u, \zeta+u)_c\right)
\]
Let 
\[ 
h_{c,a,d}: \Lambda^\vee/ c\Lambda \times \Lambda^\vee/c\Lambda \to \IC 
\]
be defined by 
\be\label{eq:gfunct}
h_{c,a,d}(\gamma, \zeta) = \sum_{u \in \Lambda/c\Lambda} 
{ f'_{c,a,d}}(\gamma, \zeta,  u). 
\ee
Then by analogy with the above argument for $g_{c,a,d}$, the function 
$h_{c,a,d}$ also descends to a function
\[
{\bar h}_{c,a,d} : \Lambda^\vee/\Lambda \times \Lambda^\vee/\Lambda \to \IC.
\]
Furthermore it will be shown below that 
\[ 
h_{c,a,d}( \gamma, \zeta) = g_{c,a,d}(\gamma, \zeta) 
\]
for any $(\gamma, \zeta)$. 

By analogy with equation 
\eqref{eq:congruences} one has 
\[
\bal 
& a (\gamma, \gamma)_c - 2(\gamma,\zeta+u)_c 
+ d(\zeta+u, \zeta+u)_c \equiv a (\gamma-du, \gamma-du)_c - 2(\gamma-du,\zeta)_c 
+ d(\zeta, \zeta)_c \quad {\rm mod}\ 2\IZ.
\eal 
\]
Hence 
\[
h_{c,a,d}( \gamma, \zeta) = \sum_{u\in \Lambda/ c\Lambda} 
f_{c,a,d}(\gamma, \zeta, - du). 
\]
Now note that $c,d$ are coprime since $ad-bc =1$. 
Since $\Lambda/ c\Lambda \simeq \left(\IZ/c\IZ\right)^{\times \ell}$, 
this implies that multiplication by $d$ gives an isomorphism 
$\Lambda/c\Lambda {\buildrel d\over \longto} 
\Lambda/ c \Lambda$. Therefore 
\[
\sum_{u\in \Lambda/ c\Lambda} 
f_{c,a,d}(\gamma, \zeta, - du) = \sum_{u\in \Lambda/ c\Lambda} 
f_{c,a,d}(\gamma, \zeta,u),
\]
which proves the claim. 

Finally, note that a closely related function 
${\bar h}_{c,d,\gamma} : \Lambda^\vee/\Lambda \to \IC$
is also used in Section \ref{proof} . For any 
 fixed element $\gamma \in \Lambda^\vee/c\Lambda$, define 
\[
h_{c,d,\gamma}: \Lambda^\vee/c\Lambda \to \IC 
\]
by
\[
h_{c,d,\gamma}(\zeta) = \sum_{u \in \Lambda/c\Lambda} 
e\left( {d\over 2}(\zeta+u, \zeta+u)_c -(\zeta+u, \gamma)_c\right)
\]
Then it follows again that $h_{c,d,\gamma}(\zeta+y) = h_{c,d,\gamma}(\zeta)$ for any $\zeta \in \Lambda^\vee/c\Lambda$, $y \in \Lambda/c\Lambda$. Hence $h_{c,d,\gamma}$ 
descends to a function 
${\bar h}_{c,d,\gamma} : \Lambda^\vee/\Lambda \to \IC$.

\bibliography{FM_ref.bib}
\bibliographystyle{abbrv}
\end{document}